\documentclass[journal]{IEEEtran}

\IEEEoverridecommandlockouts

\makeatletter
\def\markboth#1#2{\def\leftmark{\@IEEEcompsoconly{\sffamily}\MakeUppercase{\protect#1}}%
\def\rightmark{\@IEEEcompsoconly{\sffamily}\MakeUppercase{\protect#2}}}
\makeatother

\usepackage[english]{babel}
\usepackage{color}
\usepackage{lipsum}
\usepackage{cite}
\usepackage[pdftex]{graphicx}
\usepackage{amsmath}
\usepackage{amsfonts}
\usepackage{array}
\usepackage{verbatim}
\usepackage{listings}
\usepackage{url}
\usepackage{enumerate}
\usepackage{multirow}

\usepackage{epsfig}
\usepackage{epstopdf}
\usepackage{multicol}
\usepackage[font=footnotesize]{caption}
\usepackage[font=scriptsize]{subcaption}
\usepackage{tikz}
\usepackage{pgfplots}
\usepackage{tikz}
\usepackage{pgfplots}
\pgfplotsset{compat=newest}
\pgfplotsset{plot coordinates/math parser=false}
\newlength\fheight
\newlength\fwidth
\usetikzlibrary{patterns,decorations.pathreplacing,backgrounds,calc}
\definecolor{SchoolColor}{RGB}{0.71, 0, 0.106}
\definecolor{chaptergrey}{rgb}{0.61, 0, 0.09} 
\definecolor{midgrey}{rgb}{0.4, 0.4, 0.4}
\definecolor{chaptergreen}{rgb}{0.09, 0.612, 0}
\definecolor{chapterpurple}{rgb}{0.522, 0, 0.612}
\definecolor{chapterlightgreen}{rgb}{0, 0.612, 0.522}

\usepackage{algorithm}
\usepackage[noend]{algpseudocode}

\usepackage{lscape}
\usepackage{setspace}
\usepackage{textcomp}
\usepackage{dsfont}
\addto\captionsenglish{}

\renewcommand{\arraystretch}{2}

\newcommand{\bi}{\begin{itemize}}
\newcommand{\ei}{\end{itemize}}
\newcommand{\be}{\begin{equation}}
\newcommand{\ee}{\end{equation}}

\def\beq{\begin{equation}}
\def\eeq{\end{equation}}
\def\beqa{\begin{eqnarray}}
\def\eeqa{\end{eqnarray}}
\def\beqan{\begin{eqnarray*}}
\def\eeqan{\end{eqnarray*}}

\usepackage{threeparttable}
\usepackage{tabularx}
   \usepackage{multirow}
   \usepackage{booktabs}

   \usepackage{array, blindtext}
   \usepackage{wrapfig}
\usepackage{pdfpages}

\usepackage{etoolbox}
\usepackage{etex}

\usepackage[acronyms,nonumberlist,nopostdot,nomain,nogroupskip]{glossaries}
\usepackage{glossary-mcols}
\makeglossaries

\begin{document}

\newacronym{lsm}{LSM}{Link-to-System Mapping}
\newacronym{cb}{CB}{Code Block}
\newacronym{bler}{BLER}{Block Error Rate}
\newacronym{miesm}{MIESM}{Mutual Information Based Effective SINR}
\newacronym[plural=MMEs,firstplural=Mobility Management Entities (MMEs)]{mme}{MME}{Mobility Management Entity}
\newacronym{ran}{RAN}{Radio Access Network}
\newacronym{cn}{CN}{Core Network}
\newacronym{m2m}{M2M}{Machine to Machine}
\newacronym{nfv}{NFV}{Network Function Virtualization}
\newacronym{vm}{VM}{Virtual Machine}
\newacronym{son}{SON}{Self-Organizing Network}
\newacronym{iot}{IoT}{Internet of Things}
\newacronym{rss}{RSS}{Received Signal Strength}
\newacronym{ttt}{TTT}{Time-to-Trigger}
\newacronym{enb}{eNB}{evolved Node Base}
\newacronym{scoot}{SCOOT}{Split Cycle Offset Optimization Technique}
\newacronym{utc}{UTC}{Urban Traffic Control}
\newacronym{tfl}{TfL}{Transport for London}
\newacronym{ue}{UE}{User Equipment}
\newacronym{hetnet}{HetNet}{Heterogeneous Network}
\newacronym{snr}{SNR}{Signal to Noise Ratio}
\newacronym{sinr}{SINR}{Signal to Interference plus Noise Ratio}
\newacronym{los}{LOS}{Line of Sight}
\newacronym{nlos}{NLOS}{Non Line of Sight}
\newacronym{mec}{MEC}{Mobile Edge Cloud}
\newacronym{cc}{CC}{Congestion Control}
\newacronym{aqm}{AQM}{Active Queue Management}
\newacronym{ecn}{ECN}{Explicit Congestion Notification}
\newacronym{sack}{SACK}{Selective Acknowledgment}
\newacronym{aimd}{AIMD}{Additive Increase Multiplicative Decrease}
\newacronym{bdp}{BDP}{Bandwidth-Delay Product}
\newacronym{rtt}{RTT}{Round Trip Time}
\newacronym{rlc}{RLC}{Radio Link Control}
\newacronym{mss}{MSS}{Maximum Segment Size}
\newacronym{mtu}{MTU}{Maximum Transmission Unit}
\newacronym{rr}{RR}{Round Robin}
\newacronym{cdf}{CDF}{Cumulative Distribution Function}
\newacronym{harq}{HARQ}{Hybrid Automatic Repeat reQuest}
\newacronym[firstplural=Radio Access Technologies (RATs)]{rat}{RAT}{Radio Access Technology}
\newacronym{hh}{HH}{Hard Handover}
\newacronym{dc}{DC}{Dual Connectivity}
\newacronym{am}{AM}{Acknowledged Mode}
\newacronym{um}{UM}{Unacknowledged Mode}
\newacronym{pgw}{PGW}{Packet Gateway}
\newacronym{rs}{RS}{Remote Server}
\newacronym{es}{ES}{Edge Server}
\newacronym{5g}{5G}{5th generation}
\newacronym{phy}{PHY}{Physical}
\newacronym{mac}{MAC}{Medium Access Control}
\newacronym{ofdm}{OFDM}{Orthogonal Frequency Division Multiplexing}
\newacronym{lte}{LTE}{Long Term Evolution}
\newacronym{nr}{NR}{New Radio}
\newacronym{e2e}{E2E}{End-to-End}
\newacronym{sap}{SAP}{Service Access Point}
\newacronym{epc}{EPC}{Evolved Packet Core}
\newacronym{3gpp}{3GPP}{3rd Generation Partnership Project}
\newacronym{mimo}{MIMO}{Multiple Input, Multiple Output}
\newacronym{tti}{TTI}{Transmission Time Interval}
\newacronym{fdma}{FDMA}{Frequency Division Multiple Access}
\newacronym{tdma}{TDMA}{Time Division Multiple Access}
\newacronym{sdma}{SDMA}{Spatial Division Multiple Access}
\newacronym{tcp}{TCP}{Transmission Control Protocol}
\newacronym{dce}{DCE}{Direct Code Execution}
\newacronym{uml}{UML}{Unified Modeling Language}
\newacronym{tm}{TM}{Transparent Mode}
\newacronym{sm}{SM}{Saturation Mode}
\newacronym{rrc}{RRC}{Radio Resource Control}
\newacronym{dl}{DL}{Downlink}
\newacronym{ul}{UL}{Uplink}
\newacronym{tdd}{TDD}{Time Division Duplexing}
\newacronym{mi}{MI}{Mutual Information}
\newacronym{tx}{TX}{Transmitter}
\newacronym{rx}{RX}{Receiver}
\newacronym{bf}{BF}{Beamforming}
\newacronym{tb}{TB}{Transport Block}
\newacronym{pdcch}{PDCCH}{Physical Downlink Control Channel}
\newacronym{pucch}{PUCCH}{Physical Uplink Control Channel}
\newacronym{pusch}{PUSCH}{Physical Uplink Shared Channel}
\newacronym{pdsch}{PDSCH}{Physical Downlink Shared Channel}
\newacronym{dci}{DCI}{Downlink Control Information}
\newacronym{cqi}{CQI}{Channel Quality Information}
\newacronym{mcs}{MCS}{Modulation and Coding Scheme}
\newacronym{amc}{AMC}{Adaptive Modulation and Coding}
\newacronym{pf}{PF}{Proportional Fair}
\newacronym{edf}{EDF}{Earliest Deadline First}
\newacronym{mr}{MR}{Maximum Rate}
\newacronym{hol}{HOL}{Head-of-Line}
\newacronym{awgn}{AGWN}{Additive White Gaussian Noise}
\newacronym{pdcp}{PDCP}{Packet Data Convergence Protocol}
\newacronym{red}{RED}{Random Early Detection}
\newacronym{pdu}{PDU}{Packet Data Unit}
\newacronym{fs}{FS}{Fast Switching}
\newacronym{sch}{SCH}{Secondary Cell Handover}
\newacronym{dmr}{DMR}{Deadline Miss Ratio}
\newacronym{mtd}{MTD}{Machine-Type Device}
\newacronym{ftp}{FTP}{File Transfer Protocol}
\newacronym{rw}{RW}{Receive Window}
\newacronym{mptcp}{MPTCP}{Multipath TCP}
\newacronym{balia}{BALIA}{Balanced Link Adaptation}
\newacronym{pss}{PSS}{Primary Synchronization Signal}
\newacronym{bsr}{BSR}{Buffer Status Report}

\flushbottom
\setlength{\parskip}{0ex plus0.1ex}

\bstctlcite{IEEEexample:BSTcontrol}

\title{End-to-End Simulation of 5G mmWave Networks}

\author{\IEEEauthorblockN{{Marco Mezzavilla}, \textit{Member, IEEE}, {Menglei Zhang}, {Michele Polese}, \textit{Student Member, IEEE}, {Russell Ford}, \\
 {Sourjya Dutta}, \textit{Student Member, IEEE}, {Sundeep Rangan}, \textit{Fellow, IEEE}, {Michele Zorzi}, \textit{Fellow, IEEE}}
\thanks{Marco Mezzavilla, Menglei Zhang, Russell Ford, Sourjya Dutta and Sundeep Rangan are with NYU WIRELESS, NYU Tandon School of Engineering, Brooklyn, NY, USA (email: \{mezzavilla, russell.ford, menglei, sdutta, srangan\}@nyu.edu).}
\thanks{Michele Polese and Michele Zorzi are with the Department of Information Engineering, University of Padova, Padova, Italy (email: \{polesemi, zorzi\}@dei.unipd.it)}
}

\tikzstyle{startstop} = [rectangle, rounded corners, minimum width=2cm, minimum height=0.5cm,text centered, draw=black]
\tikzstyle{io} = [trapezium, trapezium left angle=70, trapezium right angle=110, minimum width=3cm, minimum height=1cm, text centered, draw=black]
\tikzstyle{process} = [rectangle, minimum width=2cm, minimum height=0.5cm, text centered, draw=black, align=center]
\tikzstyle{decision} = [ellipse, minimum width=2cm, minimum height=1cm, text centered, draw=black]
\tikzstyle{arrow} = [thick,<->,>=stealth]
\tikzstyle{line} = [thick,>=stealth]
\tikzstyle{darrow} = [thick,<->,>=stealth,dashed]
\tikzstyle{sarrow} = [thick,->,>=stealth]

\makeatletter
\patchcmd{\@maketitle}
{\addvspace{0.5\baselineskip}\egroup}
{\addvspace{-1\baselineskip}\egroup}
{}
{}
\makeatother

\maketitle
\begin{abstract}
Due to its potential for multi-gigabit and low latency wireless links, 
millimeter wave (mmWave) technology is expected to play a central role 
in \gls{5g} cellular systems. 
While there has been considerable progress in understanding the mmWave physical layer,
innovations will be required at all layers of the protocol stack, 
in both the access and the core network. Discrete-event network simulation is essential for end-to-end, cross-layer research and development. This paper provides a tutorial on a recently developed full-stack mmWave module integrated into the widely used open-source ns--3 simulator.
The module includes a number of detailed statistical channel models as well as
the ability to incorporate real measurements or ray-tracing data. 
The \gls{phy} and \gls{mac} layers are modular and highly customizable, making it easy to integrate algorithms or compare \gls{ofdm} numerologies, for example. The module is interfaced with the core network of the ns--3 \gls{lte} module for full-stack simulations of end-to-end connectivity, and advanced architectural features, such as dual-connectivity, are also available. To facilitate the understanding of the module, and verify its correct functioning, we provide several examples that show the performance of the custom mmWave stack as well as custom congestion control algorithms designed specifically for efficient utilization of the mmWave channel.
\end{abstract}

\begin{picture}(10, -350)(10,-360)
\put(0,0){
\put(0,0){\scriptsize This work has been submitted to IEEE Communication Surveys and Tutorials for possible publication.} 
\put(0,-10){\scriptsize Copyright may be transferred without notice, after which this version may no longer be accessible.}}
\end{picture}


\section{Introduction}

Millimeter Wave (mmWave) communications are emerging as a central technology in 
\gls{5g} cellular wireless systems due to their potential 
to achieve the massive throughputs required by future networks
~\cite{KhanPi:11-CommMag,rappaportmillimeter,RanRapE:14,BocHLMP:14,Rappaport2014-mmwbook}. 
In particular, mmWave has become a key focus of the \gls{3gpp} \gls{nr} effort 
currently under development~\cite{3GPP38913}.
Due to the unique propagation characteristics of mmWave signals and the need to transmit in beams with much greater directionality than previously used in cellular systems, much of the recent work in mmWave communications
has focused on channel modeling, beamforming and other physical layer procedures.
However, the design of \gls{e2e} cellular systems that can fully exploit the 
high-throughput, low-latency capabilities of mmWave links will 
require innovations not only at the physical layer, but also across all layers of the communication protocol stack.
For mmWave systems, \gls{e2e} design and analysis are at a much earlier stage of research
\cite{shokri2015millimeter,niu2015survey,zhang2016transport}.

Discrete-event network simulators are fundamental and widely used tools for developing 
new protocols and analyzing complex networks. 
Importantly, most network simulators 
enable \textit{full-stack simulation}, meaning that they model all layers of the protocol
stack as well as applications running over the network. This full-stack capability will 
play a critical role in the development of \gls{5g} mmWave systems.
The unique characteristics of the underlying mmWave channel have wide ranging effects
throughout the protocol stack. For example, the use of highly directional beams increases the complexity of  
a number of basic \gls{mac}-layer procedures such as 
synchronization, control signaling, cell search and initial access, which in turn affect 
delay and robustness~\cite{shokri2015millimeter}.  
MmWave signals are also highly susceptible to blockage~\cite{Allen:94,singh2007millimeter,KhanPi:11-CommMag,LuSCP:12}, which 
results in high variability of the channel quality.  This erratic behavior complicates the design of rate adaptation algorithms and
signaling procedures, requiring advanced solutions for multi-connectivity, fast handover and 
connection re-establishment~\cite{ghosh2014millimeter,poleseHo,giordaniMC2016,tesemaMC2015}.
New transport layer mechanisms may also be required 
in order to utilize the large capacity, when available, and to react promptly to rapid fading to avoid congestion~\cite{zhang2016transport,mengleiBufferbloat,polese2017mptcp,polese2017tcp}.
The need for ultra-low latency applications~\cite{ford2016achieving,KhanPi:11-CommMag,ford2017provisioning} may 
require solutions based on edge computing and distributed architectures that will determine a considerable departure from current cellular core network designs.

%

To better capture these design challenges, this work presents a comprehensive tutorial on the open-source mmWave simulation tool developed by New York University and the University of Padova for \gls{lte}-like \gls{5g} mmWave cellular networks, which can be used to evaluate cross-layer and end-to-end performance.
This mmWave simulation tool is developed as a new module within the widely used ns--3 network simulator \cite{henderson2008network}. ns--3 is an open-source platform,
that currently implements a wide range of protocols in C++,
making it useful for cross-layer design and analysis.
The new mmWave module presented here 
is based on the architecture and design patterns of the \gls{lte} LENA module \cite{lena,ns3Lena} and implements all the necessary \glspl{sap} needed to leverage the robust suite of \gls{lte}/\gls{epc} protocols provided by LENA. The code (publicly available at GitHub \cite{github-ns3-mmwave}, along with examples and test configurations) is highly modular and customizable to facilitate researchers to design and test novel \gls{5g} protocols.

The ns--3 mmWave module was first presented in~\cite{MezzavillaNs3:15,ford2016framework}. The \gls{3gpp} channel model implementation is introduced in~\cite{zhang20173gpp}, and the dual connectivity functionality is described in~\cite{simutoolsPolese,poleseHo}. This paper extends these works by presenting the ns--3 mmWave module from a single and organic point of view, and is intended as a tutorial for any researcher that plans to use the simulator. In addition to its comprehensive description and discussion, we provide in Sec.~\ref{sec:results} a brief guide on how to set up a simulation, followed by a number of representative examples. 

The rest of the paper is organized as follows. In Section~\ref{sec:mmwave_background}, we provide some background on mmWave cellular communications and highlight some key problems at the higher protocol layers to motivate the need for a robust full-stack simulator. We also describe the main challenges related to the design of a mmWave cellular networks simulator. Then, in Section~\ref{sec:ns3}, we introduce ns-3, the network simulator on which our mmWave module is developed, and in Section~\ref{sec:module} we present the overall architecture of the mmWave module. We then take a closer look at each component, starting with the suite of \gls{mimo} channel models in Section~\ref{sec:channel_model}. In addition to an implementation of the latest \gls{3gpp} ``above 6 GHz'' model~\cite{38900}, several custom channel models are also provided. Section~\ref{sec:phy} discusses the features of the \gls{ofdm}-based \gls{phy} layer, which has a customizable frame structure for evaluating different numerologies and parameters. In Section~\ref{sec:mac}, we provide a \gls{mac}-layer discussion that includes our proposed flexible/variable \gls{tti} \gls{tdma} \gls{mac} scheme, which is supported by several scheduler implementations. Section~\ref{sec:rlc} presents the enhancements that we introduced to the \gls{lte} \gls{rlc} layer. The dual-connectivity architecture is reported in Sec.~\ref{sec:dc}. In Section \ref{sec:results}, we show how the module can be used for cross-layer evaluation of multi-user cellular networks through a number of representative examples, and provide pointers to a large set of general results that have been obtained so far with this module. The integration of native Linux \gls{tcp} implementations, performed through the ns--3 \gls{dce} framework, is discussed in Section \ref{sec:dce}. In Section~\ref{sec:uses}, we provide details on our future plans for the simulator and suggest possible research topics in which it could be used. Finally, we conclude this tutorial paper in Section \ref{sec:conclusions}.

\section{Millimeter Wave Cellular Background}
\label{sec:mmwave_background}
Millimeter wave communications is an advanced \gls{phy} layer technology, which has recently come to the forefront of research interest and may be able to rise to the challenge of providing high-rate mobile broadband services, in addition to offering opportunities for reducing over-the-air latency for \gls{5g} New Radio. 


MmWave makes use of the radio frequency spectrum roughly between 30 and 300 GHz, even though the research challenges extend also to lower frequencies (i.e., above 6 GHz) which are considered for \gls{3gpp} \gls{nr}.
Systems that can operate in these bands are attractive
because of the large quantities of available spectrum at these higher frequency ranges and the spatial degrees of
freedom afforded by very
high-dimensional antenna arrays, which are possible thanks to the smaller size of antenna elements at higher frequencies. 
Most current commercial wireless systems operate below 6 GHz, where lower frequencies allow for long-range propagation and low penetration loss (i.e., attenuation by walls and other obstacles), which makes them well-suited for radio communications. As a result, the sub-6 GHz spectrum has become heavily congested and individual bands are generally not available in contiguous chunks wider than 200 MHz. However, large swaths of spectrum are available at the higher mmWave frequencies, which offer the possibility of very wide bandwidths of over 1 GHz, in some cases. 

Although the mmWave bands are already used by a variety of commercial applications, such as satellite and point-to-point backhaul communications, until recently they were considered impractical for mobile access networks
due to the poor isotropic propagation and the vulnerability to shadowing at these higher frequencies.
However, it has now been shown that the limitations of the mmWave channel can be overcome with the help of high-gain, directional antennas so that this vast region of spectrum can now be exploited to provide an order of magnitude or more increase in throughput for mobile devices \cite{RanRapE:14,AkdenizCapacity:14}. 


Directional smart antennas are the major technology enabler that will make it possible for mmWave devices to overcome the poor propagation effects and unlock this high-frequency spectrum. The theoretical free space path loss (as governed by Friis' Equation) is proportional to the square of the frequency, resulting in the magnitude of received power for a mmWave signal being over 30 dB (1000x) less than conventional cellular systems at equivalent distances between transmitter and receiver \cite{Rappaport:02}. 
Multi-element antenna arrays and \gls{mimo} beamforming techniques offer a means of compensating for this high attenuation. With millimeter waves, the antenna size and spacing shrinks to be on the order of millimeters, making it possible to pack hundreds of elements onto a small cell base station and dozens onto a handheld device. Smaller antenna size also allows for multiple arrays to be integrated onto mobile devices to provide diversity and maintain connectivity even if the signal from one array is blocked (for instance, by the user's hand) \cite{RanRapE:14}.


It is clear that mmWave will be highly disruptive in the wireless space thanks to the prospect of massive bandwidth and high-gain antennas. Nevertheless, before mmWave technology can be effectively realized in \gls{5g} networks, there are numerous challenges to be addressed, not only at the physical layer, but also at higher layers of the radio stack, namely:
\begin{itemize}
	\item \textit{Adaptive beamforming and beam tracking:}
	The requirement of directionality introduces new challenges for supporting mobility in mmWave networks. The transmitter and receiver must continually track the channel as the mobile user moves in order to align their antenna arrays to achieve the maximum directional gain. 
	MmWave signals are also known to be particularly susceptible to shadowing and can be completely blocked by many materials such as brick, tinted glass and even the human body \cite{Rappaport:28NYCPenetrationLoss,MacCartneyHuman:16}. Fortunately, recent field measurements have demonstrated that reflected power can be sufficient for \gls{nlos} communications to be possible.
	A blocked link may therefore be able to recover by steering the beam from the primary \gls{los} path to an alternate \gls{nlos} path. The UE and BS must then jointly initiate a procedure to search for and select another path to reestablish the link.
	
	\item \textit{Directional synchronization and broadcast channels:}
	Directionality also complicates the design of many control channels and procedures. The cell discovery and initial access procedures, where the UE must search for nearby base stations to which it can attach, will require an innovative approach to be handled efficiently. Traditional cells periodically broadcast out synchronization signals (known as the \gls{pss} in LTE systems) omnidirectionally, which are received by all devices within the cell's coverage range and used to initially connect to the cell. If a \gls{5g} mmWave \gls{enb} were to broadcast the \gls{pss} in an omnidirectional antenna pattern, the signal would not benefit from the directional gain and might not have adequate range to be detected by many UEs. Therefore, the \gls{enb} and \gls{ue} must perform an angular search in order for users to detect the \gls{pss} and hone in on the optimal \gls{tx}/\gls{rx} beamforming angles \cite{barati2015directional}.
	A similar problem also arises for other control signals, such as the \gls{dci} assignments, which indicate the resources assigned to each user for \gls{dl}/\gls{ul} transmission in the subframe or slot.
		
	\item \textit{Issues for the MAC, Network and Transport Layers:}
	The rapid channel dynamics and vulnerability of mmWave links to shadowing will require frequent, near instantaneous handovers between neighboring 5G or 4G cells. \textit{Dual-connectivity}, where mobiles are continuously connected to both the 5G and legacy 4G network, may therefore be essential to recover from an abrupt failure of the primary 5G link \cite{simutoolsPolese,poleseHo}. Additionally, at the transport layer, the congestion control and avoidance mechanisms provided by \gls{tcp} must be able to quickly adapt to sudden fluctuations in capacity to maximally utilize the link bandwidth while avoiding overwhelming the network by sending too many packets, resulting in congestion and affecting other flows in the network. Current versions of \gls{tcp} may not be optimized for mmWave channel dynamics~\cite{zhang2016transport,polese2017mptcp}, so new algorithms may be called for to provide high rates for \gls{e2e} sessions~\cite{mengleiBufferbloat,azzino2017xtcp,tcp-asilomar}.

\end{itemize}

\subsection*{Potentials and Challenges of System-level Simulations of mmWave Networks}
An \acrlong{e2e} network simulator for mmWave cellular networks is an invaluable tool that can help address these challenges by allowing the evaluation of the impact of the channel and of the \gls{phy} layer technology on the whole protocol stack. 
However, given the characteristics of mmWave communications described in the previous paragraphs, in order to have accurate results it is of paramount importance to model in detail the behavior of the different elements that interact in a cellular system. In the following paragraphs we will introduce and discuss some of the most important elements that need to be considered when designing a mmWave cellular systems simulation, and show how they depend on one another:
\begin{itemize}
	\item The channel model is the fundamental component of every wireless simulation. Given the harsh propagation conditions at mmWaves, the channel is one the main elements that affect the end-to-end network performance. Firstly, it has to account for the different \gls{los} and \gls{nlos} states for the propagation loss and the fading~\cite{38900}. Moreover, beamforming should be applied on top of the channel to accurately model directional transmissions, which have an impact on the link budget, the interference and the control procedures. Finally, the Doppler effect is particularly relevant at mmWave frequencies, especially with high mobility~\cite{RanRapE:14}. An important consideration related to the channel model is the trade off between the accuracy and the computational complexity: very accurate models that require the computation of the complete channel matrix are usually also computationally intensive~\cite{zhang20173gpp}.
	\item The users' mobility and the network deployment have an important impact on the communication performance, intertwined with that of the channel model. Given the small range of the mmWave cells, the deployment will be dense and will require frequent access point updates, which should be simulated for a realistic performance assessment~\cite{tcp-mobicom}. Moreover, mobility affects the performance of beam tracking algorithms~\cite{giordaniMC2016}. Therefore, when simulating a mmWave network it is important to use realistic deployments and mobility models. 
	\item The level of detail when modeling the protocol stack of the mmWave links and of the end devices is another important parameter for network simulations. A simplified model of the protocol stack can be accurate enough for studies that do not involve complex interplay between different layers, but cannot capture the behaviors that emerge from the interaction of the different layer, and therefore could not generate realistic results for end-to-end performance evaluations. For example, at mmWave frequencies, it has been shown that the channel behavior has an impact on the \gls{tcp} performance~\cite{zhang2016transport,polese2017mptcp}, therefore a model of the TCP/IP stack is needed when analyzing the data rate that an application can reach in an end-to-end mmWave network.
\end{itemize}

To the best of our knowledge, there are no open source simulators capable of thoroughly modeling the mmWave channel along with the cellular network protocol stack as well as other protocols (e.g., the \gls{tcp}/IP stack), realistic scenarios and mobility. There exists an ns--3-based simulator for IEEE 802.11ad in the 60 GHz band~\cite{assasa2016implementation,assasa2017extending}, which however cannot be used to simulate cellular and \gls{3gpp}-like scenarios. Other papers~\cite{kim2013tens,zheng2015hetsnets,dehos2014millimeter,taori2015point} report results from system level simulations, with custom (often MATLAB-based and not publicly available) simulators which are not able to capture the complexity of the whole stack with a very high level of detail. This is what motivated us to develop an open source cellular mmWave module for the ns--3 simulator, which we will describe in the following sections.

\begin{figure*}[ht!]
	\centering
	\includegraphics[width=0.8\textwidth] {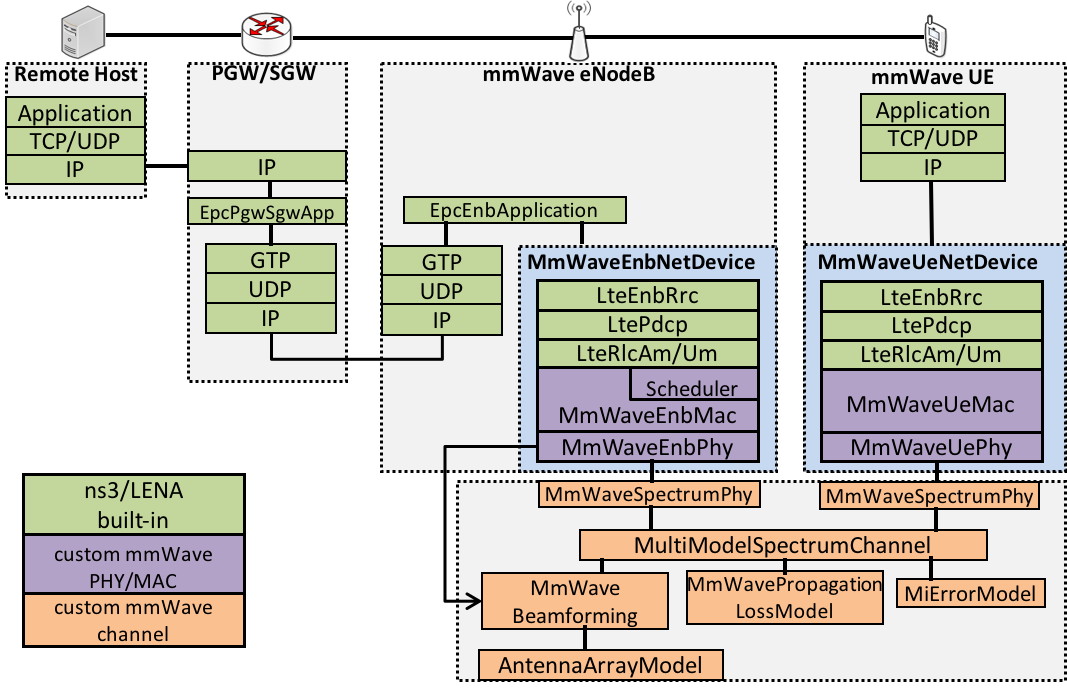}
	\caption{Class diagram of the end-to-end mmWave module.}
	\label{fig:class_diagram}
\end{figure*}

\begin{figure*}[h!]
	\centering
	\includegraphics[width=1.1\textwidth,trim={0mm -10mm 120mm 10mm},clip] {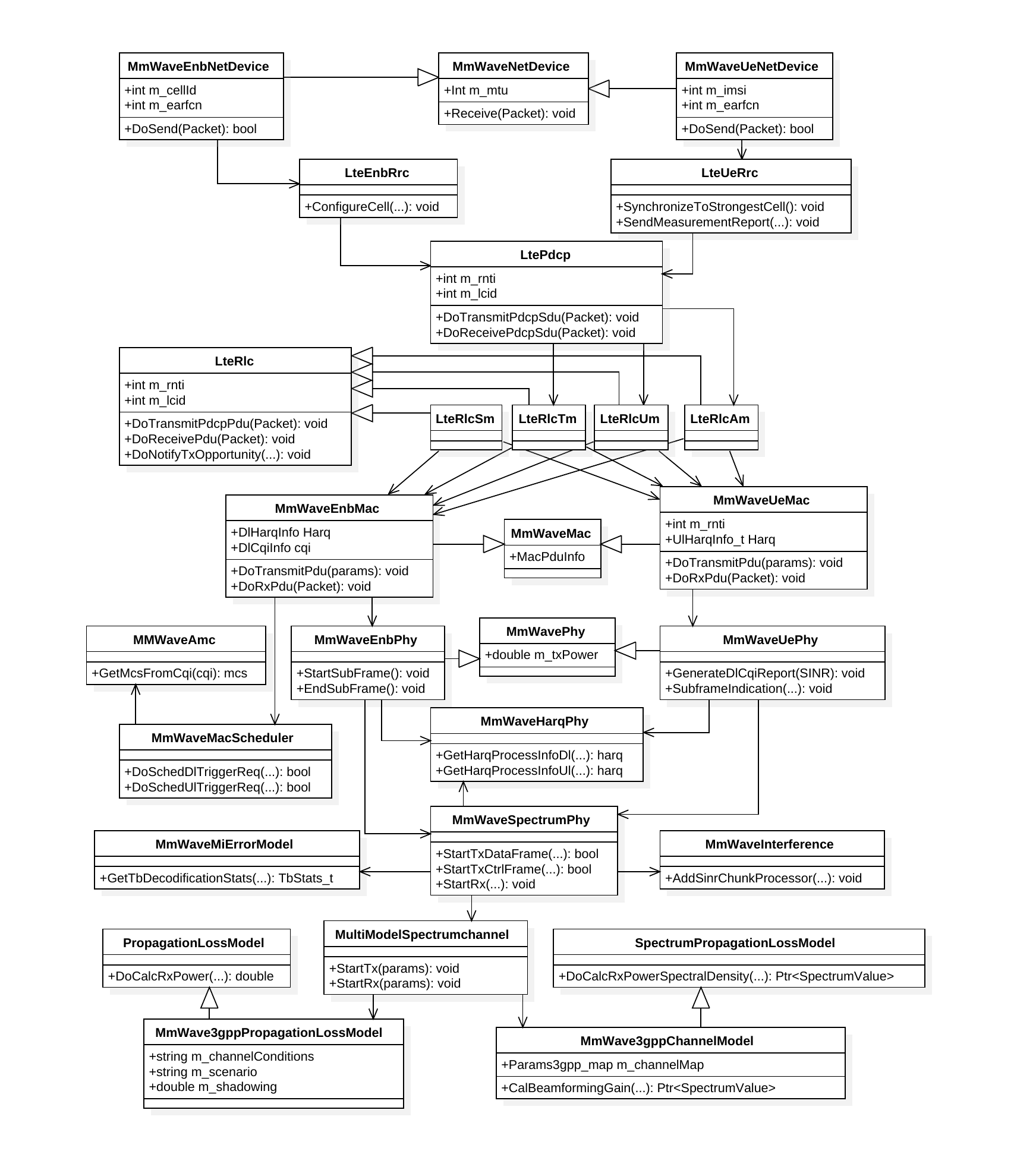}
	\caption{\gls{uml} class diagram for the end-to-end mmWave module.}
	\label{fig:uml}
\end{figure*}

\section{ns--3}
\label{sec:ns3}
The ns--3 discrete-event network simulator \cite{henderson2006ns,henderson2008network} is a very powerful tool available to communication and networking researchers for developing new protocols and analyzing complex systems. It is the successor to ns--2, a tried and tested tool that has been in use by the networking community for over a decade in the design and validation of network protocols. ns--3 is open source, and can be downloaded from the website of the project\footnote{\url{http://www.nsnam.org}}. An active community of researchers from both industry and academia has enriched the basic core of the simulator with several modules, and ns--3 can be now used to simulate a wide variety of wireless and wired networks, protocols and algorithms. There is a complete documentation\footnote{\url{https://www.nsnam.org/documentation/}} on the models in the ns--3 website, in terms of both the design of the models and what a user can do with the models. Moreover, a complete tutorial on how to install ns--3, set up ns--3 scenarios and topologies, handle the collection of statistics and log useful messages is provided in the documentation\footnote{\url{https://www.nsnam.org/docs/tutorial/html/}}. The tutorial is a good starting point for a researcher who approaches ns--3 for the first time.

The ns--3 simulator is organized into multiple folders. The \texttt{src} folder provides a collection of C++ classes, which implement a wide range of modular simulation models and network protocols. The different modules can be aggregated and instantiated to build diverse simulated network scenarios, making ns-3 especially useful for cross-layer design and analysis. The modularity and use of object-oriented design patterns also allows for new algorithms to easily be incorporated into the network stack and experimented with. Each module is itself organized into multiple subfolders, which contain the documentation and the source code of the model itself, the helpers, the examples and the tests. The helpers associated with each model have a very important role. They are classes which hide to the final user the complexity involved in setting up a complete scenario, for example by automatically assigning IP addresses, or connecting the different classes of a protocol stack. The \texttt{build} folder contains the binaries of the simulator. Finally, the \texttt{scratch} folder is a special folder in which scripts with examples and scenarios can be built on-the-fly.

Besides the core module, which provides the basic structure of the simulator, there are modules for networking protocols (e.g., the \gls{tcp}/IP stack protocols~\cite{casoni2015implementation}), wireless protocols (\gls{lte}~\cite{lena}, Wi-Fi~\cite{pei2009validation}, WiMAX~\cite{farooq2009ieee}), routing algorithms~\cite{narra2011destination}, mobility, embedding obstacles and buildings in the simulation scenarios, and data collection. All the modules are listed in the model library\footnote{\url{https://www.nsnam.org/docs/release/3.27/models/html/index.html} for ns--3 version 3.27}.

In the following sections, we will describe in detail the mmWave module for ns--3, following the same approach which is used for the other ns--3 modules. We will first describe the model in terms of implementation of the different components of a mmWave cellular network and protocol stack, and then the examples and scenarios that can be simulated with it and how they can be set up.

\section{mmWave Module Overview}
\label{sec:module}

The ns--3 mmWave module is designed to perform end-to-end simulations of \gls{3gpp}-style cellular networks. The architecture builds upon the ns--3 \gls{lte} module (LENA)~\cite{lena,ns3Lena}. It leverages the detailed implementation of \gls{lte}/\gls{epc} protocols, and implements custom \gls{phy} and \gls{mac} layers. Additionally, it is possible to connect the module to a patched version of \acrlong{dce}~\cite{dce}, a tool that allows the Linux stack \gls{tcp}/IP implementation to run as the \gls{tcp}/IP stack of ns--3 nodes, as well as to execute POSIX socket-based applications (i.e., wget, iPerf, etc).
Figure \ref{fig:class_diagram} depicts the high-level composition of the \texttt{MmWaveEnbNetDevice} and \texttt{MmWaveUeNetDevice} classes, which represent the mmWave \gls{enb}\footnote{Recently, \gls{3gpp} has proposed the term next generation Node Base (gNB) for the \gls{5g} \gls{nr} base station.} and \gls{ue} radio stacks, respectively, along with a perspective on the end-to-end structure of the simulator. A more detailed \gls{uml} class diagram is provided in Figure \ref{fig:uml}.

The ns--3 mmWave module also includes a \texttt{McUeNetDevice}, which is a \texttt{NetDevice} with a dual stack (\gls{lte} and mmWave), i.e., a device capable of connecting to both technologies. More details will be given in Section~\ref{sec:dc}.

The \texttt{MmWaveEnbMac} and \texttt{MmWaveUeMac} \gls{mac} layer classes implement the \gls{lte} module Service Access Point (\gls{sap}) \textit{provider} and \textit{user} interfaces, which enable the inter-operation with the \gls{lte} \gls{rlc} layer. Support for \gls{rlc} \gls{tm}, \gls{sm}, \gls{um},  \gls{am} is built into the \gls{mac} and scheduler classes (i.e., \texttt{MmWaveMacScheduler} and derived classes). The \gls{mac} scheduler also implements a \gls{sap} for configuration at the \gls{lte} \gls{rrc} layer (\texttt{LteEnbRrc}). Hence, every component required to establish Evolved Packet Core (\gls{epc}) connectivity is available.

 \begin{figure*}
 	\centering
 	\includegraphics[width=0.9\textwidth,trim={65mm 0mm 160mm 5mm},clip] {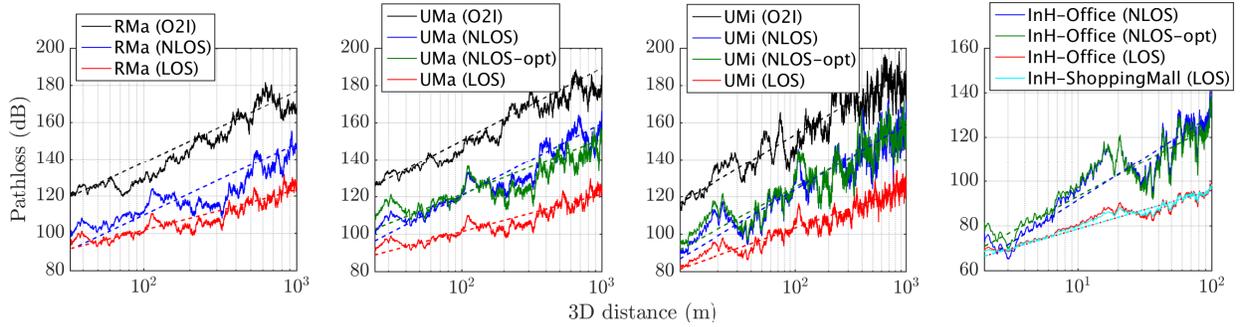}
 	\caption{Typical realization of the \gls{3gpp} pathloss model~\cite{38900} using spatial consistency for 3 outdoor scenarios (RMa stands for Rural Macro, UMa for Urban Macro and UMi for Urban Micro) and 2 indoor scenarios (InH stands for Indoor, either in the office or in a shopping mall). The channel can be \gls{los}, \gls{nlos} with or without the optional equations for the propagation loss and outdoor to indoor (O2I).}
 	\label{fig:channel3gpp}
 \end{figure*}

The \texttt{MmWavePhy} classes handle directional transmission and reception of the \gls{dl} and \gls{ul} data and control channels based on control messages from the \gls{mac} layer. Similar to the \gls{lte} module, each \gls{phy} instance communicates over the channel (i.e., \texttt{SpectrumChannel}) via an instance of the \texttt{MmWaveSpectrumPhy} class, which is shared for both the \gls{dl} and the \gls{ul} (since our design of the mmWave \gls{phy} layer is based on \gls{tdd}, as detailed in Section~\ref{subsec3.1}). Instances of \texttt{MmWaveSpectrumPhy} encapsulate all \gls{phy}-layer models: interference calculation (\texttt{MmWaveInterference}), \gls{sinr} calculation (\texttt{MmWaveSinrChunkProcessor}), the \gls{mi}-based error model (\texttt{MmWaveMiErrorModel}), which computes the packet error probability, as well as the \gls{harq} \gls{phy}-layer entity (\texttt{MmWaveHarqPhy}) to perform soft combining.

Since the structure, high-level functions and naming scheme of each class closely follow the \gls{lte} LENA module, the reader is also referred to the LENA project documentation for more information \cite{ns3LenaDoc}.

\section{Channel and \gls{mimo} Modeling}
\label{sec:channel_model}

\subsection{Channel Models}

The ns--3 mmWave module allows the user to choose among different channel models, which provide a trade-off between computational complexity, flexibility and accuracy of the results. The most flexible and detailed channel model is the one described in detail in~\cite{zhang20173gpp}, which is based on the official \gls{3gpp} channel model for the 6-100 GHz frequency band~\cite{38900}. It accounts also for spatial consistency of mobility-based simulations and provides a random blockage model, as well as the modeling of outdoor to indoor communications. The second model is based on traces from measurements or third-party ray-tracing software. This makes the channel model detailed and realistic, but constrains the simulation to limited measurements/ray-tracing routes. The third is the statistical channel model introduced in~\cite{MezzavillaNs3:15} and based on MATLAB traces, which makes the computation less demanding, but is available only for the 28 and 73 GHz frequencies. In the following paragraphs we will provide architectural details of all the available channel models. 

\subsubsection{\gls{3gpp} Statistical Channel Model}
The \gls{3gpp} model for the 6-100 GHz band, described in~\cite{38900}, is applicable for bandwidths up to 10\% of the carrier frequency and accounts for mobility. It provides several optional features that can be plugged into the basic model, in order to simulate, for example, spatial consistency (i.e., the radio environment conditions of close-by users are correlated) and random blockage. The model defines different scenarios, which describe different possible cellular network deployments: urban (with macrocells and microcells), rural and indoor.

\textbf{Pathloss:} The pathloss of the propagation channel is implemented in the \texttt{MmWave3gppPropagationLossModel} class. The model provides a statistical \gls{los}/\gls{nlos} condition characterization, as well as pathloss computation considering outdoor to indoor penetration loss, as described in~\cite[Sec.~7.4]{38900}. The \texttt{MmWave3gppBuildingPropagationLoss\-Model} class, instead, determines the \gls{los} condition according to the reciprocal position of the \gls{ue} and the \gls{enb} and to the presence of buildings or obstacles in the scenario. These classes also optionally apply an additional shadowing component to the pathloss. For a moving \gls{ue}, the shadowing is correlated in space. Given the distance $\Delta d_{2D} > 0$ on the horizontal plane from the last position in which the shadowing was computed, the exponential correlation parameter is computed as $R(\Delta d_{2D}) = e^{-\Delta d_{2D}/d_{cor}}$, where $d_{cor}$ is the correlation distance. In our implementation, pathloss and shadowing (if enabled) are updated at every transmission. Figure~\ref{fig:channel3gpp} shows the pathloss in dB for the 3D distance from the smallest value supported in each scenario to $10^3$ m for outdoor and $10^2$ m for indoor.

\textbf{Small-scale fading:} The small-scale fading model is implemented in the \texttt{MmWave3gppChannel} class, and follows the step by step approach of~\cite[Sec.~7.5]{38900}. Small-scale fading is the bottleneck of this channel model implementation, 
since it is very detailed and computationally demanding.
The fading is generated following the 3D statistical spatial approach originally proposed in~\cite{25996}. The channel is described by a channel matrix $\mathbf{H}(t, f)$, where $t$ is the time and $f$ is the frequency, of size $U\times S$, where $U$ and $S$ are the number of antennas at the receiver and the transmitter. Each entry depends on $N \le 20$ different multipath components, called \textit{clusters}, which have different delays and received powers, according to an exponential power delay profile. A cluster is itself a combination of $M = 20$ \textit{rays}, each with a slightly different arrival and departure angle in the vertical and horizontal planes.

The \texttt{MmWave3gppChannel} class has a method that generates the channel matrix, and stores the coefficient for each transmit element $s$, receive element $u$ and cluster $n$ in a data structure, that can be accessed by other methods in order to update the channel matrix or compute the beamforming gain. We introduced some assumptions with respect to the \gls{3gpp} model, in order to decrease the computational overhead introduced by the high level of detail of the channel. For example, we consider only antennas with vertical polarization, and the speed-dependent Doppler effect is not computed for each ray, but only for the central angle of each cluster. Further details on this implementation are given in~\cite{zhang20173gpp}.

\textbf{Spatial consistency:} The basic channel model described in the previous paragraphs can be used for drop-based simulations with limited mobility, i.e., for \glspl{ue} that move in an area in which the channel is very correlated and the fading parameters do not change. However, for simulations in which mobility is an important factor, the spatial consistency of the channel throughout the path on which the \gls{ue} moves can be simulated by enabling this option in the \texttt{MmWave3gppChannel} class. In the current implementation, we support spatial consistency with Procedure A of~\cite[Sec~7.6.3.2]{38900} for both \gls{los} and \gls{nlos} communications. It is possible to set the period of update $t_{PER}$, and every $t_{PER}$ the cluster delays, powers and departure and arrival angles are updated with a transformation that accounts for the speed of the \gls{ue} and for the distance traveled on the horizontal plane.

\textbf{Blockage:} This optional feature can be used to model the attenuation in certain clusters, according to their angle of arrival. The attenuation can be caused by the human body that holds the \gls{ue}, or by external elements such as for example cars, other human bodies, trees. The blockage model is implemented in the \texttt{MmWave3gppChannel} class and can be optionally activated. In our implementation we consider blockage model A, which only distinguishes between self-blocking and non-self-blocking, and is generic and computationally efficient~\cite{38900}. In particular, this model randomly generates $K+1$ blocking regions, one for self-blocking, with different parameters according to the orientation of the \gls{ue} (i.e., portrait or landscape mode), and $K$ for non-self-blocking. The attenuation is of 30 dB for self-blocking, and dependent on the scenario and on horizontal and vertical arrival angles for non-self-blocking. Moreover, the blocking of a certain cluster is correlated in both space and time, according to the \gls{ue} mobility, the blocker speed and the simulation scenario. Notice that, if both the blockage and the spatial consistency options are used, then the update of the channel with both features is synchronized, i.e., the cluster blockage is updated before the channel coefficients are recomputed with the spatial consistency procedure.

\subsubsection{Ray-tracing or Measurement Trace Model}
\texttt{MmWaveCh\-annelRaytracing} uses software-generated or measurement traces to model the channel in ns--3, for pathloss and fading. The trace samples need to contain the number of paths and the propagation loss, delay, angle of arrival and angle of departure for each path. The following trace files have been tested in our channel and are available in \texttt{mmwave/model/Raytracing/}.

\begin{figure*}[ht!]
\centering
\begin{subfigure}[t]{0.32\textwidth}
\setlength{\belowcaptionskip}{0cm}
    \centering
    \includegraphics[width=0.9\textwidth,trim = 0.5cm 4.75cm 1.25cm 2.5cm,clip]{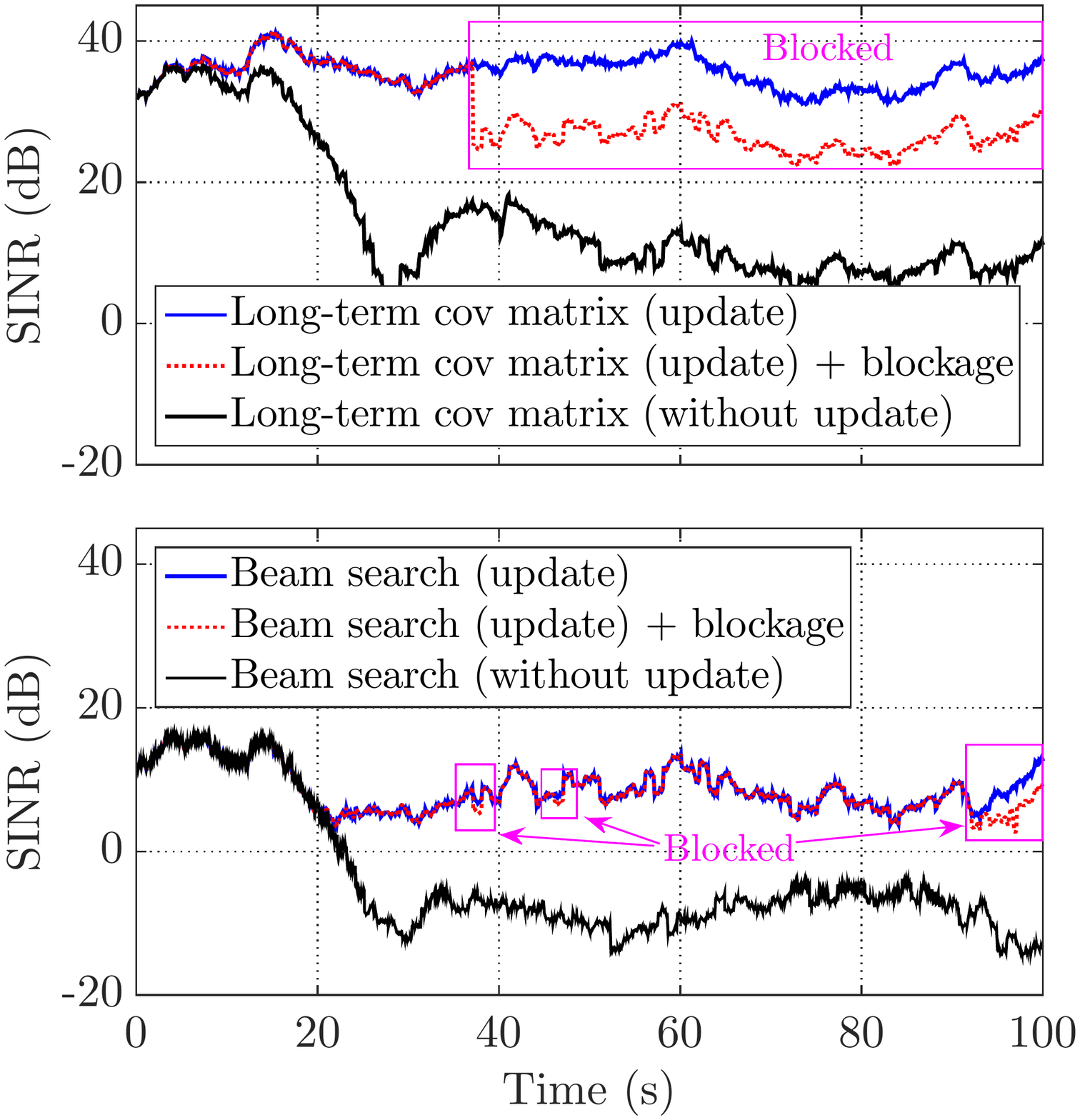}
    \caption{\gls{3gpp} statistical channel model}
    \label{fig:beamforming}
\end{subfigure}
\begin{subfigure}[t]{0.33\textwidth}
\setlength{\belowcaptionskip}{0cm}
    \centering
    \includegraphics[width=0.9\textwidth,trim = 0cm 0cm 0cm 0cm,clip]{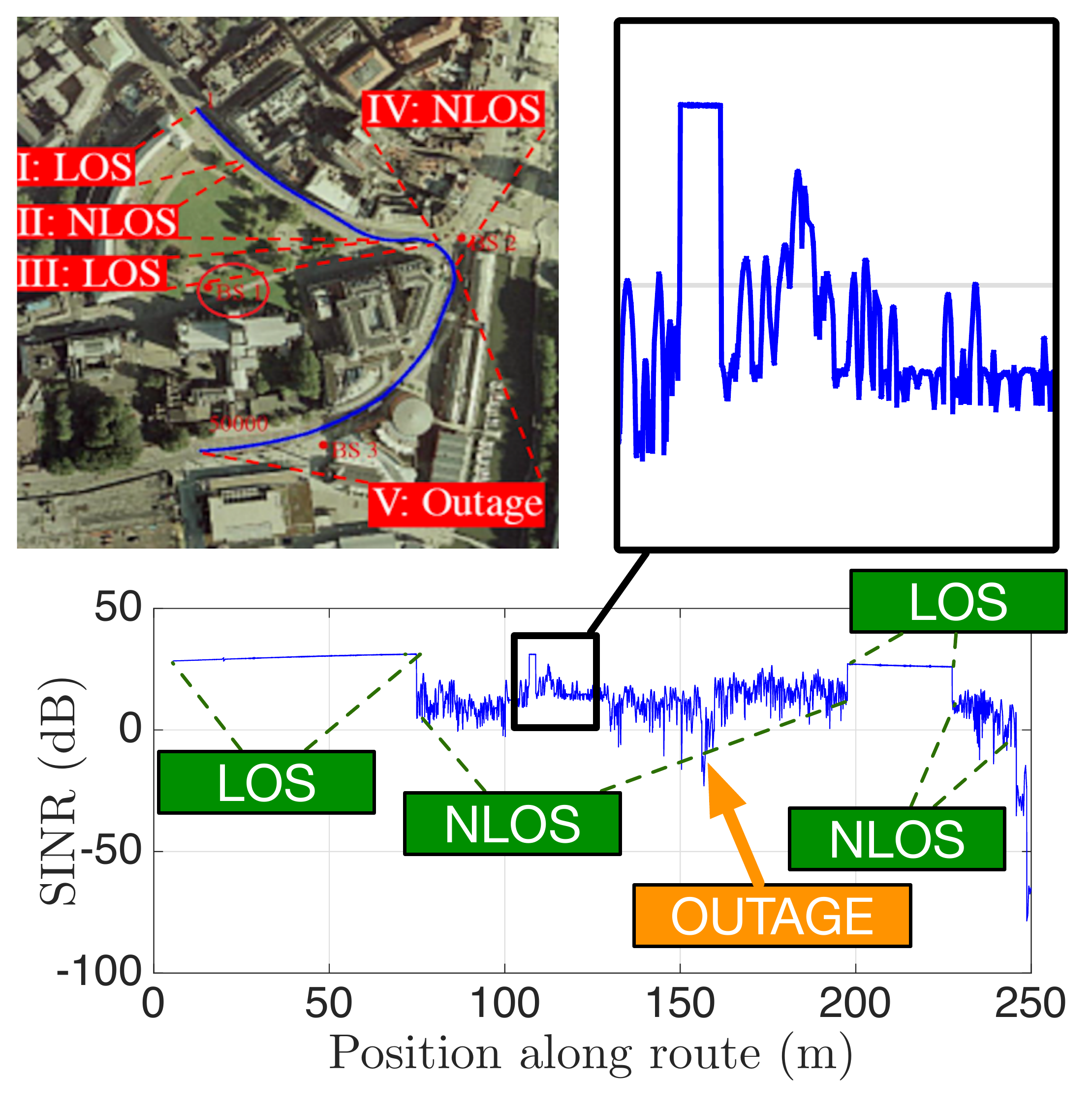}
    \caption{Ray-tracing Trace Model}
    \label{fig:raytra1}
\end{subfigure}
\begin{subfigure}[t]{0.32\textwidth}
\setlength{\belowcaptionskip}{0cm}
    \centering
    \includegraphics[width=0.9\textwidth,trim = 0cm 2.5cm 1cm 0cm,clip]{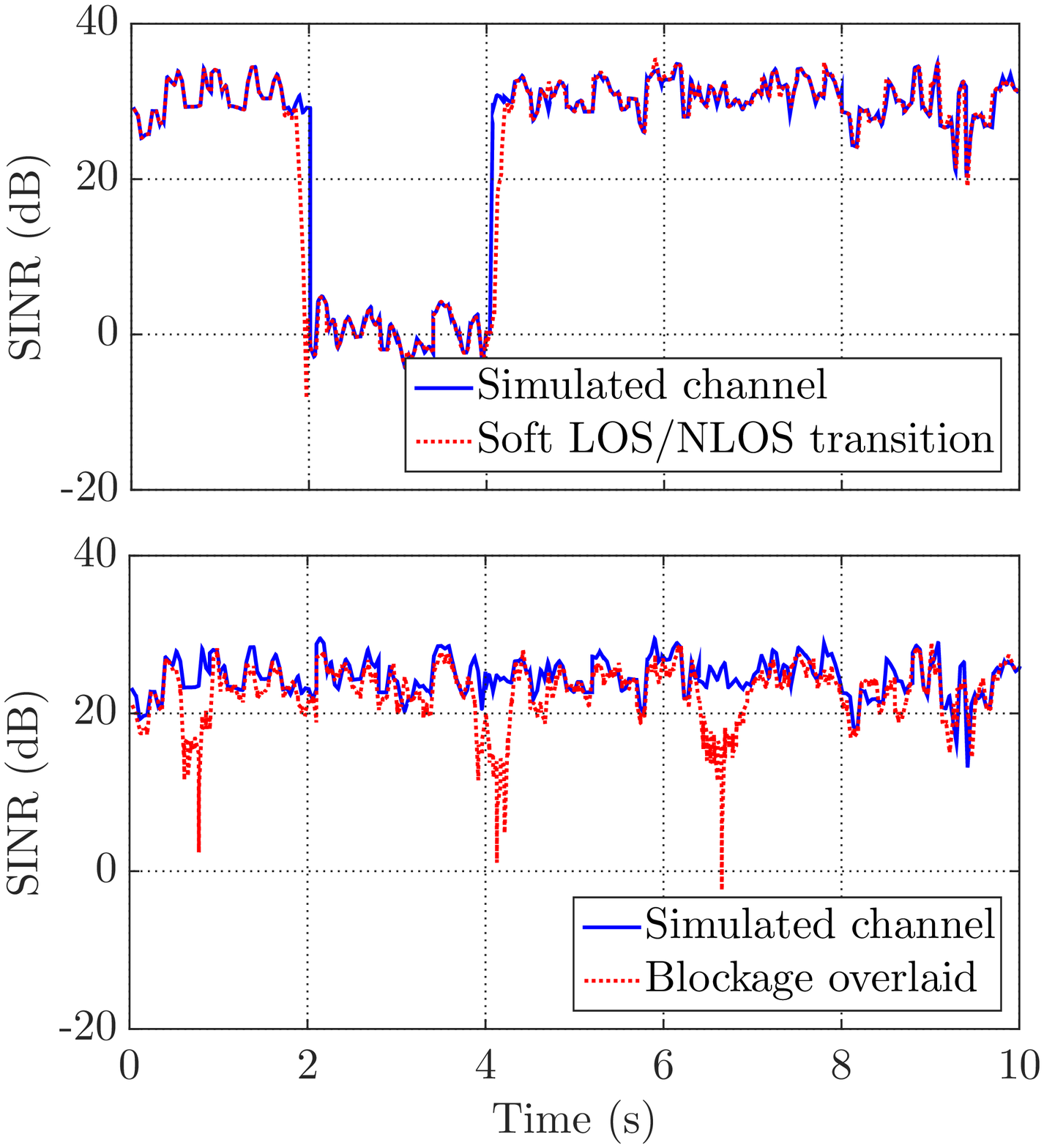}
    \caption{NYU Statistical Model}
    \label{fig:simSINR}
\end{subfigure}
\caption{Example of average \gls{sinr} plots for the three channel models.}
\label{fig:sinr}
\end{figure*}

\textbf{Ray-tracing:} Any ray-tracing software (e.g., WinProp~\cite{WinProp}) can be used to generate the channel information for a specific route. This means that the simulation scenario must be chosen \textit{a priori}, and cannot be random since it has to be given as input to the ray-tracing software. An example of ray-tracing route\footnote{The ray tracing data was provided by the Communication Systems and Networks Group, University of Bristol, UK~\cite{nix2017model,zhang2016transport}.} is shown in Figure~\ref{fig:raytra1}.

\textbf{QuaDRiGa:} The Quasi Deterministic Radio Channel Generator model~\cite{jaeckel2014quadriga}, supports consistent user mobility and massive \gls{mimo} at several frequencies (10, 28, 43, 60, 82 GHz). It also adds some time evolution characterization on top of the statistical channel to capture user mobility, which makes it suitable for system level simulations.

\subsubsection{NYU Statistical Model}
\label{sec:statistical_model}
This channel model is based on the approach described in~\cite{AkdenizCapacity:14} and implemented in our previous work~\cite{MezzavillaNs3:15}. A MATLAB implementation of the same channel model is also available in~\cite{Sun17b,Samimi16a,NYUSIM}. It provides two pathloss models, which differ in how they capture the \gls{los}/\gls{nlos} condition. The first, \texttt{MmWavePropagationLossModel}, is based on a statistical characterization of the \gls{los} state, while the second, \texttt{BuildingsObstaclePropagationLossModel}, leverages the ns--3 buildings module in order to decide whether there is an obstacle between the \gls{ue} and the \gls{enb} or not. In particular, it is possible to deploy -- deterministically or randomly -- objects of different sizes to mimic humans, cars, and buildings. A virtual line is drawn between the transmitter and the receiver: If this line intersects any object, the state is \gls{nlos}, otherwise it is \gls{los}. In both classes, once the channel state is selected, the propagation loss is computed as in~\cite{AkdenizCapacity:14}.

%

\textbf{Channel configuration:}
Since the channel matrices and optimal beamforming vectors do not depend on the distance between the \gls{ue} and the \gls{enb}, they are pre-generated in MATLAB to reduce the computational overhead in ns--3. At the beginning of each simulation we load 100 instances of the spatial signature matrices, along with the beamforming vectors. Moreover, in order to simulate realistic channels with large-scale fading, the channel matrices are updated periodically and independently (\textit{block fading}). Currently, no results are available for modeling how the large-scale statistics of the mmWave channel change over time for a mobile user, thus it should be noted that the accuracy of this method is not verified at this time. The matrix update can take place at some fixed intervals, specified by the \texttt{LongTermUpdatePeriod} attribute of the \texttt{MmWaveBeamforming} class. The small-scale fading, instead, is calculated at every transmission, where we obtain the speed of the user directly from the mobility model. The remaining parameters that depend on the environment are assumed to be constant over the entire simulation time.

\textbf{Semi-empirical feature:} Finally, as shown in Fig. \ref{fig:simSINR}, the soft transition between \gls{los}/\gls{nlos} conditions can be modeled in a ``semi-empirical" fashion, meaning that we overlay the statistical channel with blockage measurements performed in our lab \cite{giordani-tracking}: Waving a hand in front of the receiver (hand blockage), walking between the transmitter and the receiver (human blockage), and placing a metal plate between the transmitter and the receiver to emulate an obstacle, like a car or a building.

\subsection{Beamforming Gain}
For the long-term statistical channel model, the beamforming vectors are directly loaded from MATLAB generated files. For the other channel models, two methods are implemented to compute beamforming vectors, i.e., the \textit{long-term covariance matrix method} and the \textit{beam search method}.
Currently, the only available beamforming architecture for data transmission is \emph{analog}, meaning that devices can transmit or receive in only one direction at a time. As part of our future work, we plan to integrate \emph{hybrid} and \emph{digital} transceiver designs. 

In the long-term covariance matrix method, we assume that the transmitter 
estimates the spatial correlation matrix $\mathbf{Q}_{tx} = \mathbb{E}[\mathbf{H}^{\dagger}(t,f)\mathbf{H}(t,f)]$,
where the expectation is taken over the frequencies and some interval of time.  
An analogous operation is done for the receiver. In practice, the \gls{tx} and \gls{rx} would estimate the spatial covariance matrix from reference or synchronization signals and beam scanning. 
Estimation of this covariance matrix is discussed in \cite{parisa}. We do not, however, model the covariance estimation directly; instead we simply assume that the \gls{tx} and \gls{rx}
know the correct long-term channel with some configurable delay.  Beamforming vectors can then be computed
from the maximal eigenvectors of the covariance matrices~\cite{love2003equal}.  A computationally simple procedure is to use
the power method~\cite{wilkinson1965algebraic}. The algorithm selects a random initial beamforming vector and iteratively multiplies it with the spatial correlation matrix $\mathbf{Q}_{tx}$, normalizing the results at each iteration. Finally, the output will converge to the correct eigenvector. The computation for the receiver is done in the same way, starting from $\mathbf{Q}_{rx} = \mathbb{E}[\mathbf{H}(t,f)\mathbf{H}^{\dagger}(t,f)]$.


In the beam search method, we assume that the \gls{tx} and the \gls{rx} scan a discrete number of beams
from a pre-designed codebook~\cite{wang2009beam}.  Codebook design is discussed in detail in \cite{beamforming}. The beamforming vector is selected as the one with the
highest power, possibly with some time-averaging.


\subsection{Interference}
\label{sec:interference}

\begin{figure}[t]
\centering
\begin{subfigure}[t]{0.49\columnwidth}
\setlength{\belowcaptionskip}{0cm}
    \centering
\includegraphics [width=\columnwidth] {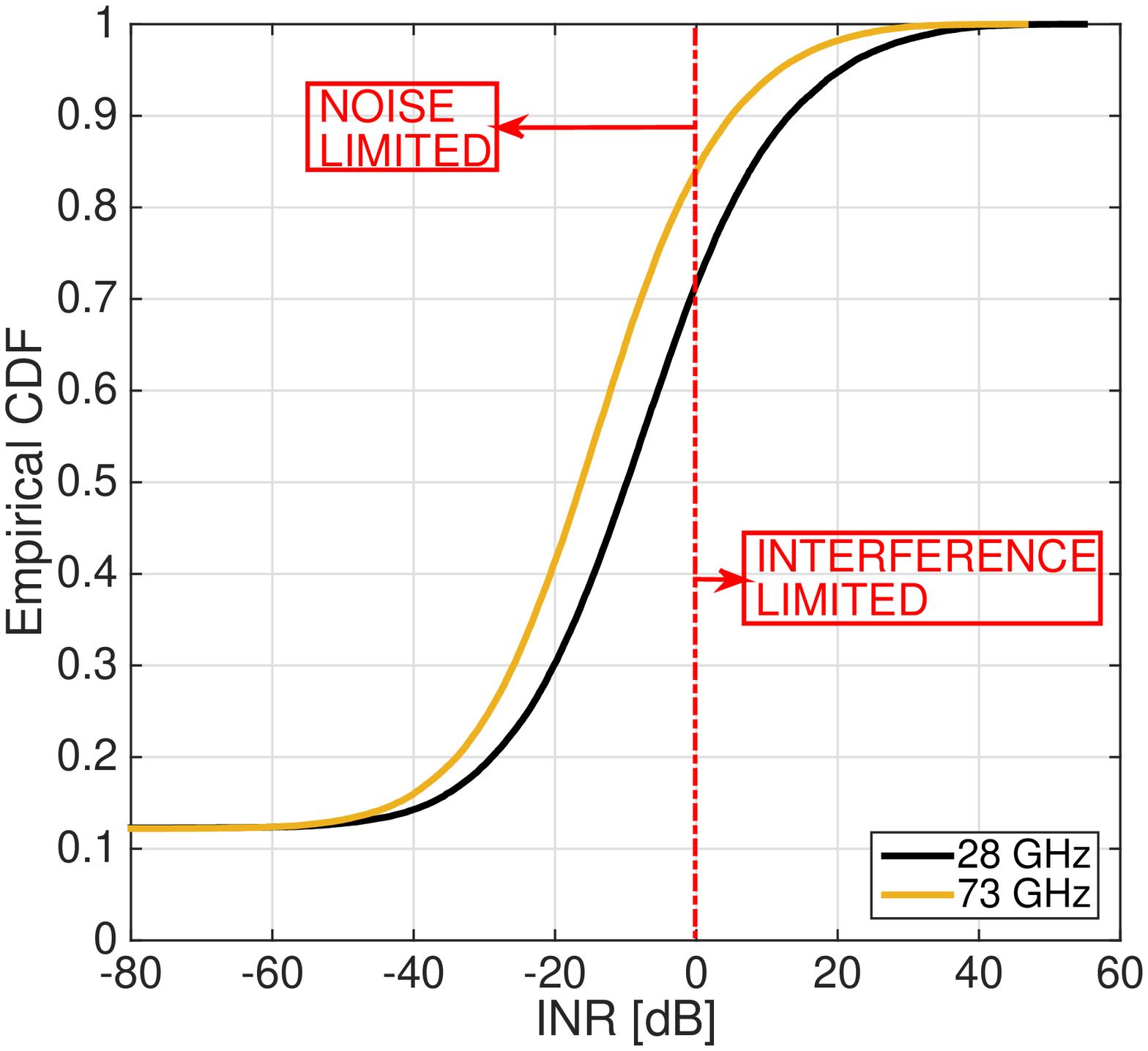} 
\caption{Empirical CDF of the INR for $\lambda_{UE}~=~300$  UEs/km\textsuperscript{2} and $\lambda_{BS}~=~30 $ BSs/km\textsuperscript{2}.}
    \label{fig:300lambda}
\end{subfigure}
\begin{subfigure}[t]{0.49\columnwidth}
\setlength{\belowcaptionskip}{0cm}
    \centering
\includegraphics [width=\columnwidth] {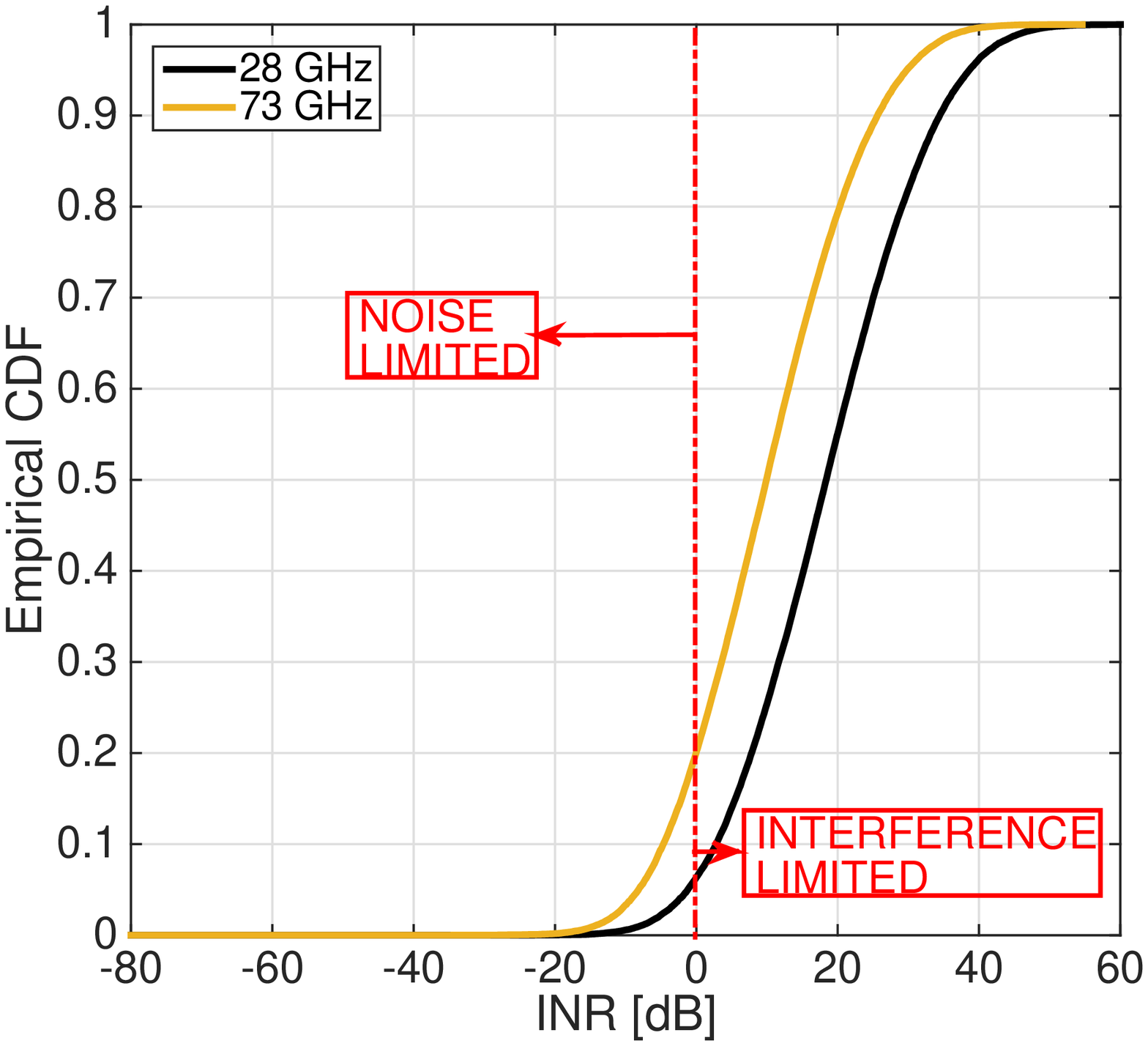}    
\caption{Empirical CDF of the INR for $\lambda_{UE}~=~1200$  UEs/km\textsuperscript{2} and $\lambda_{BS}~=~120 $ BSs/km\textsuperscript{2}.}
    \label{fig:1200lambda}
\end{subfigure}
\caption{INR trends at different user and base station density levels~\cite{rebato2016understanding}.}
\label{fig:inr}
\end{figure}

MmWave systems may be interference- or power-limited. Albeit potentially less significant for directional mmWave signals, which are generally assumed to be power-limited, there are still some cases where interference is non-negligible \cite{rebato2017stochastic}. We report in Figure \ref{fig:inr} some representative results obtained in \cite{rebato2016understanding}, where, by plotting the Interference to Noise Ratio (INR), we show that the majority of the links are still interference-limited for some dense topologies. Additionally, although intra-cell interference (i.e., from devices of the same cell) can be neglected in \gls{tdma} or \gls{fdma} operation, it does need to be explicitly calculated in the case of \gls{sdma}/Multi-User \gls{mimo}, where users are multiplexed in the spatial dimension but operate in the same time-frequency resources. Therefore, we propose an interference computation scheme that takes into account the beamforming vectors associated with each link.

As an example, we compute the \gls{sinr} between nodes $\gls{enb}_1$ and $\gls{ue}_1$ in the presence of an interferer, $\gls{enb}_2$. To do so, we first need to obtain the beamforming gains associated with both the desired and interfering signals, i.e.,
\begin{equation}
\begin{aligned}
G_{11}= |{\bf w}^{\dagger}_{rx_{11}}\mathbf{H}(t,f)_{11}{\bf w}_{tx_{11}}|^{2}, &
\\ G_{21}= |{\bf w}^{\dagger}_{rx_{11}}\mathbf{H}(t,f)_{21}{\bf w}_{tx_{22}}|^{2}.
\end{aligned}
\end{equation}
The \gls{sinr} is then computed as:
\begin{equation}
SINR_{11}= \frac{\frac{P_{Tx,11}}{PL_{11}}G_{11}}{\frac{P_{Tx,22}}{PL_{21}}G_{21}+BW\times N_0},
\end{equation}
where $P_{Tx,ii}$ is the transmit power of $\gls{enb}_i$, $PL_{ij}$ is the pathloss between $\gls{enb}_i$ and $\gls{ue}_j$, and $BW\times N_0$ is the thermal noise.

\subsection{Error Model}
\label{sec:error_model}
The mmWave module exploits the error model introduced in the ns-3 \gls{lte} LENA project, which follows a link abstraction technique for simulating \gls{tb} errors in the downlink of an LTE system. In a nutshell, the model described in \cite{mezzavilla-miesm} defines an accurate and lightweight procedure for the computation of the residual errors after PHY layer processing. This is achieved by exploiting:
\begin{itemize}
\item Mutual Information-based multi-carrier compression metrics to derive a unique SINR value of the channel, known as \emph{effective SINR}, which is represented as $\gamma_i$ in Eq. \ref{eq:cbler}, and
\item Link-Level performance curves obtained with a MATLAB bit-level LTE PHY simulator \cite{vienna}, which have been used to match a mathematical approximation of the \gls{bler}, as reported in Eq.~\eqref{eq:cbler}. 
\end{itemize}
The ultimate goal is to let the receiver derive the error probability of each \gls{tb} to determine whether the packet can be decoded or not. Because each \gls{tb} can be composed of multiple \glspl{cb}, whose size depends on the channel capacity, the \gls{bler} can be formulated as follows:
\begin{equation}
C_{BLER,i}(\gamma_i)=\frac{1}{2}\left [ 1-\textup{erf}\left ( \frac{\gamma_i -b_{C_{SIZE},MCS}}{\sqrt{2} c_{C_{SIZE},MCS}} \right ) \right ],
\label{eq:cbler}
\end{equation}
where $\gamma_i$ corresponds to the mean mutual information per coded bit of code block $i$, as explained earlier, and $b_{C_{SIZE},MCS}$ and $c_{C_{SIZE},MCS}$ represent the mean and standard deviation of the Gaussian cumulative distribution, respectively, which have been obtained from the link level performance curves mentioned above. Finally, the \gls{tb} block error rate is given by: 
\begin{equation}
T_{BLER}=1-\prod_{i=1}^{C}(1-C_{BLER,i}(\gamma_i)).
\end{equation}

\subsection{Examples}

An example of \gls{sinr} plots for the three channel models is presented in Figure~\ref{fig:sinr}. An example related to the setup of the channel model can be found in the \texttt{examples} folder of the mmWave module, in the file \texttt{mmwave-3gpp-channel-example.cc}.

Figure \ref{fig:beamforming} shows an example of a rural scenario with an \gls{enb} at coordinates $(0,0)$~m and at the height of 35 m, and a \gls{ue} in position $(100, 0)$~m, at the height of 1.5 m and moving at 1 m/s along the y axis, maintaining \gls{los} connectivity. The channel is updated consistently every 100~ms. The top figure shows the \gls{sinr} when the \gls{bf} vector is updated with the long-term covariance matrix method, while in the bottom one it is updated with the beam search method. Notice that the current implementation of the beam search method uses a fixed elevation angle of 90 degrees and sweeps only the horizontal plane. Therefore, the beam search method cannot align with the \gls{los} cluster and the power is reduced by 20 dB. Moreover, after enabling the blockage model, the \gls{sinr} achieved by the long-term covariance matrix method dropped by 20 dB when the \gls{los} cluster was blocked. However, the beam search method experienced less blockage impact, as it did not align with the \gls{los} cluster. In the other case, without update, the \gls{bf} vector is computed at $t=0$~s but never updated, and this causes the \gls{sinr} to drop as the \gls{ue} moves. Comparing the blue and black curves, it is possible to observe that for the first 20 s the performance with and without \gls{bf} update is similar, because of the consistency of the channel and of the low mobility of the \gls{ue}, but after $t = 20$ s the \gls{sinr} without update degrades by nearly 30 dB. The last observation is that the long-term covariance matrix method finds the optimal \gls{bf} vector whenever the channel is changed, therefore the \gls{sinr} is very stable. On the other hand, the beam search method shows an \gls{sinr} drop after 20 s even with update, because when the \gls{ue} moves both the \gls{ue} and the \gls{enb} are unable to optimally adapt the \gls{bf} vector and just select one of the available sectors.

Figure \ref{fig:raytra1} plots the average \gls{sinr} of a ray tracing channel indicating both \gls{los} intervals and \gls{nlos} channel states. The ray tracing data contains 5000 samples along a 500 meter route. The \gls{sinr} has a sudden change when the channel state switches. We note that the \gls{sinr} curve within \gls{los} is relatively stable, whereas more random variations are introduced for \gls{nlos}.

Finally, Figure~\ref{fig:simSINR} shows the average \gls{sinr} trace generated with the NYU channel model~\cite{AkdenizCapacity:14} in two cases, a walking user blocked by a building (top) or by other humans (bottom). The main difference is that, with buildings, the link capacity drops rapidly and the blocking interval lasts seconds; on the other hand, with humans, the channel deteriorates slowly and the blockage lasts only for a short interval. From the top figure, we can observe that with soft \gls{los}/\gls{nlos} transition enabled, the \gls{sinr} curve changes less suddenly when the channel condition switches. In the bottom graph, three human blockage events, at 1, 4 and 7 seconds, are added on top of the statistical channel.

\section{Physical Layer}
\label{sec:phy}

In this section, we discuss the key features of the mmWave \gls{phy} layer. Specifically, we have implemented a \gls{tdd} frame and subframe structure, which has similarities to \gls{tdd}-\gls{lte}, but allows for more flexible allocation and placement of control and data channels within the subframe and is suitable for the \textit{variable \gls{tti}} \gls{mac} scheme described in Section \ref{sec:mac}. Moreover, we implemented an error model and \gls{harq} model based on those in LENA, but compatible with our custom mmWave \gls{phy} and numerology (for instance, they support larger \gls{tb} and codeword sizes as well as multi-process stop-and-wait \gls{harq} for both \gls{dl} and \gls{ul}).

\begin{table*}[ht!]
\centering
\small

\renewcommand{\arraystretch}{1}
\begin{tabular}{@{}lll@{}}
\toprule
 Parameter Name &  Default Value & Description \\
\midrule
SubframePerFrame 			&	  10 			& Number of subframes in one frame \\
SubframeLength 				&	 $100$			& Length of one subframe in $\mu s$ \\
SymbolsPerSubframe 			&	 24 			& Number of \gls{ofdm} symbols per slot \\
SymbolLength 				&	 $4.16$ 		& Length of one \gls{ofdm} symbol in $\mu s$ \\
NumSubbands 				&	 72 			& Number of subbands \\
SubbandWidth 				&	 13.89			& Width of one subband in MHz \\
SubcarriersPerSubband 		&	 48 			& Number of subcarriers in each subband \\
CenterFreq 					&	 [6-100]		& Possible carrier frequencies in GHz$^*$ \\
NumRefScPerSymbol 			&	  864 (25\% total) & Reference subcarriers per symbol \\
NumDlCtrlSymbols 			&	  1 			& Downlink control symbols per subframe \\
NumUlCtrlSymbols 			&	  1 			& Uplink control symbols per subframe \\
GuardPeriod 				&	  $4.16$ 		& Guard period for \gls{ul}-to-\gls{dl} mode switching in $\mu$s\\
MacPhyDataLatency 			&	  2 			& Subframes between \gls{mac} scheduling request and scheduled subframe  \\
PhyMacDataLatency 			&	  2 			& Subframes between \gls{tb} reception at \gls{phy} and delivery to \gls{mac} \\
NumHarqProcesses 			&	  20 			& Number of \gls{harq} processes for both \gls{dl} and \gls{ul} \\
\bottomrule
\end{tabular}
\caption{Parameters for configuring the mmWave \gls{phy}.\\\scriptsize $^*$The NYU channel model~\cite{AkdenizCapacity:14} supports only 28 and 73 GHz.}
\label{tab:phy_mac_params}
\end{table*}

\subsection{Frame Structure}
\label{subsec3.1}
It is widely contended that \gls{5g} mmWave systems will target Time Division Duplex (\gls{tdd}) operation because it offers improved utilization of wider bandwidths and the opportunity to take advantage of channel reciprocity for channel estimation \cite{RanRapE:14,PiSysDes:11,AGhosh:14,ktRlc,cudak2014experimental}. In addition, shorter symbol periods and/or slot lengths have been proposed in order to reduce radio link latency \cite{levanen2014radio,Dutta:15,dutta2016mac}. The ns--3 mmWave module therefore implements a \gls{tdd} frame structure which is designed to be configurable and supports short slots in the hope that it will be useful for evaluating different potential designs and numerologies. These parameters, shown in Table {\ref{tab:phy_mac_params}}, are accessible through the attributes of the common \texttt{MmwavePhyMacCommon} class, which stores all user-defined configuration parameters used by the \gls{phy} and \gls{mac} classes. Examples related to the setup of the \gls{phy} layer parameters can be found in the \texttt{mmwave-tdma.cc} and \texttt{mmwave-epc-tdma.cc} files.

The frame and subframe structures share some similarities with \gls{lte} in that each frame is subdivided into a number of subframes of fixed length~\cite{astely2009lte}. However, in this case, the user is allowed to specify the subframe length in multiples of \gls{ofdm} symbols\footnote{Though many waveforms are being considered for \gls{5g} systems, \gls{ofdm} is still viewed as a possible candidate. In \cite{verizon,ktRlc}, Verizon and the consortium led by Korea Telecom propose a frame structure and \gls{ofdm} numerology. However, this is still under debate in \gls{3gpp}~\cite{38802}. We naturally chose to adopt \gls{ofdm}, at least initially, for the mmWave module, which allows us to leverage the existing \gls{phy} models derived for \gls{ofdm} from the \gls{lte} LENA module. As soon as the \gls{3gpp} \gls{nr} will be standardized, the protocol stack in our module can be adapted to the updated parameters.}.
Within each subframe, a variable number of symbols can be assigned by the \gls{mac} scheduler and designated for either control or data channel transmission. The \gls{mac} entity therefore has full control over multiplexing of physical channels within the subframe, as discussed in Section \ref{sec:mac}. Furthermore, each variable-length time-domain data slot can be allocated by the scheduler to different users for either the uplink or the downlink.

\begin{figure}[t!]
\includegraphics [width=0.8\columnwidth,trim={4.5cm 4cm 5cm 3cm},clip] {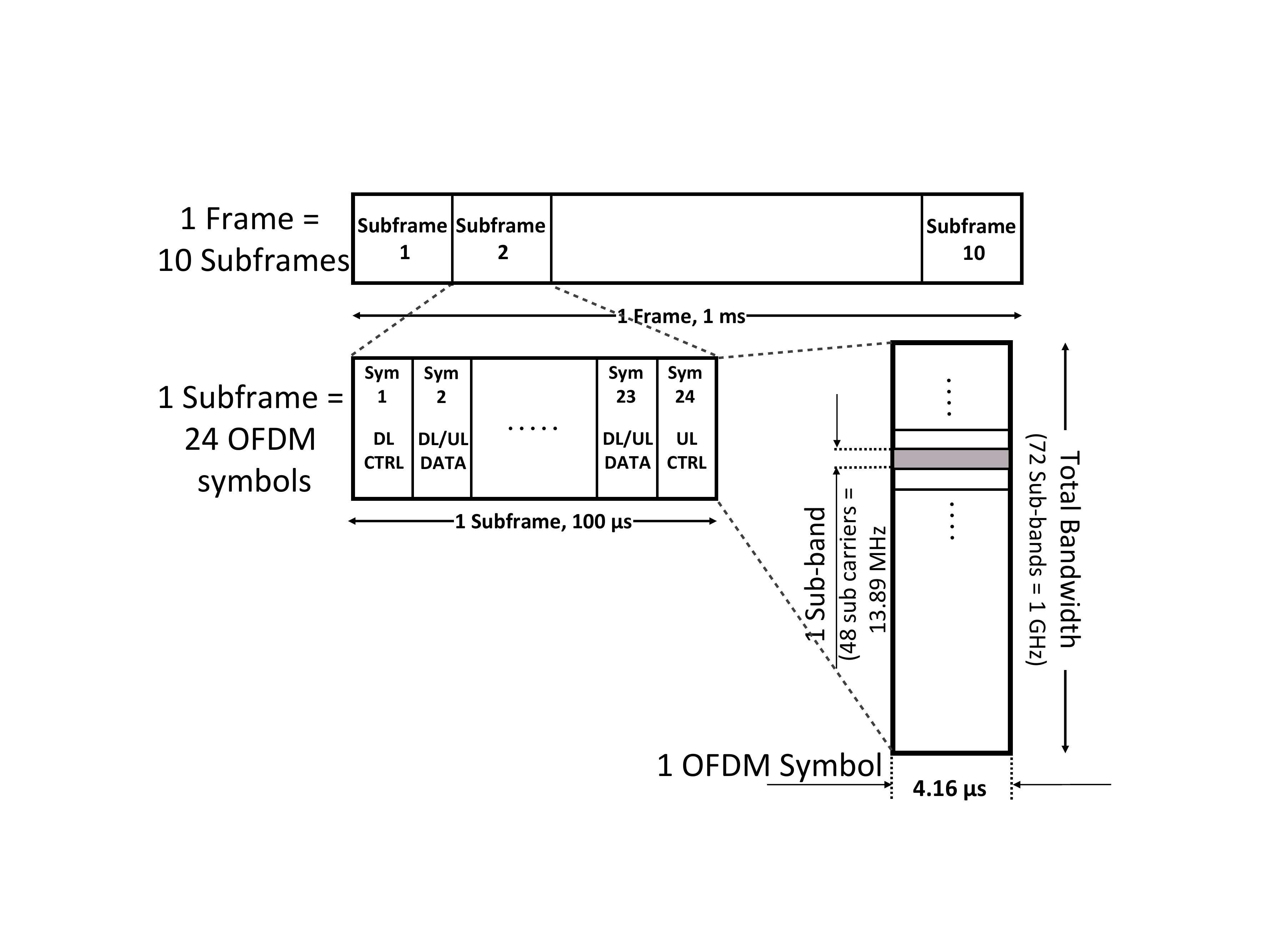}
\caption{Proposed mmWave frame structure.}
\label{fig:frame_structure}
\end{figure}

Figure \ref{fig:frame_structure} shows an example of frame structure with the numerology taken from our proposed design in \cite{Dutta:15}. Each frame of length 1 ms is split in time into 10 subframes, each of duration $100~\mu$s, representing 24 symbols of approximately  $4.16~\mu$s. In this particular scheme, the downlink and uplink control channels are always fixed in the first and last symbol of the subframe, respectively. A switching guard period of one symbol period is introduced each time the direction changes from \gls{ul} to \gls{dl}. In the frequency domain, the entire bandwidth of 1 GHz is divided into 72 subbands of width 13.89 MHz, each composed of 48 subcarriers. It is possible to assign \gls{ue} data to each of these subbands, as is done with Orthogonal Frequency-Division Multiple Access (\gls{ofdm}A) in \gls{lte}, however only \gls{tdma} operation is currently supported for reasons we shall explain shortly.

\subsection{\gls{phy} Transmission and Reception}
The \texttt{MmWaveEnbPhy} and \texttt{MmWaveUePhy} classes model the physical layer for the mmWave \gls{enb} and the \gls{ue}, respectively, and encapsulate similar functionalities as the \texttt{LtePhy} classes from the \gls{lte} module. Broadly, these objects (i) handle the transmission and reception of physical control and data channels (analogous to the \gls{pdcch}/\gls{pucch} and \gls{pdsch}/\gls{pusch} channels of \gls{lte}), (ii) simulate the start and the end of frames, subframes and slots, and (iii) deliver received and successfully decoded data and control packets to the \gls{mac} layer.

In the \texttt{MmWaveEnbPhy} and \texttt{MmWaveUePhy} classes, calls to \texttt{StartSubFrame()} and \texttt{EndSubFrame()} are scheduled at fixed periods, based on the user-specified subframe length, to mark the start and end of each subframe. The timing of variable-\gls{tti} slots, controlled by scheduling the \texttt{StartSlot()} and \texttt{EndSlot()} methods, is dynamically configured by the \gls{mac} via the \gls{mac}-\gls{phy} \gls{sap} method \texttt{SetSfAllocInfo()}, which enqueues an \texttt{SfAllocInfo} allocation element for some future subframe index specified by the \gls{mac}. A \textit{subframe indication} to the \gls{mac} layer triggers the scheduler at the beginning of each subframe to allocate a future subframe. For the \gls{ue} \gls{phy}, \texttt{SfAllocInfo} objects are populated after reception of \gls{dci} messages. At the beginning of each subframe, the current subframe allocation scheme is dequeued, which contains a variable number of \texttt{SlotAllocInfo} objects. These, in turn, specify contiguous ranges of \gls{ofdm} symbol indices occupied by a given slot, along with the designation as either \textit{\gls{dl}} or \textit{\gls{ul}} and control (\textit{CTRL}) or data (\textit{DATA}).

The data packets and the control messages generated by the \gls{mac} are mapped to a specific subframe and slot index in the \textit{packet burst map} and \textit{control message map}, respectively. Presently, in our custom subframe design, certain control messages which must be decoded by all \gls{ue}s, such as the \gls{dci}s, are always transmitted in fixed \gls{pdcch}/\gls{pucch} symbols at the first and last symbol of the subframe, but this static mapping can be easily changed by the user\footnote{As in \cite{levanen2014radio,Dutta:15}, we assume either \gls{fdma} or \gls{sdma}-based multiple access in the control regions. However, we do not currently model these modulation schemes nor the specific control channel resource mapping explicitly. We intend for this capability to be available in later versions, which will enable more accurate simulation of the control overhead.}.
Other \gls{ue}-specific control and data packets are recalled at the beginning of each allocated \gls{tdma} data slot and are transmitted to the intended device.

To initiate transmission of a data slot, the \gls{enb} \gls{phy} first calls \texttt{AntennaArrayModel::ChangeBeamforming\-Vector()} to update the transmit and receive beamforming vectors for both the \gls{enb} and the \gls{ue}. In the case of control slots, no beamforming update is applied since we currently assume an ``ideal'' control channel. For both \gls{dl} and \gls{ul} transmissions, either the \texttt{MmWaveSpectrumPhy} method \texttt{StartTxDataFrame()} or \texttt{StartTxCtrlFrame()} is called to transmit a data or control slot, respectively. The functions of \texttt{MmWaveSpectrumPhy}, which is similar to the corresponding LENA class, are as follows. After the reception of data packets, the \gls{phy} layer calculates the \gls{sinr} of the received signal in each subband, taking into account the path loss, \gls{mimo} beamforming gains and frequency-selective fading. This triggers the generation of \gls{cqi} reports, which are fed back to the base station in either \gls{ul} data or control slots. The error model instance is also called to probabilistically compute whether a packet should be dropped by the receiver based on the \gls{sinr} and, in the case of an \gls{harq} retransmission, any soft bits that have been accumulated in the \gls{phy} \gls{harq} entity (see Section \ref{sec:harq}). Uncorrupted packets are then received by the \texttt{MmWavePhy} instance, which forwards them up to the \gls{mac} layer \gls{sap}.

\section{\gls{mac} Layer}
\label{sec:mac}
\gls{tdma} is widely assumed to be the de-facto scheme for mmWave access because of the dependence on analog beamforming, where the transmitter and receiver align their antenna arrays to maximize the gain in a specific direction (rather than with a wide angular spread or omni-directionally, as in conventional systems). Many early designs and prototypes have been \gls{tdma}-based \cite{PiSysDes:11,AGhosh:14,cudak2014experimental}, with others incorporating \gls{sdma} for the control channel only \cite{levanen2014radio}. \gls{sdma} or \gls{fdma} schemes (as in \gls{lte}) are possible with \textit{digital beamforming}, which would allow the base station to
transmit or receive in multiple directions at the same time.

Furthermore, one of the foremost considerations driving innovation for the \gls{5g} \gls{mac} layer is latency. Specifically, the Key Performance Indicator of 1 ms over-the-air latency has been proposed as one of the core \gls{5g} requirements by such standards bodies as the International Telecommunication Union~\cite{itu2014framework}, as well as by recent pre-standardization studies such as those carried out under the METIS 2020 project \cite{popovsk2013eu}. However, a well-known drawback of \gls{tdma} is that fixed slot lengths or \gls{tti}s can result in poor resource utilization and latency, which can become particularly severe in scenarios where many intermittent, small packets must be transmitted to/received from many devices~\cite{Dutta:15}.

Based on these considerations, variable \gls{tti}-based \gls{tdma} frame structures and \gls{mac} schemes have been proposed in \cite{levanen2014radio, Dutta:15, kela2015novel,ford2016achieving,dutta2016mac}. This approach allows for slot sizes that can vary according to the length of the packet or \gls{tb} to be transmitted and are well-suited for diverse traffic since they allow bursty or intermittent traffic with small packets as well as high-throughput data like streaming and file transfers to be scheduled efficiently.

The \gls{mac} layer implementation can be found in the \texttt{MmWaveEnbMac} and \texttt{MmWaveUeMac} classes, whose main role is the coordination of procedures such as scheduling and retransmission. Moreover, they interact with the \gls{rlc} layer to receive periodic reports on the buffer occupancy, i.e., the \glspl{bsr}, and with the physical layer classes for the transmission and reception of packets. To carry out their functionalities, the \gls{mac} classes interact with several other classes, that we will describe in the following paragraphs. 

\subsection{Adaptive Modulation and Coding}
\label{sec:amc}

The role of the \gls{amc} mechanism is to adapt the modulation scheme and the coding applied on top to the channel quality, measured using \glspl{cqi}. In the simulator, this translates into (i) mapping the \gls{cqi} into the \gls{mcs}, using the error model implemented in the \texttt{MmWaveMiErrorModel} and described in Sec.~\ref{sec:error_model}, and (ii) computing the available \gls{tb} size for a subframe given the \gls{mcs}. This information is then used by the scheduler to perform radio resource management.

The \gls{amc} is implemented in the \texttt{MmWaveAmc} class, which uses most of the code of the corresponding LENA module class. Some minor modifications and additional methods were necessary to accommodate the dynamic \gls{tdma} \gls{mac} scheme and frame structure. For instance, the \texttt{GetTbSizeFromMcsSymbols()} and \texttt{GetNumSymbolsFromTbsMcs()} methods are used by the scheduler to compute the \gls{tb} size from the number of symbols for a given \gls{mcs} value, and vice versa. Also the \texttt{CreateCqiFeedbackWbTdma()} method is added to generate wideband \gls{cqi} reports for variable-\gls{tti} slots.

\begin{figure}[!t]
\centering
\includegraphics [width=0.8\columnwidth] {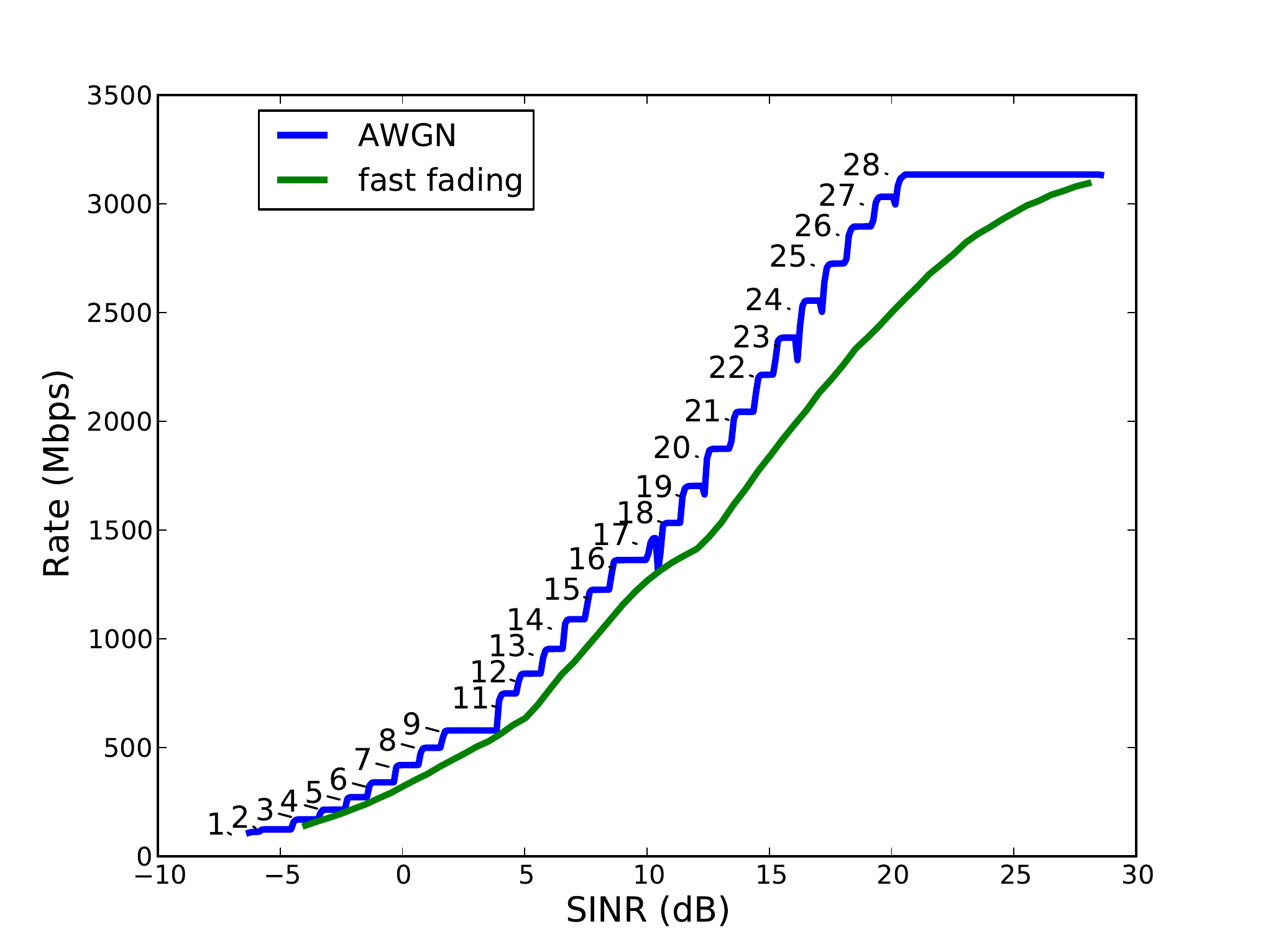}
\caption{Rate and \gls{mcs} vs. \gls{sinr} for a single user under \gls{awgn} and fast-fading mmWave channels.}
\label{fig:amc_sinr_rate}
\end{figure}

Figure \ref{fig:amc_sinr_rate} shows the results of the test case provided in \texttt{mmwave-amc-test.cc}. This simulation serves to demonstrate the performance of the \gls{amc} and \gls{cqi} feedback mechanisms for a single user in the uplink (although a multi-user scenario could easily be configured as well). The default \gls{phy}/\gls{mac} parameters in Table \ref{tab:phy_mac_params} are used along with the default scheduler and default parameters for the statistical path loss, fading and beamforming models (i.e., \texttt{MmWavePropagationLossModel} and \texttt{MmWaveBeamforming}).

We compute the rate versus the average \gls{sinr} over a period of 12 seconds (long enough for the small-scale fading to average out). The average \gls{phy}-layer rate is computed as the average sum of the sizes of successfully-decoded \gls{tb}s per second.
Every 12 seconds we artificially increase the path loss while keeping the \gls{ue} position fixed.  As the \gls{sinr} decreases, the \gls{mac} will select a lower \gls{mcs} level to encode the data. The test is performed for the \gls{awgn} case (i.e., no fading) as well as for small-scale fading. Although the \gls{ue} position relative to the base station is constant, we can generate time-varying multi-path fading through the \texttt{MmWaveBeamforming} class by setting a fixed speed of 1.5 m/s to artificially generate Doppler, which is a standard technique for such an analysis. Also, we assume that the long-term channel parameters do not change for the duration of the simulation.

Figure~\ref{fig:amc_sinr_rate} therefore shows the data rate that it is possible to achieve with a certain \gls{sinr} and with a certain modulation and coding scheme. If this plot is compared to the one generated from a similar test in Figure 3.1 of the LENA documentation \cite{ns3LenaDoc}, we notice that the \gls{awgn} curve from the mmWave test is shifted by approximately 5 dB to the left, indicating that the LENA version is transitioning to a lower \gls{mcs} at a much higher \gls{sinr}. This is because the LENA test is using the more conservative average \gls{sinr}-based \gls{cqi} mapping, which targets a much smaller \gls{tb} error probability. In our test, we use the Mutual Information-Based Effective \gls{sinr} scheme described in Sec.~\ref{sec:error_model} with a target maximum \gls{tb} error of 10\% in order to maximize the rate for a given \gls{sinr} \cite{mezzavilla-miesm}.

\subsection{Hybrid ARQ Retransmission}
\label{sec:harq}
Full support for \gls{harq} with soft combining is now included in the mmWave module. 
\gls{harq} is a technique introduced in~\cite{harq2001class} and extensively used in \gls{lte} networks~\cite{ford2016achieving}, which enables fast retransmissions with incremental redundancy in order to increase the probability of successful decoding and the efficiency of the transmissions. In \gls{lte}, the \gls{harq} mechanism is based on multiple stop and wait retransmission processes, and a maximum of 8 simultaneous \gls{harq} processes can be active at any given time~\cite{3GPP36.321}. The \gls{harq} retransmissions have priority with respect to new transmissions, thus the available resources are given first to \gls{harq} processes and then to the data queued in the \gls{rlc} buffers. Despite being fundamental in protecting from the losses of packets due to rapid variations in the channel quality, the \gls{harq} mechanism introduces additional latency~\cite{ford2016achieving,polese2017mptcp}, therefore the optimization of its performance is necessary to enable the target of sub-1-ms latency for ultra-low-latency communications.

The \texttt{MmWaveHarqPhy} class along with the functionalities within the different scheduler classes are based heavily on the LENA module code. The scheduler at the \gls{enb} uses the information provided by \gls{harq} feedback messages to assign new resources to the \gls{harq} processes that require retransmissions. Each transport block is granted a maximum number of transmission attempts, which is set to 3, as in \gls{lte}. However, some novelties are introduced in \texttt{MmWaveHarqPhy} in order to account for the more challenging channel conditions of the mmWave scenario. First, multiple \gls{harq} processes per user can be created not only for the downlink but also for the uplink. Second, the number of processes per user is not fixed to 8, but can be configured through the \texttt{NumHarqProcesses} attribute in \texttt{MmWavePhyMacCommon}. This makes it possible to control (and, if needed, increase) the number of the simultaneous stop and wait retransmission processes and optimize the bandwidth utilization. Third, additional modifications were needed to support larger codeword sizes in both the \texttt{MmWaveHarqPhy} and \texttt{MmWaveMiErrorModel} classes. Finally, the integration with the flexible \gls{tti} physical layer allows a reduction in the latency of the retransmissions, as discussed in~\cite{ford2016achieving}.

\subsection{Schedulers}

We now present the implementations of four scheduler classes for the variable \gls{tti} scheme. These differ significantly from the \gls{ofdm}A-based schedulers available in ns--3 LENA \cite{ns3LenaDoc} as, instead of allocating Resource Blocks/Resource Block Groups of frequency-domain resources, these \gls{tdma}-based schedulers allocate time-domain symbols within a periodic subframe to different users in the \gls{dl} or \gls{ul} direction.

Before scheduling new data, Buffer Status Report and \gls{cqi} messages are first processed. The \gls{mcs} is computed by the \gls{amc} model for each user based on the \gls{cqi}s for the \gls{dl} or \gls{sinr} measurements for the \gls{ul} data channel. The \gls{mcs} and the buffer length of each user are used to compute the minimum number of symbols required to schedule the data in the user's \gls{rlc} buffers. This procedure for estimating the optimal \gls{mcs} and determining the minimum number of symbols is common to each of the schedulers described in the following.

\textbf{\gls{rr} Scheduler:}
The \texttt{MmWaveFlexTtiMacScheduler} class is the default scheduler for the mmWave module. It supports the variable \gls{tti} scheme previously described in Section \ref{sec:phy} and assigns \gls{ofdm} symbols to user flows in \textit{Round-Robin} order. Upon being triggered by a subframe indication, any \gls{harq} retransmissions are automatically scheduled using the available \gls{ofdm} symbols. While the slot allocated for a retransmission does not need to start at the same symbol index as the previous transmission of the same \gls{tb}, it does need the same number of contiguous symbols and \gls{mcs}, since an adaptive \gls{harq} scheme (where the re-transmission can be scheduled with a different \gls{mcs}) has not yet been implemented.

To assign symbols to users, the total number of users with active flows is first calculated. Then the total available data symbols in the subframe are divided evenly among users. If a user requires fewer symbols to transmit its entire buffer, the remaining symbols (i.e., the difference between the available and required slot length) are distributed among the other active users. 

One also has the option to set a fixed number of symbols per slot by enabling the \textit{fixed \gls{tti}} mode. Although the same general subframe structure is maintained, slots will then be allocated in some multiple of \texttt{SymPerSlot} symbols. Setting the \texttt{SymPerSlot} attribute of the scheduler class to the number of slots per subframe, for instance, will result in only one \gls{ue} being scheduled per subframe, which would be highly inefficient in a multi-user cell.

\textbf{\gls{pf} Scheduler:}
\textit{Proportional Fair} is another well-known discipline, and is provided by the \texttt{MmWave\-FlexTti\-Pf\-Mac\-Scheduler} class.  The \gls{pf} scheduler attempts to prioritize traffic for high-\gls{sinr} users while maintaining some measure of fairness by ensuring that low-\gls{sinr}, cell-edge users are also scheduled \cite{SesiaLTE}. 

\textbf{\gls{edf} Scheduler:}
The \texttt{MmWave\-FlexTti\-Edf\-Mac\-Scheduler} class implements an \textit{Earliest Deadline First} policy, which is a priority queue-based policy that weighs flows by their relative deadlines for packet delivery. The deadlines are initially set according to the delay budget of the QoS Class Indicator (QCI) configured by the \gls{rrc} layer \cite{Dahlman:11,3GPP23.203}. The deadline of the \gls{hol} packets of each \gls{rlc} buffer is then compared, and that with the earliest deadline is scheduled first. Any remaining symbols in the subframe are allocated to the packet with the next smallest relative deadline and so forth until all $N_{sym}$ symbols are assigned. The \gls{edf} scheduler is the only deliberately delay-sensitive scheme included in the mmWave module and can be useful for evaluating the latency performance of mmWave links, as in the simulations presented in Sec.~\ref{sec:results}.


\textbf{\gls{mr} Scheduler:}
The \textit{Maximum Rate} policy realized in the \texttt{MmWave\-FlexTti\-Mr\-Mac\-Scheduler} class schedules only the users with the highest SINRs to maximize cell throughput. Initially, \gls{ue}s are sorted based on their optimal \gls{mcs} values. Symbols are distributed in round-robin fashion among \gls{ue}s at the highest \gls{mcs} level until the minimum number of symbols required to transmit the entire buffers of these users has been assigned. This is then repeated for \gls{ue}s at the second highest level, and so forth until all symbols of the subframe are allocated.

The \gls{mr} scheduler potentially suffers from extremely poor fairness when there are both high- and low-rate users, and some users may not be scheduled at all, thus making it impractical for any real-world multi-user system. However, it may still be useful for testing system capacity and performance.

\section{\gls{rlc} Layer}
\label{sec:rlc}
The \gls{rlc} layer is inherited directly from the \gls{lte} module described in \cite{lena}, and therefore all the \gls{lte} \gls{rlc} entities are included. Moreover, the \gls{rlc} \gls{am} retransmission entity is modified to be compatible with the mmWave \gls{phy} and \gls{mac} layers, and \gls{aqm} for the \gls{rlc} buffers is introduced as a new optional feature.

\subsection{Modified \gls{rlc} \gls{am} Retransmission }
Reordering and retransmission play an important role in \gls{rlc} \gls{am}. Due to the shortened mmWave frame structure, the timers of the \gls{rlc} entity should also be reduced accordingly, e.g., the \texttt{PollRetransmitTimer} is changed to 2 ms from 20 ms. Moreover, the original \gls{lte} module does not perform re-segmentation for retransmissions, and the \gls{rlc} segment waits in the retransmission buffer until the transmission opportunity advertised by the lower layers is big enough. This becomes problematic when the transmission is operated over an intermittent channel, as a sudden channel capacity drop would halt the retransmission entirely. Therefore, we added to the \gls{rlc} \gls{am} layer implementation the capability of performing segmentation also for the retransmission process, in order to support an intermittent mmWave channel. The re-segmentation process deployed in our \gls{rlc} \gls{am} class works as follows: If the number of bytes that can be transmitted in the next opportunity is smaller than the bytes of the segment that should be retransmitted, then the segment will be split into smaller subsegments with a re-segment flag set to be true. The \gls{rlc} layer at the receiver side will check the flags of the subsegments, and wait until the final one if the flag is set to be true. Finally, the subsegments are assembled to construct the original segment and forwarded to the upper \gls{pdcp} layer if all subsegments are received correctly. Otherwise, all subsegments are discarded and another retransmission is triggered.

\subsection{Active Queue Management}
\acrlong{aqm} techniques are used in the buffers of routers, middleboxes and base stations in order to improve the performance of \gls{tcp} and avoid the manual tuning of the buffer size. Different strategies have been defined in the literature~\cite{abbas2016fairness,rfc7567}, and several of them are implemented in ns--3~\cite{imputato2016design,imputato2017traffic,deepak2017design}. \gls{aqm} strategies allow the network to avoid congestion at the buffers, because they react early to the increase in the buffer occupancy by dropping some packets before the buffer is full. With respect to the default Drop-tail approach, in which no packet is dropped until the buffer is full, \gls{aqm} techniques make \gls{tcp} aware of possible congestion earlier, avoiding the latency increase which is typical of the bufferbloat phenomenon~\cite{gettys2011bufferbloat}.

Some early \gls{aqm}, such as \gls{red} \cite{floyd1993random,ott1999sred}, were widely studied in the literature, but failed to find market traction because of the intrinsic complexity of their tuning parameters. Recently, a simpler \gls{aqm} technique, namely Controlled Delay (CoDel)~\cite{nichols2012controlling}, was proposed to replace \gls{red} queues. CoDel adapts to dynamic link rates without parameter configuration, and is able to discriminate ``good" and ``bad" queues: good queues can quickly empty the buffer, whereas ``bad" queues persistently buffer packets. CoDel works by monitoring the minimum queue delay in every 100 ms interval, and only drops packets when the minimum queue delay is more than 5 ms.

In the \gls{rlc} layer of the \gls{lte} module, the default queue management is Drop-tail. In the mmWave module, the \gls{rlc} layer can use either the default Drop-tail approach or more sophisticated \gls{aqm} techniques, that can be enabled by setting the \texttt{Enable\gls{aqm}} attribute to true. The default \gls{aqm} is the CoDel scheme, however it is possible to use any of the queues available in ns--3 by modifying the queue attribute in the \texttt{LteRlcAm} class. The evaluation of the \gls{aqm} scheme is further discussed in Section \ref{sec:tcp}.

\section{Dual Connectivity Extension}
\label{sec:dc}

The ns--3 mmWave module is also capable of performing simulations with dual-stack \gls{ue}s connected both to an \gls{lte} \gls{enb} and to a mmWave \gls{enb}. This feature, partially described in~\cite{simutoolsPolese}, was introduced because mmWave \gls{5g} networks will likely use multi-connectivity and inter-networking with legacy \glspl{rat} in order to increase the robustness with respect to mobility and channel dynamics~\cite{giordaniMC2016,tesemaMC2015,poleseHo,5Gmulticonnectivity2016,EuCnCBackhaul2016,osseiran2014scenarios}. The source code can be found in the \texttt{new-handover} branch of the ns--3 mmWave module repository.

The \gls{dc} implementation of this simulation module assumes that the core networks of \gls{lte} and of mmWave will be integrated, as in one of the options described in~\cite{attLte5g2016}. Therefore the \gls{lte} and the mmWave \gls{enb}s share the same backhaul network, i.e., they are connected to each other with X2 links and to the \gls{mme}/\gls{pgw} nodes with the S1 interface. As to the \gls{ran}, the \gls{dc} solution of this module is an extension of \gls{3gpp}'s \gls{lte} \gls{dc} proposal~\cite{36842}. In particular, a single bearer per \gls{dc} flow is established, with a connection from the core network to the \gls{lte} \gls{enb}, where the flow is split and forwarded either to the local stack or to the remote mmWave stack. We chose the \gls{pdcp} layer as the integration layer, since it allows a non-colocated deployment of the \gls{enb}s and a clean-slate approach in the design of the \gls{phy}, \gls{mac} and \gls{rlc} layers~\cite{dasilva}.

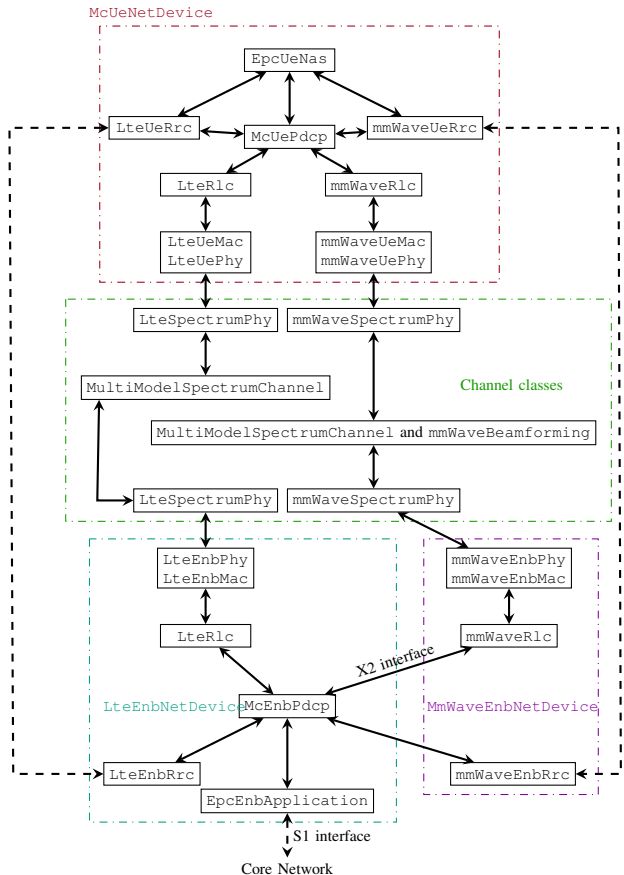
\begin{figure}[t]
\centering

\begin{tikzpicture}[node distance=1.5cm, scale=0.6, every node/.style={scale=0.6}]
  \node (epc) [process] {\texttt{EpcUeNas}};
  \node (rrc) [process, below of=epc, xshift=-3cm] {\texttt{LteUeRrc}};
  \node (mmrrc) [process, below of=epc, xshift=3cm] {\texttt{mmWaveUeRrc}};
  \node (pdcp) [process, below of=epc, yshift=-0.2cm] {\texttt{McUePdcp}};
  \node (rlc) [process, below left of=pdcp, xshift=-0.8cm] {\texttt{LteRlc}};
  \node (mmrlc) [process, below right of=pdcp, xshift=0.8cm] {\texttt{mmWaveRlc}};
  \node (ltephy) [process, below of=rlc] {\texttt{LteUeMac} \\ \texttt{LteUePhy}};
  \node (mmphy) [process, below of=mmrlc] {\texttt{mmWaveUeMac} \\ \texttt{mmWaveUePhy}};

  \draw[chaptergrey,dashdotted] ($(rrc.north west)+(-0.2,2)$) rectangle ($(mmphy.south east)+(1.5,-0.2)$);
  \node (legendMcUe) [above left of=epc, xshift=-2cm]{\textcolor{chaptergrey}{\texttt{McUeNetDevice}}};

  \node (ltephy2) [process, below of=ltephy] {\texttt{LteSpectrumPhy}};
  \node (mmphy2) [process, below of=mmphy] {\texttt{mmWaveSpectrumPhy}};
  \node (channel) [process, below of=mmphy2, yshift=-1cm] {\texttt{MultiModelSpectrumChannel} and \texttt{mmWaveBeamforming}};
  \node (ltechannel) [process, below of=ltephy2] {\texttt{MultiModelSpectrumChannel}};
  \node (enbltephy2) [process, below of=ltechannel, yshift = -1cm] {\texttt{LteSpectrumPhy}};
  \node (enbmmphy2) [process, below of=channel] {\texttt{mmWaveSpectrumPhy}};

  \draw[chaptergreen,dashdotted] ($(ltephy2.north west)+(-1.5,0.2)$) rectangle ($(enbmmphy2.south east)+(3.35,-0.2)$);
  \node (legendChannel) [above right of=channel, xshift=2cm] {\textcolor{chaptergreen}{Channel classes}};

  \node (lteEnbLl) [process, below of=enbltephy2] {\texttt{LteEnbPhy} \\ \texttt{LteEnbMac}};
  \node (mmEnbLl) [process, below of=enbmmphy2, xshift=3cm] {\texttt{mmWaveEnbPhy} \\ \texttt{mmWaveEnbMac}};
  \node (lteEnbRlc) [process, below of=lteEnbLl] {\texttt{LteRlc}};
  \node (mmEnbRlc) [process, below of=mmEnbLl] {\texttt{mmWaveRlc}};
  \node (enbPdcp) [process, below right of=lteEnbRlc, xshift=0.75cm, yshift=-0.5cm] {\texttt{McEnbPdcp}};
  \node (enbrrc) [process, below of=enbPdcp, xshift=-3cm] {\texttt{LteEnbRrc}};
  \node (enbmmrrc) [process, below of=enbPdcp, xshift=5cm] {\texttt{mmWaveEnbRrc}};
  \node (epcEnb) [process, below of=enbPdcp,yshift=-0.6cm] {\texttt{EpcEnbApplication}};
  \node (cn) [below of=epcEnb] {Core Network};

  \draw[chapterpurple,dashdotted] ($(mmEnbLl.north west)+(-0.5,0.2)$) rectangle ($(enbmmrrc.south east)+(0.5,-0.2)$);
  \node (legendChannel) [left of=enbPdcp, xshift=6.5cm] {\textcolor{chapterpurple}{\texttt{MmWaveEnbNetDevice}}};

  \draw[chapterlightgreen,dashdotted] ($(lteEnbLl.north west)+(-1.5,0.2)$) rectangle ($(epcEnb.south east)+(0.5,-0.2)$);
  \node (legendChannel) [left of=enbPdcp, xshift=-1cm] {\textcolor{chapterlightgreen}{\texttt{LteEnbNetDevice}}};


  \draw[arrow] (epc) -- (rrc);
  \draw[arrow] (epc) -- (mmrrc);
  \draw[arrow] (epc) -- (pdcp);
  \draw[arrow] (mmrrc) --  (pdcp);
  \draw[arrow] (rrc) --  (pdcp);
  \draw[arrow] (pdcp) -- (rlc);
  \draw[arrow] (pdcp) -- (mmrlc);
  \draw[arrow] (rlc) -- (ltephy);
  \draw[arrow] (mmrlc) -- (mmphy);
  \draw[arrow] (mmphy) -- (mmphy2);
  \draw[arrow] (ltephy) -- (ltephy2);
  \draw[arrow] (ltephy2) -- (ltechannel);
  \draw[arrow] (mmphy2) -- (channel);
  \draw[arrow] (channel) -- (enbmmphy2);
  \draw[arrow] ([xshift=0.4cm]ltechannel.south west) -- node[anchor=east] {} ([xshift=-0.8cm]enbltephy2.west) -- (enbltephy2.west);
  \draw[arrow] (enbmmphy2) -- (mmEnbLl);
  \draw[arrow] (enbltephy2) -- (lteEnbLl);
  \draw[arrow] (mmEnbLl) -- (mmEnbRlc);
  \draw[arrow] (lteEnbLl) -- (lteEnbRlc);
  \draw[arrow] (lteEnbRlc) -- (enbPdcp);
  \draw[arrow] (mmEnbRlc) -- node[sloped, anchor=center, above] {X2 interface} (enbPdcp);
  \draw[arrow] (enbmmrrc) --  (enbPdcp);
  \draw[arrow] (enbrrc) --  (enbPdcp);
  \draw[arrow] (enbPdcp) -- (epcEnb);
  \draw[darrow] (rrc.west) -- node[anchor=east] {} ([xshift=-2.2cm]rrc.west) -- node[anchor=east] {} ([xshift=-2cm]enbrrc.west) -- (enbrrc.west);
  \draw[darrow] (mmrrc.east) -- node[anchor=east] {} ([xshift=+3cm]mmrrc.east) -- node[anchor=east] {} ([xshift=+1cm]enbmmrrc.east) -- (enbmmrrc.east);
  \draw[darrow] (epcEnb) -- node[anchor=west] {S1 interface} (cn);

\end{tikzpicture}
\caption{Block diagram of a dual-connected device, an \gls{lte} \gls{enb} and a mmWave \gls{enb} \cite{simutoolsPolese}.}
\label{fig:mcdevice}
\end{figure}

A basic diagram for a \gls{dc} \gls{ue} device, an \gls{lte} \gls{enb} and a mmWave \gls{enb} is shown in Figure~\ref{fig:mcdevice}. The core of the \gls{dc} implementation is the \texttt{McUeNetDevice} class, which is a subclass of the ns--3 \texttt{NetDevice} and provides an interface between the ns--3 \gls{tcp}/IP stack and the custom lower layers. The \texttt{McUeNetDevice} holds pointers to the custom lower layer stack classes, and has a \texttt{Send} method that forwards packets to the \gls{tcp}/IP stack. This method is linked to a callback on the \texttt{DoRecvData} of the \texttt{EpcUeNas} class, which as specified by the \gls{3gpp} standard acts as a connection between the \gls{lte}-like protocol stack and the \gls{tcp}/IP stack.

The \texttt{McUeNetDevice} describes a dual connected \gls{ue} with a single \texttt{EpcUeNas}, but with a dual stack for the lower layers, i.e., there are separate \gls{lte} and mmWave \gls{phy} and \gls{mac} layers. Moreover, there is an instance of the \gls{rrc} layer for both links. This grants a larger flexibility, because the functionalities and the implementation of the two layers may differ. Besides, the \gls{lte} \gls{rrc} manages both the \gls{lte} connection and the control plane features related to \gls{dc}, while the mmWave \gls{rrc} handles only the mmWave link. The usage of a secondary \gls{rrc}, dedicated to the mmWave link, avoids latency in control commands (i.e., the mmWave \gls{enb} does not have to encode and transmit the control \gls{pdu}s to the master \gls{lte} \gls{enb}). The \texttt{EpcUeNas} layer has an interface to both \gls{rrc} entities to exchange information between them.

The \gls{lte} \gls{rrc} manages also the data plane for the \gls{dc} devices. In particular, for each bearer, a dual connected \gls{pdcp} layer is initialized and stored in the \gls{lte} \gls{rrc}. The classes describing the \gls{dc} \gls{pdcp} layer are \texttt{McEnbPdcp} and \texttt{McUePdcp}, respectively at the \gls{enb} side and at the \gls{ue} side. They both extend the \texttt{LtePdcp} class with a second interface to the \gls{rlc}. However, while \texttt{McUePdcp} simply has to communicate with a local \gls{rlc} in the \gls{ue}, the implementation of \texttt{McEnbPdcp} requires new interfaces to the class describing the X2 links between \gls{enb}s (i.e., \texttt{EpcX2}). In particular, in downlink the \gls{enb} \gls{pdcp} has to send packets to the X2 link and the mmWave \gls{rlc} layer has to receive them, and vice versa in uplink.

The \gls{dc} module can be used to simulate different dual connected modes, i.e., it can support both \gls{fs} and throughput-oriented dual connectivity, according to which \gls{rrc} and X2 procedures and primitives are implemented. With \gls{fs}, the \gls{ue} is in the \gls{rrc}\_CONNECTED state with respect to both \gls{enb}s, but only transmits data to one of the two. With the other option, the \gls{ue} can transmit data simultaneously on both \gls{rat}s, and different flow control algorithms can be plugged in and tested.

As to the physical layer, the two stacks rely on the mmWave and \gls{lte} channel models. Notice that since the two systems operate at different frequencies, modeling the interference between the two \gls{rat}s is not needed. Each of the two channel models can therefore be configured independently.

In order to use an \texttt{McUeNetDevice} as a mobile User Equipment in the simulation, the helper class of the mmWave module was extended with several features, such as (i) the initialization of the objects related to the \gls{lte} channel; (ii) the installation and configuration of the \gls{lte} \gls{enb}s, so that they can be connected to the \gls{lte} stack of the \texttt{McUeNetDevice}; and (iii) the methods to set up a \texttt{McUeNetDevice} and link its layers as shown in Fig~\ref{fig:mcdevice}. An example on how to set up a dual-connectivity based simulation is provided in the file \texttt{mc-twoenbs.cc}.

\textbf{\gls{rrc} Layer for Dual Connectivity and Mobility.}
The \gls{rrc} layer implementation of the original \gls{lte} ns--3 module was extended in order to account for \gls{dc}-related control procedures. In particular, the multi-connectivity uplink-based measurement framework described in~\cite{giordaniMC2016} was added with changes to the \texttt{MmWaveEnbPhy}, \texttt{EpcX2} and \texttt{LteEnbRrc} classes. The \texttt{MmWaveEnbPhy} instance simulates the reception of uplink reference signals (which are accounted for as overhead in the data bearers resource allocation), computes the \gls{sinr} for each \gls{ue} in the scenario\footnote{The framework assumes that the optimal beam is always chosen, so the actual directional scan procedure described in~\cite{giordaniMC2016} is not simulated}, and sends this information to the \gls{lte} \gls{enb} on the X2 link. This also allows the simulation of a delay in the reporting, since the control packets with the \gls{sinr} values must be transmitted on an ns--3 \texttt{PointToPointLink}, which adds a certain latency and has a certain bitrate.

Thanks to this framework, the \gls{lte} \gls{enb} is able to act as a coordinator for the surrounding mmWave \gls{enb}s, and learns which is the best association (in terms of \gls{sinr}) between \gls{ue}s and mmWave secondary \gls{enb}s. This enables automatic cell selection for mmWave \gls{enb}s at the beginning of a simulation, and the control of mobility-related operations. The \gls{dc} module is indeed capable of simulating \gls{fs} procedures between mmWave and \gls{lte} links and \gls{sch} (i.e., handovers between mmWave \gls{enb}s that do not involve the \gls{mme} in the core network) initiated by the central controller in the \gls{lte} \gls{enb}. It is also possible to use the \gls{dc} module to simulate X2-based \gls{rat} handovers between the \gls{lte} and mmWave \gls{enb}s, i.e., to use standalone \gls{ue}s based on \texttt{McUeNetDevice} that can perform handovers from the \gls{lte} to the mmWave \gls{enb}s, and vice versa.

Different handover (either inter-\gls{rat} or \gls{sch}) algorithms can be tested, by implementing them in the \texttt{LteEnbRrc} class. In order to make the handover simulation more compliant with the \gls{3gpp} specifications, the lossless handover option implemented for ns--3 in~\cite{llho} was adapted to the \gls{dc} module in order to forward the \gls{rlc} buffer content to the target \gls{rat}/\gls{enb} \gls{rlc} layer for both the \gls{sch} and the \gls{fs}. Moreover, in order to model the additional latency given by the interaction with the \gls{mme} for inter-\gls{rat} handovers for standalone \gls{ue}s, the link between the \gls{enb}s and the \gls{mme} is modeled in this module as a \texttt{PointToPointLink}, while in the original ns--3 \gls{lte} module it is an ideal connection.

\section{Use Cases}
\label{sec:results}

In this section, we illustrate various examples of scenarios\footnote{The simulations in this section are all configured with the basic \gls{phy} and \gls{mac} parameters in Table \ref{tab:phy_mac_params}, with other notable parameters given in the sequel.} that can be simulated to show the utility of the module for the analysis of novel mmWave protocols and for testing higher-layer network protocols, such as \gls{tcp}, over \gls{5g} mmWave networks. After each particular use case example, we also provide to the interested readed some references to recent papers that report additional results obtained using the ns--3 mmWave simulation module.

\subsection{Simulation Setup Walk-through}

\begin{table*}[t!]
	\centering
	\small
	\renewcommand{\arraystretch}{1}
	\begin{tabular}{@{}llllll@{}}
		\toprule 
		& & \multicolumn{4}{c}{\gls{phy}-layer throughput [Mbit/s]} \\ \midrule
		 & Policy & Cell & Mean \gls{ue} & Mean 5\% Worst \gls{ue} & Max \gls{ue} \\
		\midrule
		
		{\multirow{4}{*}{\parbox[t]{1.2cm}{70 \gls{ue}/\\10 Mbps}}}& \gls{rr} & 1815.92 & 	 25.94 & 	 1.11 & 	 48.80 \\
		 & \gls{pf} & 1494.61 & 	 21.35 & 	 3.26 & 	 43.94 \\
		 & \gls{mr} & 2273.18 & 	 32.47 & 	 0.00 & 	 151.36 \\ 
		 & \gls{edf} & 925.80 & 	 13.23 & 	 7.31 & 	 31.02 \\  
		\midrule		
		\multirow{4}{*}{\parbox[t]{1.2cm}{7 \gls{ue}/\\100 Mbps}}& \gls{rr} & 715.26 & 	 102.18 & 	 49.18 & 	 134.28 \\
		& \gls{pf} & 758.32 & 	 108.33 & 	 52.32 & 	 158.16 \\
		& \gls{mr} & 766.26 & 	 109.47 & 	 47.26 & 	 158.22 \\ 
		& \gls{edf} & 647.98 & 	 92.57 & 	 63.89 & 	 121.76 \\  \bottomrule
	\end{tabular}
	\caption{\gls{dl} \gls{phy} throughput for \gls{rr}, \gls{pf}, \gls{mr} and \gls{edf} scheduling policies.}
	\label{tab:multiuser_phy_rate}
\end{table*}

\begin{table*}[t!]
\centering
	\begin{minipage}{0.45\textwidth}
		\centering
		\small
		\renewcommand{\arraystretch}{1}
		\begin{tabular}{@{}lllll@{}}
			\toprule	
			& & \multicolumn{3}{c}{IP-layer latency [ms]} \\ \midrule
			 & Policy & Mean \gls{ue} & Mean 5\% Worst \gls{ue} & Max \gls{ue} \\
			\midrule
			\multirow{4}{*}{\parbox[t]{1.2cm}{70 \gls{ue}/10 Mbps}}& \gls{rr} & 7.47 & 	 69.35 & 	 118.62 \\ 
			& \gls{pf} & 2.83 & 	 34.48 & 	 106.54\\
			& \gls{mr} & 0.65 & 	 1.89 & 	 3.07 \\
			& \gls{edf} & 1.63 & 	 7.91 & 	 30.65 \\
			\midrule
			
			\multirow{4}{*}{\parbox[t]{1.2cm}{7 \gls{ue}/100 Mbps}}& \gls{rr} & 0.67 & 	 2.01 & 	 2.37 \\ 
			& \gls{pf} & 0.55 & 	 0.68 & 	 0.77 \\ 
			& \gls{mr} & 0.56 & 	 0.78 & 	 1.09 \\ 
			& \gls{edf} & 0.69 & 	 1.41 & 	1.44 \\ 
			\bottomrule
		\end{tabular}
		\caption{IP-layer latency for \gls{rr}, \gls{pf}, \gls{mr} and \gls{edf} scheduling policies.}
		\label{tab:multiuser_lat}	
	\end{minipage}%
	\qquad\qquad
	\begin{minipage}{0.45\textwidth}
		\centering
	\small
	\renewcommand{\arraystretch}{1}
	\begin{tabular}{@{}lllll@{}}
		\toprule	
		& Policy & Fairness & Utilization \\
		\midrule
		\multirow{4}{*}{\parbox[t]{1.2cm}{70 \gls{ue}/10 Mbps}}& \gls{rr} & 0.71 & 	 0.53 \\ 
		& \gls{pf} & 0.76 & 	 0.73 \\
		& \gls{mr} & 0.28 & 	 0.39 \\
		& \gls{edf} & 0.96 & 	 0.87 \\  
		\midrule
		\multirow{4}{*}{\parbox[t]{1.2cm}{7 \gls{ue}/100 Mbps}}& \gls{rr} & 0.95 & 	 0.74 \\
		& \gls{pf} & 0.90 & 	 0.77 \\
		& \gls{mr} & 0.91 & 	 0.76 \\
		& \gls{edf} & 0.99 & 	 0.84 \\  \bottomrule
	\end{tabular}
	\caption{Fairness index and utilization (received IP-layer rate/allocated \gls{phy} rate) for \gls{rr}, \gls{pf}, \gls{mr} and \gls{edf} scheduling policies.}
	\label{tab:multiuser_fairness_util}
	\end{minipage}
\end{table*}

In order to proficiently use the mmWave ns--3 module, a basic knowledge of ns--3 is required. We therefore advise the interested users to first study the extensive documentation referenced in Sec.~\ref{sec:ns3}. Moreover, we provide some basic ns--3 scripts in the \texttt{examples} folder of the mmWave module, that can be a basis for the design of any simulation script that uses the mmWave module. In the following paragraphs, we will describe the basic structure of a typical example in simple steps.

The first step is to configure all the attributes needed in a simulation. A complete list of attributes related to the mmWave module can be found in the \texttt{mmWaveAttributesList} file in the module repository. The second step involves the setup of the \texttt{MmWaveHelper} object, which provides methods to create the entities involved in the simulation (e.g., the channel-related objects and the \texttt{MmWavePhyMacCommon} object), to install the mmWave stack over ns--3 nodes (for both \gls{ue}s and \gls{enb}s), to perform the initial attachment of a \gls{ue} to the closest \gls{enb} and to enable or disable the generation of simulation traces. Moreover, if the scenario of interest is an end-to-end scenario, the core network and the internet must be set up as well. The first is created by the \texttt{MmWavePointToPointEpcHelper}, which also provides a pointer to the Packet Gateway (PGW) node. This is then usually connected to a remote host, and the internet stack (i.e., the \gls{tcp}/IP protocol suite) is added to the \gls{ue}s and to the remote host.

In the third step, the positions and velocities of the \gls{enb}s and \gls{ue}s are specified using one or more \texttt{MobilityHelper} objects and different mobility models. Moreover, buildings and obstacles can be added to the scenario using the ns--3 buildings module and the \texttt{Buildings} and \texttt{BuildingsHelper} objects. The fourth step requires the setup of applications in the \gls{ue}s and in the remote host (if an end-to-end scenario is considered), in order to simulate downlink and uplink traffic. ns--3 provides a wide range of different applications, and helpers that take care of their setup. They can run on either UDP or \gls{tcp} sockets, and several \gls{tcp} congestion control versions are available. Finally, the simulation can be run using the \texttt{Simulator} object of ns--3, and traces are generated.

\begin{figure}[t]
\setlength{\belowcaptionskip}{0cm}
	\centering
	\begin{subfigure}[t]{0.7\columnwidth}
		\centering
		\includegraphics[trim={4cm 8.5cm 4cm 8.5cm},width=.9\textwidth]{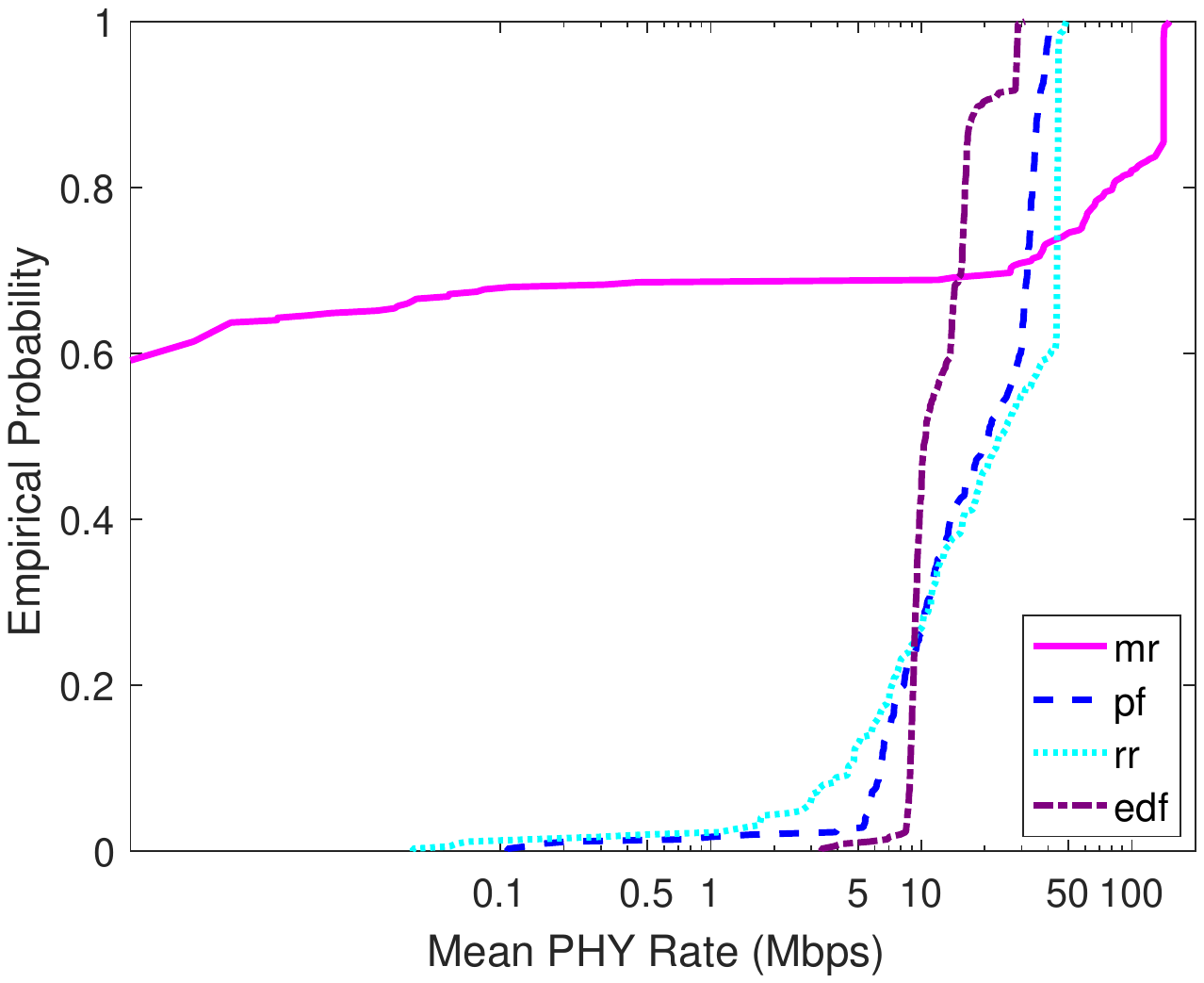}
		\caption{Empirical CDF of rate}
		\label{fig:multiuser_rate_10mbps}
	\end{subfigure}
	\begin{subfigure}[t]{0.7\columnwidth}
		\centering
		\includegraphics[trim={4cm 8.5cm 4cm 8.5cm},width=.9\textwidth]{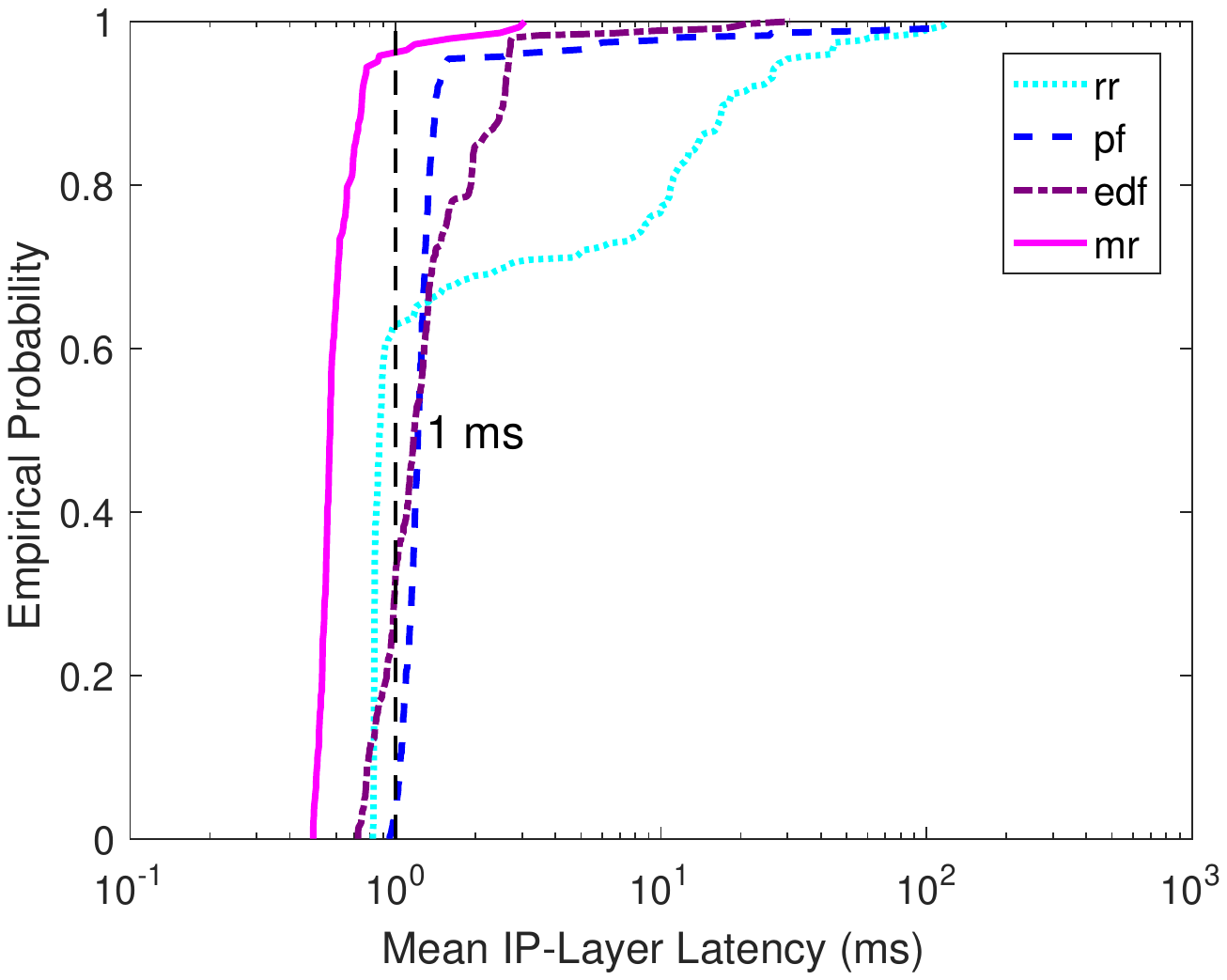}
		\caption{Empirical CDF of latency}
		\label{fig:multiuser_lat_10mbps}
	\end{subfigure}
	\caption{Distributions of \gls{phy}-layer throughput and IP-layer latency for 70 \gls{ue}s, 10 Mbps/\gls{ue} arrival rate}
	\label{fig:multiuser_10mbps}
\end{figure}

\begin{figure}[t]
\setlength{\belowcaptionskip}{0cm}
	\centering
	\begin{subfigure}[t]{0.7\columnwidth}
		\centering
		\includegraphics[trim={4cm 8.5cm 4cm 8.5cm},width=.9\textwidth]{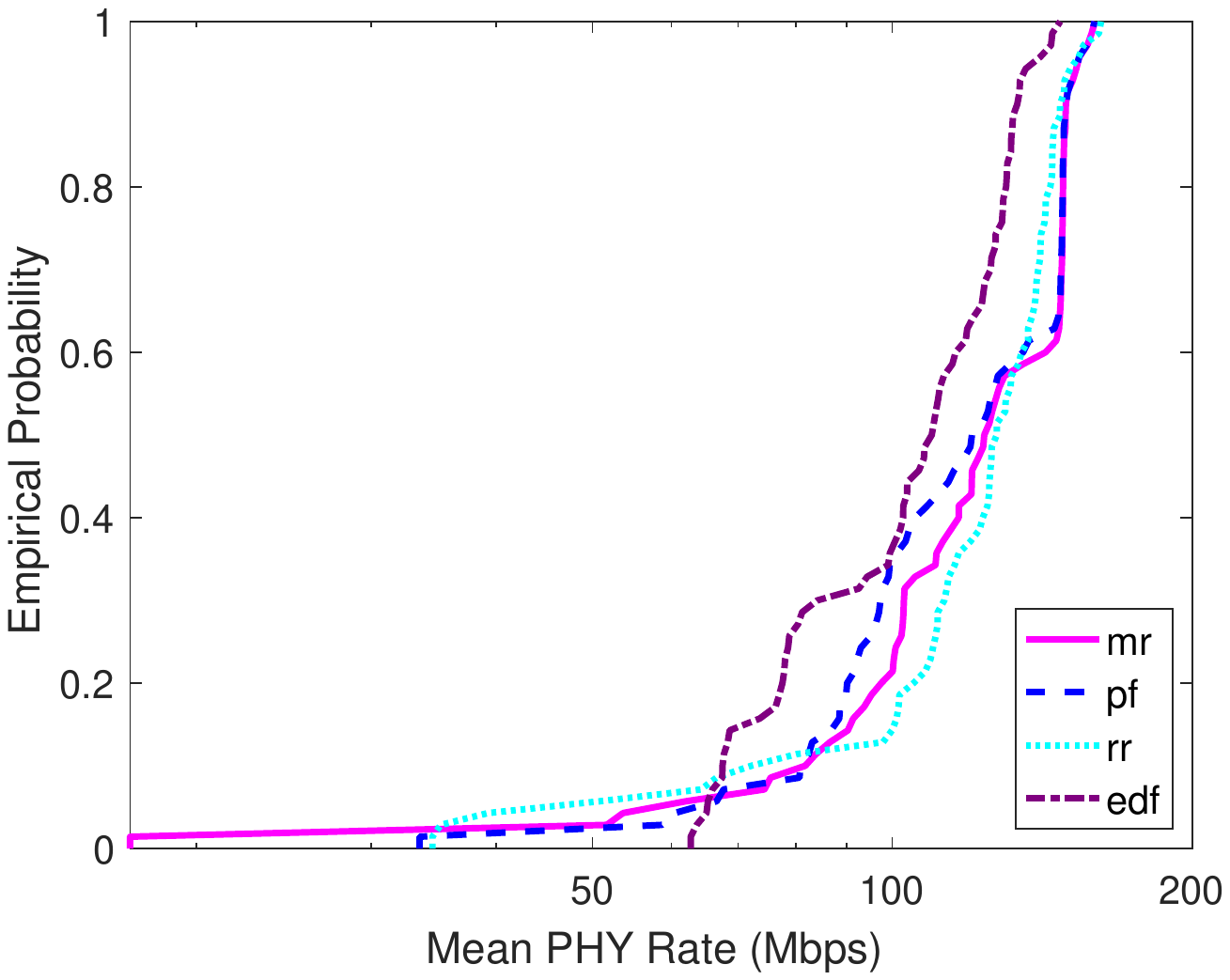}
		\caption{Empirical CDF of rate}
		\label{fig:multiuser_rate_100mbps}
	\end{subfigure}
	\begin{subfigure}[t]{0.7\columnwidth}
		\centering
		\includegraphics[trim={4cm 8.5cm 4cm 8.5cm},width=.9\textwidth]{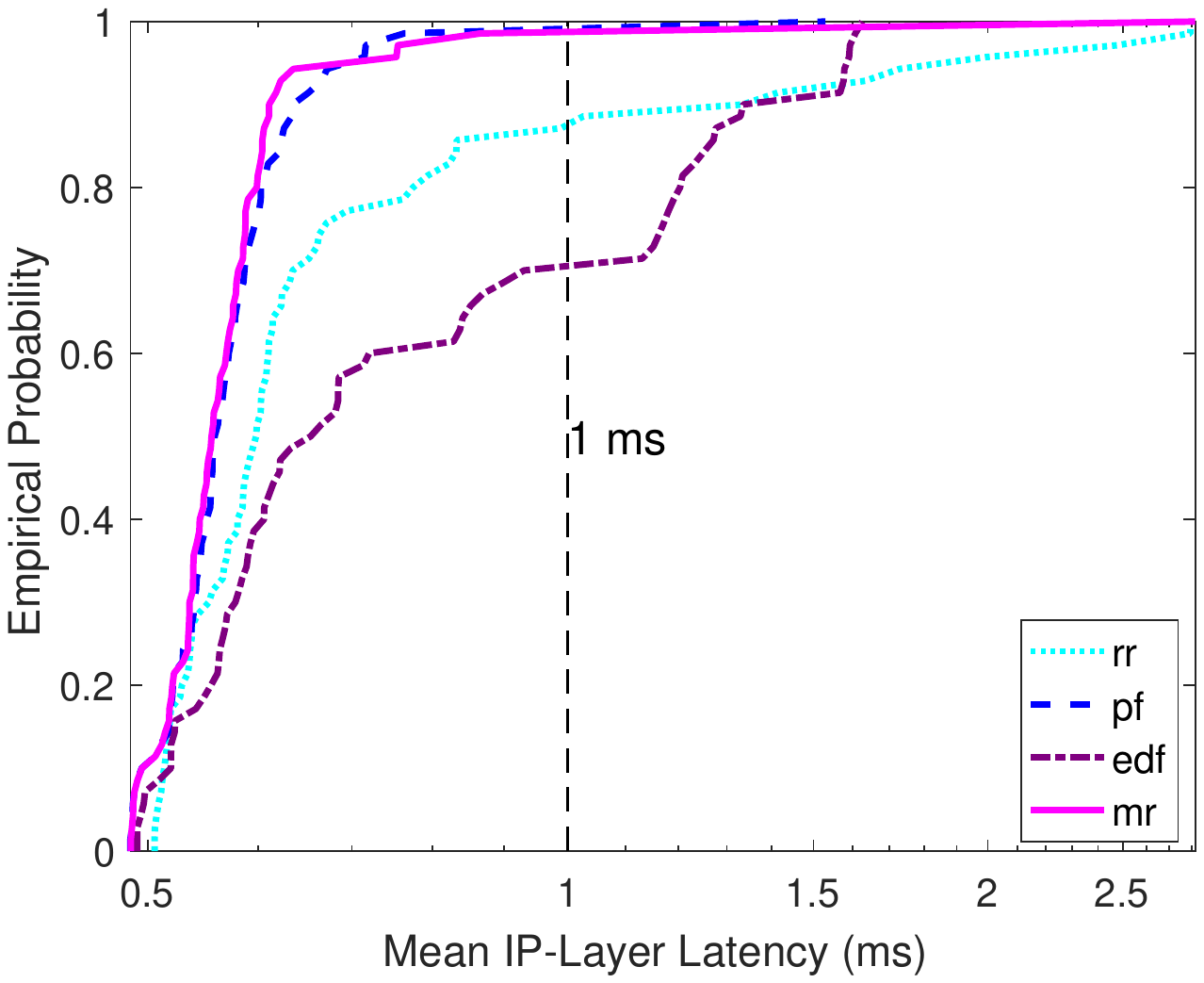}
		\caption{Empirical CDF of latency}
		\label{fig:multiuser_lat_100mbps}
	\end{subfigure}
	\caption{Distributions of \gls{phy}-layer throughput and IP-layer latency for 7 \gls{ue}s, 100 Mbps/\gls{ue} arrival rate}
	\label{fig:multiuser_100mbps}
\end{figure}

\subsection {Multi-User Scheduling Simulation}
In this experiment, the throughput and latency of users of a mmWave cell with 1 GHz of bandwidth are simulated for variable \gls{tti} and each of the scheduling policies described in Section~\ref{sec:mac}.\footnote{The multi-user scheduling experiment can be reproduced by running the the \texttt{mmwave-epc-tdma.cc} example simulation \cite{github-ns3-mmwave}.} We shall see how the choice of the scheduler has a significant impact on the subframe utilization and latency of the multi-user cell. In these scenarios, \gls{ue}s have similar distances from the \gls{enb} but are assigned the constant speed of 25 m/s (typical of vehicular users), which results in a lower achievable rate, on average, as well as increased packet errors compared to walking users due to the more rapid variation in the channel.

The simulation is again run over 10 drops for each of two scenarios and using default parameters from Table~\ref{tab:phy_mac_params}. In the first scenario, 70 \gls{ue}s are simulated with each \gls{ue} generating IP-layer traffic at an average arrival rate of 10 Mbps. In the second scenario, 7 \gls{ue}s are simulated with a 100 Mbps arrival rate per \gls{ue}.

These specific combinations of users and rates are deliberately chosen because they illustrate the cut-off point at which the system is no longer able to service most users at the requested rate, leading to backlogged queues and increased latency. That is, we wish to analyze the performance at the knee in the curve of the delay taken as a function of the system utilization. In the variable \gls{tti} system, this bottleneck effect has the following potential causes: (i) the number of users that must be serviced exceeds the number of available slots (ultimately limited by the number of time-domain symbols), independently of the total throughput requested by the users; (ii) the number of users that are connected to \gls{enb} is smaller than the number of available slots, but the total throughput they request exceeds the available resources in the given time period; or (iii) a combination of the previous cases.

These effects are demonstrated in Figures~\ref{fig:multiuser_10mbps} and \ref{fig:multiuser_100mbps} for the 70 \gls{ue}/10 Mbps and 7 \gls{ue}/100 Mbps arrival rate scenarios, respectively. The mean, maximum and cell-edge (i.e., 5\% worst-case) user \gls{phy} rates and IP-to-IP layer latencies are also provided in Tables~\ref{tab:multiuser_phy_rate} and \ref{tab:multiuser_lat} along with the utilization and Jain's Fairness Index in Table~\ref{tab:multiuser_fairness_util}.

	

For the 70 \gls{ue} case, Figure~\ref{fig:multiuser_rate_10mbps} shows the distribution of the mean rate experienced by each \gls{ue} over the simulation duration. It can be seen that the \gls{mr} and \gls{rr} policies exhibit the greatest disparity between users scheduled with high and low rates.

\begin{figure*}[ht!]
\setlength{\belowcaptionskip}{0cm}
	\centering
	\begin{subfigure}[t]{.43\textwidth}
		\includegraphics[width=.65\textwidth, trim={4cm 8.5cm 6.5cm 8cm}]{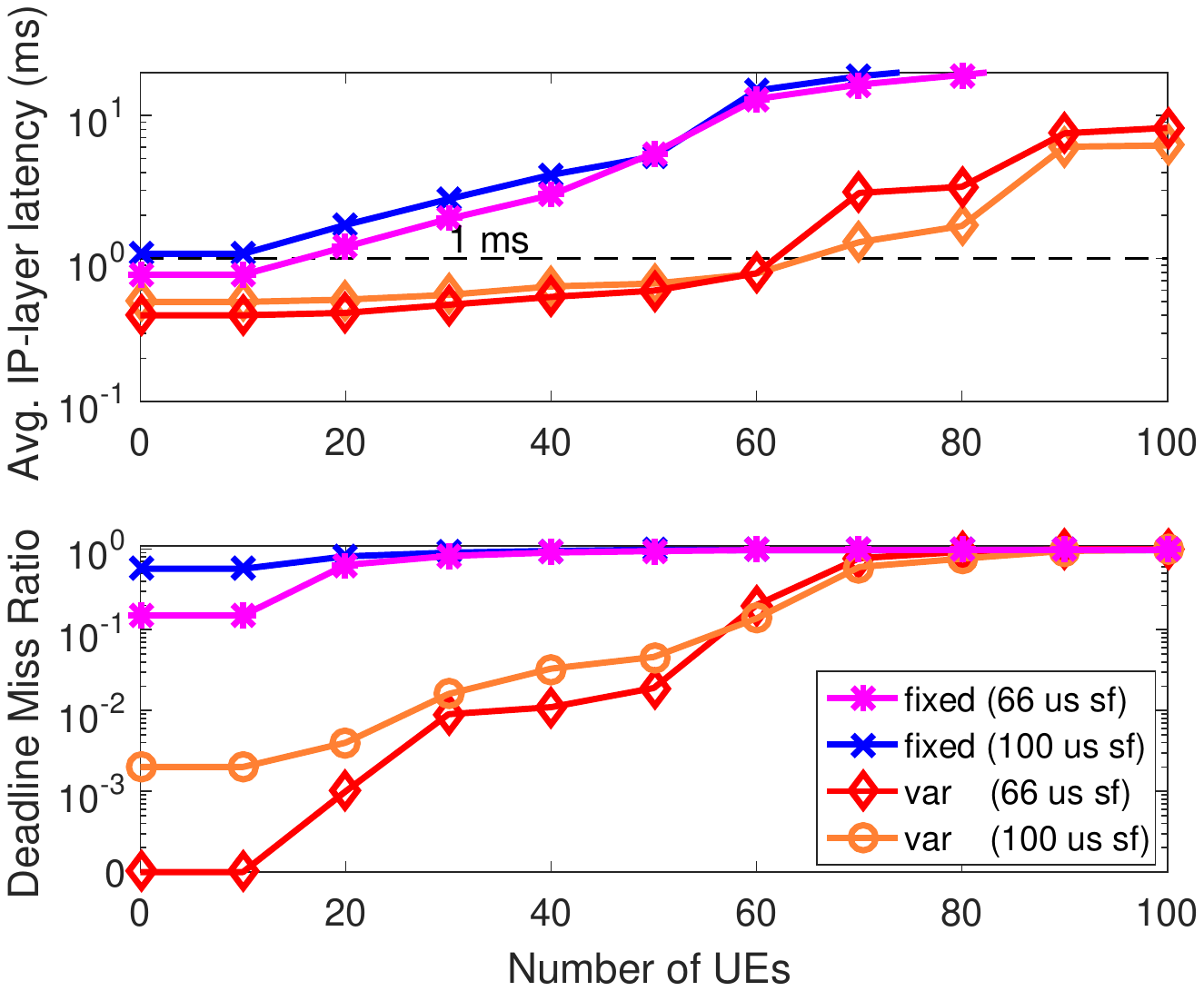}
		\caption{10 Mbps per \gls{ue} arrival rate (100-bytes packets)}
		\label{fig:lat_dmr_10mbps}
	\end{subfigure}
	\begin{subfigure}[t]{.43\textwidth}
		\centering
		\includegraphics[width=.65\textwidth, trim={4cm 8.5cm 6.5cm 8cm}]{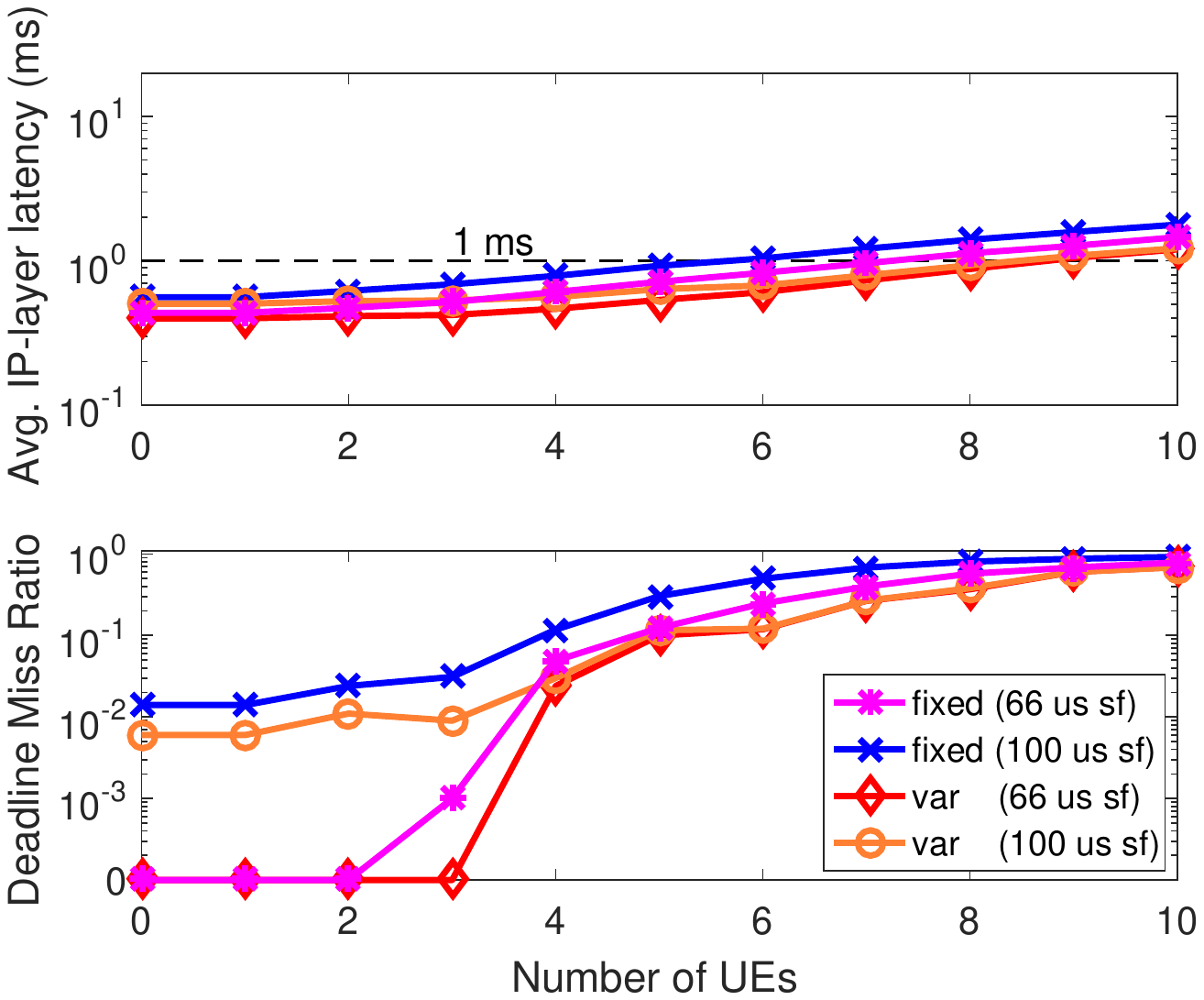}
		\caption{100 Mbps per \gls{ue} arrival rate (1200-bytes packets)}
		\label{fig:lat_dmr_100mbps}
	\end{subfigure}\quad
	\caption{Latency and Deadline Miss Ratio as a function of the downlink IP-layer arrival rate for fixed and variable \gls{tti} radio frame structures.}
	\label{fig:lat_dmr}
\end{figure*}

\begin{table*}[t]
\centering
\small
\renewcommand{\arraystretch}{1}
\begin{tabular}{@{}ll@{}}
\toprule
 Description &  Value  \\
\midrule
Subframe length in \textmu s & 100/66.67    \\
\gls{ofdm} symbols per slot & 24/16 \\
\gls{harq} processes (\gls{dl} and \gls{ul}) & 20 \gls{dl}/20 \gls{ul} \\
Number of \gls{ue}s & Case 1: $\{10,20,30,40,50,60,70,80,100\}$ \\
Number of \gls{ue}s & Case 2: $\{1,2,3,4,5,6,7,8,10\}$ \\
Traffic model & Case 1: Poisson, $\lambda = 12.5K$  pck/s, 100 B packets \\
Traffic model & Case 2: Poisson, $\lambda = 83K$  pck/s, 1200 B packets \\
\bottomrule
\end{tabular}
\caption{Additional parameters for variable and fixed \gls{tti} latency experiment.}
\label{tab:phy_mac_params2}
\end{table*}

It can also be observed that the \gls{phy} rate significantly exceeds the 10 Mbps arrival rate for some users, which leads to the poor utilization for these two policies, as shown in Table~\ref{tab:multiuser_fairness_util}. The reason why the utilization (defined as the ratio of the received IP-layer rate to the allocated \gls{phy}-layer rate for each terminal) suffers in these cases is that the \gls{ue}s with higher achievable rates are heavily favored by the \gls{mr} and \gls{rr} schedulers. As these users are typically scheduled at a higher \gls{mcs} level, even a single 4.16 \textmu s-long time-domain symbol has the capacity to transmit kilobytes of data, which cannot be fully taken advantage of given the low 10 Mbps arrival rate. Insufficient data is buffered at the \gls{mac} layer to utilize the full slot and useless padding bits must be added. This effect is felt less by users under the \gls{pf} and \gls{edf} policies, which are inherently more fair and allow more resources to be scheduled for lower-\gls{mcs} users.

The ensuing effect of these trends on latency is shown in Figure~\ref{fig:multiuser_lat_10mbps}. Here latency is measured as the time between the arrival time of packets at the \gls{pdcp} layer of the \gls{enb} stack and the time they are delivered to the IP layer at the \gls{ue}. Naturally, the \gls{mr} scheduler offers the best delay performance because only 40\% of the users, i.e., those with the highest rates, are ever scheduled (unscheduled users with zero rate are not included in the figure). The \gls{rr} policy offers the highest worst-case delays but is able to achieve mean latencies of less than 1 ms for more than 60\% of the users. Of all the policies besides \gls{mr}, Earliest-Deadline First offers the best worst-case delays, as it attempts to balance the delay of all users by scheduling them exclusively based on their relative deadlines (not taking into account achievable rates). The \gls{edf} scheduler is able to achieve a mean \gls{ue} latency of 1.6 ms, which, as we will see from the experiment in the next section, drops below 1 ms for 60 or fewer users (with the same arrival rate).

Finally, it can be observed from Figure~\ref{fig:multiuser_lat_100mbps} that, despite having the same total packet arrival rate of 700 Mbps as in the 70 \gls{ue} case, latencies are much lower overall in the 7 \gls{ue}/100 Mbps per \gls{ue} case. This can be clearly explained by the higher utilization factor in Table~\ref{tab:multiuser_fairness_util}.
In this scenario, the number of available slots for scheduling different users is no longer the bottleneck. Though we still see that a significant number of users are scheduled at rates that exceed their 100 Mbps arrival rates, the utilization is notably better than the 10 Mbps case for all scheduling policies. Thus, the channel capacity itself is better utilized, allowing most users to be scheduled at the requested rate, thereby avoiding additional queue wait time and delay.

Additional results on the impact of the scheduling on throughput and latency can be found in~\cite{ford2017low}.

\subsection{Latency Evaluation for Variable and Fixed \gls{tti} Schemes}
The results and the discussion introduced in this section are taken from our previous article~\cite{ford2016achieving}, where the interested reader will find a more comprehensive treatment of techniques to achieve low latency in mmWave 5G cellular systems. While the qualitative benefits of variable \gls{tti} over fixed \gls{tti} may seem self-evident, in this section we quantify the performance gains for a multi-user \gls{tdma} mmWave system with 1 GHz of bandwidth. We also demonstrate that, with the low-latency scheduling loop enabled by the proposed frame structure, \gls{lte}-style Hybrid ARQ can still be employed for enhanced link-layer reliability without excessively exceeding the delay constraints. We model the subframe formats shown in Figure \ref{fig:frame_structure} for two subframe periods: the default 100 \textmu s subframe, equivalent to 24 \gls{ofdm} symbols, and a 66.67 \textmu s subframe, equivalent to 16 \gls{ofdm} symbols. The symbol length of 4.16 $\mu s$ is based on the numerology in \cite{KahnMmw:12}.
Each subframe has one fixed \gls{dl}-CTRL and one \gls{ul}-CTRL symbol, with the remaining symbols used for \gls{dl} or \gls{ul} data slots. For fixed \gls{tti} mode, the entire subframe is allocated to a single user, whereas for variable \gls{tti} mode, the scheduler may allocate any number of data symbols within the subframe to match the throughput required by each user.

We also note that \gls{ue}s are again modeled as moving at 25 m/s, typical of vehicular speeds, which causes fast channel variation and frequent packet errors from small-scale fading (it is observed that between 0.5\% and 3\% of the transport blocks are lost and require retransmission).

We consider a simple traffic model with Poisson arrivals, where each \gls{ue} sends an average of 12.5K packets per second (100-bytes packets resulting in an average rate of 10 Mbps), as well as a separate, higher-throughput case where each \gls{ue} sends an average of 83K packets per second (1200-bytes packets resulting in an average rate of 100 Mbps). Scheduling is performed based on the \gls{edf} policy where the scheduler attempts to deliver each IP packet within 1 ms from its arrival at the \gls{pdcp} layer and packets are assigned a priority based on how close they are to the deadline. Priority is therefore always given to \gls{harq} retransmissions. We simulate the performance for between 10 and 100 \gls{ue}s for a 10 Mbps (per \gls{ue}) arrival rate and between 1 and 10 \gls{ue}s for the 100 Mbps case, equivalent to a total IP-layer arrival rate of between 100 and 1000 Mbps in both cases.

In Fig. \ref{fig:lat_dmr}, the downlink radio link latency is averaged among the best 95\% of the users (i.e., the 5\% of \gls{ue}s experiencing the highest latency are not considered). The \gls{dmr}, which represents the fraction of packets delivered after the 1 ms deadline, is also given for the top 95th percentile \gls{ue}s. We see that, for a 10 Mbps arrival rate (Figure \ref{fig:lat_dmr_10mbps}), variable \gls{tti} is able to achieve sub-ms average latency and a \gls{dmr} of about 10\% with over 60 users (corresponding to a 600 Mbps total packet arrival rate) and consistently outperforms fixed \gls{tti}. Fixed \gls{tti}, despite the relatively short subframe compared to \gls{lte}, exceeds 1 ms average latency and has a \gls{dmr} of over 60\% even for the 20 \gls{ue} case and of more than 90\% for 40 or more users. This result shows that variable \gls{tti} will be essential for reliable, low-latency service, particularly when considering use cases with many lower-rate devices, such as \glspl{mtd}~\cite{BocHLMP:14}.

For the higher-load (100 Mbps arrival rate per \gls{ue}) case in Figure \ref{fig:lat_dmr_100mbps}, we expect the deviation between the variable and fixed \gls{tti} schemes to be less pronounced, as the bottleneck is now the multi-user channel capacity and not the minimum slot size. However, we do find a reduction in radio link latency of around 500 \textmu s, or 30\%, for the variable \gls{tti} scheme in some cases.

We also find that, for a smaller number of users, the shorter 66.67 \textmu s subframe offers some improvement over the longer 100 \textmu s subframe thanks to the decreased turnaround time. In particular, the \gls{dmr} is consistently less for the 100 Mbps/\gls{ue} case for both variable and fixed \gls{tti}. However, this trend reverses with more users due to the lower ratio of control to data symbols in the 100 \textmu s subframe case. We note that the control overhead could be somewhat mitigated by multiplexing data in the \gls{dl}-CTRL region. 

We also note that, in real-world implementations, there may be some additional delay related to beam tracking (i.e., for computing and applying the optimal \gls{tx}/\gls{rx} beamforming vectors), although the performance limitations of adaptive beamforming transceivers and channel tracking techniques in future implementations are still unknown. We assume that this delay can be neglected in our analysis because data is constantly being transmitted to each \gls{ue} and channel state feedback is being transmitted by the \gls{ue}s to the \gls{enb} in each subframe period (which is well within the coherence time observed in many studies), thus ensuring that the channel state information is always up-to-date at the \gls{enb}.

The performance of a mmWave cellular network with respect to the end-to-end or the \gls{ran} latency has been studied in several papers with the ns-3 mmWave module. In~\cite{ford2016achieving,ford2017low} the authors discuss architectural and protocol solution for low latency networks based on mmWave, while~\cite{zhang2016transport,polese2017mptcp,tcp-asilomar,drago2018reliable} propose and evaluate latency-reduction techniques for transport protocols and video streaming. 

\subsection{\gls{tcp} Performance over mmWave}

\begin{figure}[t!]
    \centering
    \includegraphics[trim={2.5cm 1.4cm 2cm 0cm},clip, width=0.9\columnwidth]{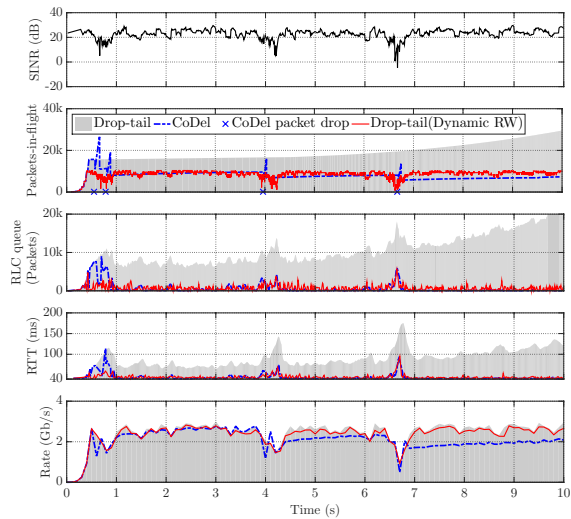}
    \caption{\gls{tcp} performance of a single \gls{ue} with human blockages}
    \label{fig:CoDelH}
\end{figure}

Another typical use case involves the simulation of end-to-end networks, in order to assess the performance of higher-layer protocols on top of mmWave links. An example of these kinds of simulations is shown in Figure~\ref{fig:CoDelH}, where we evaluated the \gls{tcp} performance with an \gls{enb} that uses Drop-tail or CoDel queue management at the \gls{rlc} layer, and a mobile \gls{ue} is experiencing blockages from other humans. The sender opens a \gls{ftp} connection and sends a large file to the \gls{ue}. The congestion control is \gls{tcp} Cubic, with delayed acknowledgment disabled. The maximum queue length is 50K packets. The core network latency is 40 ms. The \gls{ue} is walking at 1 m/s, 300 meters away from the base station, while maintaining \gls{los} connectivity, and experiencing 3 human blocking events.

\textbf{Drop-tail:} Since the \gls{rlc} queue size is large enough and all packets lost in the wireless link are recovered by means of lower layer retransmissions (\gls{rlc} \gls{am} and \gls{mac} \gls{harq}), the sender is unaware of the packet loss, thus keeping a large congestion window that results in high throughput, but also high buffer occupancy and consequent high delay.

\textbf{CoDel:} CoDel has the ability to actively drop packets when it detects high buffering delay. The CoDel packet drop events are also labeled in Figure~\ref{fig:CoDelH}. At 0.5 s, as the \gls{rlc} queue is building up, the first packet is dropped, which informs the sender to reduce the congestion window. At 0.8 s, a human blockage deteriorates the wireless link capacity and causes the \gls{rlc} queue to grow, thus triggering one more packet drop. Similarly, at 4 s and 6.7 s, two more packets are dropped. As a consequence, the congestion window will decrease and all the packets buffered in the \gls{rlc} will be delivered. Nonetheless, after the wireless link has recovered from a human blockage, the congestion window ramps up to link capacity very slowly.

\label{sec:tcp}

\textbf{Dynamic \gls{rw}:} Our previous results in \cite{mengleiBufferbloat} showed that a user may mitigate the delay by sending \gls{tcp} acknowledgments containing Dynamic \gls{rw}. The optimal \gls{rw} is fed back to the sender, and the sender takes the minimum of the receiver window and the congestion window as the sending window. As a result, the sending rate is precisely regulated so that the delay is reduced without rate degradation.

Additional results on the performance of \gls{tcp} on mmWave cellular networks obtained with ns-3 mmWave can be found in the literature. In~\cite{zhang2016transport,mengleiBufferbloat,jimenez2017analysis,kassler2017tcp,polese2017mptcp} the authors investigate the performance of different congestion control algorithms, considering in particular the throughput-latency trade off. Proxy architectures that improve the throughput~\cite{kim2017enhancing} and both throughput and latency~\cite{tcp-asilomar} have been proposed, as well as uplink-based cross layer congestion control algorithms~\cite{azzino2017xtcp}. The performance of \gls{tcp} with different mobility management schemes is studied in~\cite{tcp-mobicom}. Finally,~\cite{polese2017mptcp,polese2017tcp} use ns-3 mmWave to analyze the performance of Multipath TCP~\cite{rfc6182} in mmWave and LTE cellular networks.

\begin{figure}[t!]
\centering
\begin{tikzpicture}[font=\sffamily, scale=0.4, every node/.style={scale=0.4}]
  \centering

    \node[anchor=south west,inner sep=0] (image) at (0,0) {\includegraphics[width=0.9\textwidth]{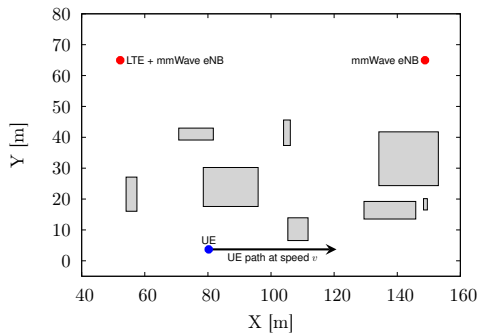}};
    \begin{scope}[x={(image.south east)},y={(image.north west)}]
        \filldraw[red,ultra thick] (0.24,0.8) circle (2pt);
        \node[anchor=west] at (0.245,0.8) (mm1label) {\gls{lte} + mmWave \gls{enb}};

        \filldraw[red,ultra thick] (0.86,0.8) circle (2pt);
        \node[anchor=east] at (0.855,0.8) (mm2label) {mmWave \gls{enb}};

        \draw[sarrow] (0.42, 0.25) -- node[anchor=north] {\gls{ue} path at speed $v$} (0.68, 0.25);

        \filldraw[blue,ultra thick] (0.42, 0.25) circle (2pt);
        \node[anchor=north] at (0.42, 0.30) (ltelabel) {\gls{ue}};
    \end{scope}
\end{tikzpicture}
\caption{Random realization of the simulation scenario. The grey rectangles are randomly deployed non-overlapping obstacles.}
\label{fig:dcscenario}
\end{figure}

\begin{figure}[t!]
\setlength{\belowcaptionskip}{0cm}
\begin{subfigure}[t]{\columnwidth}
	\centering
	\setlength\fwidth{0.9\columnwidth}
  	\setlength\fheight{0.5\columnwidth}
%
%
\definecolor{mycolor1}{rgb}{0.53003,0.74911,0.46611}%
\definecolor{mycolor2}{rgb}{0.20810,0.16630,0.52920}%
\begin{tikzpicture}
\tikzstyle{every node}=[font=\scriptsize]
\pgfplotsset{every x tick label/.append style={font=\tiny, yshift=0.5ex}}
\pgfplotsset{every y tick label/.append style={font=\tiny}}

\begin{axis}[%
width=0.9\fwidth,
height=0.4\fheight,
scale only axis,
xmin=8,
xmax=12,
separate axis lines,
every outer y axis line/.append style={mycolor1},
every y tick label/.append style={font=\color{mycolor1}\tiny},
every y tick/.append style={mycolor1},
ymin=0.9,
ymax=3.1,
ytick={1, 2, 3},
axis background/.style={fill=white},
title style={font=\bfseries},
legend style={legend cell align=left,align=left,draw=white!15!black,at={(0.01,0.03)},anchor=south west},
]
\addplot[const plot, color=mycolor1] table[row sep=crcr] {%
8		3\\
8.318	3\\
8.318	3\\
8.318	3\\
8.318	3\\
8.318	3\\
8.318	3\\
8.318	3\\
8.318	3\\
8.9704	2\\
8.9704	2\\
8.9704	2\\
8.9704	2\\
8.9704	2\\
8.9704	2\\
8.9704	2\\
8.9704	2\\
8.9704	2\\
9.0125	3\\
9.0125	3\\
9.0125	3\\
9.0125	3\\
9.0125	3\\
9.0125	3\\
9.0125	3\\
9.0125	3\\
9.0125	3\\
9.0125	3\\
9.6044	2\\
9.6044	2\\
9.6044	2\\
9.6044	2\\
9.6044	2\\
9.6044	2\\
9.6044	2\\
9.6044	2\\
9.6044	2\\
9.6044	2\\
9.6044	2\\
9.7864	3\\
9.7864	3\\
9.7864	3\\
9.7864	3\\
9.7864	3\\
9.7864	3\\
9.7864	3\\
9.7864	3\\
9.7864	3\\
9.7864	3\\
9.7864	3\\
9.7864	3\\
9.806	1\\
9.9052	2\\
9.9052	2\\
9.9052	2\\
9.9052	2\\
9.9052	2\\
9.9052	2\\
9.9052	2\\
9.9052	2\\
9.9052	2\\
9.9052	2\\
9.9052	2\\
9.9052	2\\
9.9052	2\\
10.7125	3\\
10.7125	3\\
10.7125	3\\
10.7125	3\\
10.7125	3\\
10.7125	3\\
10.7125	3\\
10.7125	3\\
10.7125	3\\
10.7125	3\\
10.7125	3\\
10.7125	3\\
10.7125	3\\
10.7125	3\\
11.2168	2\\
11.2168	2\\
11.2168	2\\
11.2168	2\\
11.2168	2\\
11.2168	2\\
11.2168	2\\
11.2168	2\\
11.2168	2\\
11.2168	2\\
11.2168	2\\
11.2168	2\\
11.2168	2\\
11.2168	2\\
11.2168	2\\
12		2\\
};
\addlegendentry{CellId};

\end{axis}

\begin{axis}[%
width=0.9\fwidth,
height=0.4\fheight,
scale only axis,
xmin=8,
xmax=12,
every outer y axis line/.append style={mycolor2},
every y tick label/.append style={font=\color{mycolor2}\tiny},
every y tick/.append style={mycolor2},
ymin=0,
ymax=500,
ytick={         0,  250, 500},
axis x line*=bottom,
axis y line*=right,
legend style={legend cell align=left,align=left,draw=white!15!black,at={(0.99,0.03)},anchor=south east},
]
\addplot [color=mycolor2]
  table[row sep=crcr]{%
8.04999999999998	290.73536\\
8.09999999999998	275.38912\\
8.14999999999998	348.07296\\
8.19999999999998	497.488\\
8.24999999999998	338.79776\\
8.29999999999998	338.79776\\
8.34999999999999	258.19584\\
8.39999999999999	338.9664\\
8.44999999999999	125.46816\\
8.49999999999999	123.61312\\
8.54999999999999	281.96608\\
8.59999999999999	319.40416\\
8.64999999999999	276.40096\\
8.69999999999999	242.50432\\
8.74999999999999	289.72352\\
8.79999999999999	276.5696\\
8.84999999999999	169.4832\\
8.89999999999999	168.30272\\
8.94999999999999	196.97152\\
8.99999999999999	279.27584\\
9.05	2.5376\\
9.1	317.0432\\
9.15	495.46432\\
9.2	428.3456\\
9.25	422.27456\\
9.3	401.02592\\
9.35	153.29376\\
9.4	191.74368\\
9.45	273.1968\\
9.5	253.12864\\
9.55	163.24352\\
9.6	148.57184\\
9.65	129.35488\\
9.7	123.61312\\
9.75000000000001	167.12224\\
9.80000000000001	179.77824\\
9.85000000000001	74.54\\
9.90000000000001	75.21344\\
9.95000000000001	364.43904\\
10	415.52896\\
10.05	354.48128\\
10.1	360.72096\\
10.15	414.0112\\
10.2	405.41056\\
10.25	289.38624\\
10.3	289.55488\\
10.35	348.2416\\
10.4	336.26816\\
10.45	412.83072\\
10.5	414.0112\\
10.55	272.3536\\
10.6	289.38624\\
10.65	438.80128\\
10.7	453.47296\\
10.75	77.5824\\
10.8	421.6\\
10.85	235.92736\\
10.9	412.3248\\
10.95	438.80128\\
11	518.23072\\
11.05	200.17568\\
11.1	198.82656\\
11.15	232.89184\\
11.2	265.94528\\
11.25	465.62304\\
11.3	503.55904\\
11.35	352.62624\\
11.4	352.4576\\
11.45	487.53824\\
11.5	452.96704\\
11.55	289.55488\\
11.6	245.87712\\
11.65	402.37504\\
11.7	414.17984\\
11.75	289.38624\\
11.8	289.55488\\
11.85	452.62976\\
11.9	424.29824\\
11.95	503.89632\\
};
\addlegendentry{Throughput [Mbits]};
\end{axis}

\end{tikzpicture}%
  	\caption{\gls{pdcp} throughput with \textit{Fast Switching}.}
  	\label{fig:dc-th}
\end{subfigure}
\begin{subfigure}[b]{\columnwidth}
	\centering
	\setlength\fwidth{0.9\columnwidth}
  	\setlength\fheight{0.5\columnwidth}
%
%
\definecolor{mycolor1}{rgb}{0.53003,0.74911,0.46611}%
\definecolor{mycolor2}{rgb}{0.20810,0.16630,0.52920}%
\begin{tikzpicture}
\tikzstyle{every node}=[font=\scriptsize]
\pgfplotsset{every x tick label/.append style={font=\tiny, yshift=0.5ex}}
\pgfplotsset{every y tick label/.append style={font=\tiny}}

\begin{axis}[%
width=0.9\fwidth,
height=0.4\fheight,
scale only axis,
xmin=8,
xmax=12,
separate axis lines,
every outer y axis line/.append style={mycolor1},
every y tick label/.append style={font=\color{mycolor1}\tiny},
every y tick/.append style={mycolor1},
ymin=0.9,
ymax=3.1,
ytick={1, 2, 3},
axis background/.style={fill=white},
title style={font=\bfseries},
legend style={legend cell align=left,align=left,draw=white!15!black,at={(0.01,0.03)},anchor=south west},
]
\addplot[const plot, color=mycolor1] table[row sep=crcr] {%
8		3\\
8.3333	3\\
8.9793	2\\
9.0353	3\\
9.6233	2\\
9.8218	3\\
9.88121	1\\
9.9403	2\\
10.7208	3\\
11.2373	2\\
12		2\\
};
\addlegendentry{CellId};

\end{axis}

\begin{axis}[%
width=0.9\fwidth,
height=0.4\fheight,
scale only axis,
xmin=8,
xmax=12,
every outer y axis line/.append style={mycolor2},
every y tick label/.append style={font=\color{mycolor2}\tiny},
every y tick/.append style={mycolor2},
ymin=0,
ymax=500,
ytick={         0,  250, 500},
axis x line*=bottom,
axis y line*=right,
legend style={legend cell align=left,align=left,draw=white!15!black,at={(0.99,0.03)},anchor=south east},
]
\addplot [color=mycolor2]
  table[row sep=crcr]{%
8.04999999999998	290.73536\\
8.09999999999998	275.7264\\
8.14999999999998	347.06112\\
8.19999999999998	497.488\\
8.24999999999998	338.79776\\
8.29999999999998	338.79776\\
8.34999999999999	45.37216\\
8.39999999999999	347.06608\\
8.44999999999999	93.42656\\
8.49999999999999	123.44448\\
8.54999999999999	281.12288\\
8.59999999999999	318.89824\\
8.64999999999999	276.23232\\
8.69999999999999	242.33568\\
8.74999999999999	289.04896\\
8.79999999999999	276.5696\\
8.84999999999999	169.4832\\
8.89999999999999	168.13408\\
8.94999999999999	197.14016\\
8.99999999999999	87.02624\\
9.05	0.1816\\
9.1	319.40912\\
9.15	495.46432\\
9.2	428.00832\\
9.25	423.2864\\
9.3	400.01408\\
9.35	152.28192\\
9.4	191.74368\\
9.45	273.70272\\
9.5	253.29728\\
9.55	162.7376\\
9.6	149.2464\\
9.65	18.05744\\
9.7	130.02144\\
9.75000000000001	166.9536\\
9.80000000000001	156.16864\\
9.85000000000001	1.01184\\
9.90000000000001	1.19488\\
9.95000000000001	0.18288\\
10	0.51088\\
10.05	506.7632\\
10.1	284.66432\\
10.15	338.9664\\
10.2	330.70304\\
10.25	214.1728\\
10.3	214.34144\\
10.35	272.01632\\
10.4	262.90976\\
10.45	337.61728\\
10.5	338.9664\\
10.55	195.6224\\
10.6	214.1728\\
10.65	363.58784\\
10.7	377.92224\\
10.75	2.54256\\
10.8	338.29184\\
10.85	159.70208\\
10.9	337.11136\\
10.95	363.58784\\
11	444.3664\\
11.05	124.96224\\
11.1	123.61312\\
11.15	157.6784\\
11.2	190.5632\\
11.25	156.50592\\
11.3	420.42448\\
11.35	276.90688\\
11.4	278.08736\\
11.45	409.28928\\
11.5	377.58496\\
11.55	214.1728\\
11.6	172.0128\\
11.65	326.99296\\
11.7	338.9664\\
11.75	214.1728\\
11.8	214.34144\\
11.85	377.41632\\
11.9	348.41024\\
11.95	430.032\\
};
\addlegendentry{Throughput [Mbits]}

\end{axis}

\end{tikzpicture}%
  	\caption{\gls{pdcp} throughput with \textit{Hard Handover}.}
  	\label{fig:hh-th}
\end{subfigure}
\caption{\gls{pdcp} throughput with multiple \gls{rat}s and \gls{enb}s. \gls{enb}s with CellId 2 and 3 are mmWave \gls{enb}s, while CellId 1 stands for the \gls{lte} \gls{enb} co-located with mmWave \gls{enb} 2.}
\label{fig:dcsim}
\end{figure}
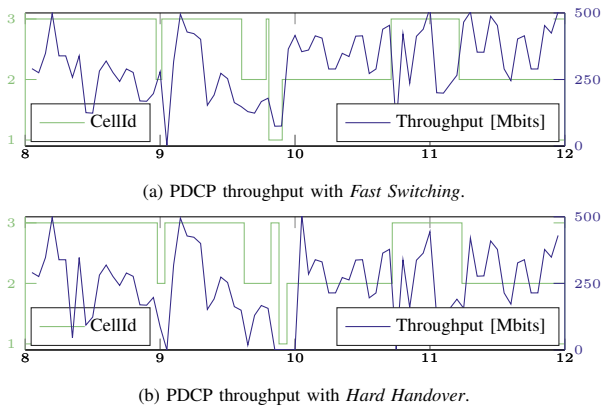

\subsection{\gls{lte}-aided Multi Connectivity}
Finally, a fourth use case regards the study of mobility management schemes for mmWave radio access networks. The example described in the following paragraphs uses the Dual Connectivity extension introduced in Section~\ref{sec:dc}.
Thanks to the \texttt{McUeNetDevice}, it is possible to simulate scenarios similar to that in Figure~\ref{fig:dcscenario}, with an \gls{lte} \gls{enb} and multiple mmWave \gls{enb}s under its coverage, either co-located, or connected via the X2 interface. The \gls{ue} can move with different patterns, according to the mobility models available in ns--3, and while moving experiences different channel conditions according to its \gls{los}/\gls{nlos} position and distance from the \gls{enb}s. These scenarios can be used to evaluate the performance of different mobility management and multi connectivity schemes.

An example can be found in the \texttt{mc-twoenbs.cc} file. The purpose of this example is to compare a system with \gls{dc} and fast switching among Radio Access Technologies (\gls{rat}s) and a stand-alone architecture, with the \gls{ue} connected to a single \gls{rat} (either \gls{lte} or mmWave) at any given time. The benefit of the first architecture is evident when the \gls{ue} connected to the mmWave \gls{rat} experiences an outage: thanks to dual connectivity, it can immediately recover the communication using the \gls{lte} link, while the stand-alone \gls{ue} must sense the link failure and then perform a handover to the \gls{lte} \gls{enb}. This takes more time and leads to a time interval in which the throughput that can be achieved at the \gls{pdcp} layer is zero. This is shown in Figure~\ref{fig:dcsim}. The green line represents the current cell to which the \gls{ue} is attached, and this information can be retrieved from the \texttt{CellIdStats.txt} and \texttt{CellIdStatsHandover.txt} files. The purple line, instead, is the instantaneous \gls{pdcp} throughput, sampled from the \texttt{DlPdcpStats.txt} trace.

Besides this specific example, more general results on dual connectivity for LTE and mmWave can be found in~\cite{poleseHo,simutoolsPolese}, where smart handover strategies are proposed and extensively evaluated with UDP as the transport protocol, and in~\cite{tcp-mobicom,jimenez2017analysis}, which instead use TCP.

\section{Integration with \gls{dce} and Examples}
\label{sec:dce}

\gls{dce} was introduced in~\cite{dce} as a powerful tool that combines the flexibility of a network simulator such as ns--3 with the robustness of the \gls{tcp}/IP stack of the Linux kernel and the authenticity of real applications. There are several benefits in using this tool. First, the Linux kernel implements protocols which are not yet available for ns--3, or which are in an early development phase and present some limitations. An example is \gls{mptcp}, the multipath extension of \gls{tcp} which makes it possible to transmit data on multiple subflows (i.e., a mobile user could simultaneously transmit on a Wi-Fi subflow and a cellular subflow)~\cite{rfc6182}. At the time of writing, it was implemented for ns--3 by different projects~\cite{chihani2011multipath,coudron2017implementation}, but none of them is completely compliant with the \gls{mptcp} specification, and they are not integrated in the main ns--3 release and validated. With \gls{dce}, instead, it is possible to use the \gls{mptcp} code developed and tested by the same \gls{mptcp} protocol designers~\cite{mptcpImpl}. Second, the Linux kernel \gls{tcp}/IP stack is the most widely used in real production environments and datacenters, besides being the basis for the Android mobile operating system. Therefore, it is a very well tested codebase, with very few bugs. Moreover, its usage in network simulations provides a higher level of realism. Finally, with \gls{dce} it is possible to use real POSIX socket-based applications. For example, the well known iPerf tool~\cite{iperf} can be used to measure the maximum achievable datarate in the network. It is also possible to simulate a website, with an http daemon in the server and wget as a client. Besides, standard ns--3 applications (\texttt{OnOffApplication}, \texttt{BulkSendApplication}) can be used with the Linux \gls{tcp}/IP stack thanks to \gls{dce} Cradle.

In order to integrate \gls{dce} with the ns--3 mmWave module, it is necessary to patch the \texttt{KernelFdSocketFactory} class so that it recognizes the \texttt{MmWaveUeNetDevice}. The patch can be found in the \texttt{utils} folder of the ns--3 mmWave module repository. Then, replace the standard ns--3 folder with our mmWave module. Notice that, if \gls{mptcp} is used as the transport protocol, the \gls{dc} extension must be used with the patch provided in the \texttt{utils} folder.


\textbf{\bf \gls{mptcp} on mmWave links:}
The latest Linux kernel implementation of \gls{mptcp} compatible with \gls{dce} can be found in the \texttt{net-next-nuse} library of the LibOS project~\cite{libos}. The standard \gls{dce} distribution already provides \gls{mptcp} examples, which can be promptly extended in order to account for mmWave and \gls{lte} subflows, as long as they operate on links with different carrier frequencies (i.e., it is possible to simulate an \gls{mptcp} connection on a 2.1 GHz \gls{lte} link and a 28 GHz mmWave link). It is possible to simulate different state of the art congestion control algorithms for \gls{mptcp}, either coupled or uncoupled, as shown in~\cite{polese2017mptcp, polese2017tcp}.

\begin{figure}[t!]
\centering
\begin{tikzpicture}[font=\sffamily, scale=0.4, every node/.style={scale=0.4}]
  \centering

    \node[anchor=south west,inner sep=0] (image) at (0,0) {\includegraphics[width=0.9\textwidth]{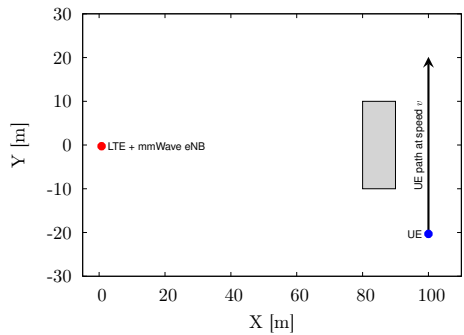}};
    \begin{scope}[x={(image.south east)},y={(image.north west)}]
        \filldraw[red,ultra thick] (0.2,0.55) circle (2pt);
        \node[anchor=west] at (0.205,0.55) (mm1label) {\gls{lte} + mmWave \gls{enb}};

        \draw[sarrow] (0.865, 0.295) -- node[sloped, anchor=south] {\gls{ue} path at speed $v$} (0.865, 0.81);

        \filldraw[blue,ultra thick] (0.865, 0.295) circle (2pt);
        \node[anchor=east] at (0.86, 0.295) (ltelabel) {\gls{ue}};
    \end{scope}
\end{tikzpicture}
\caption{Random realization of the simulation scenario. The grey rectangle represents a building.}
\label{fig:mptcpscenario}
\end{figure}

\begin{figure}[t]
\centering
\begin{subfigure}[t]{\columnwidth}
\setlength{\belowcaptionskip}{0cm}

	\centering
	\setlength\fwidth{0.85\columnwidth}
  	\setlength\fheight{0.45\columnwidth}
%
%
\begin{tikzpicture}
\tikzstyle{every node}=[font=\scriptsize]
\pgfplotsset{every x tick label/.append style={font=\tiny, yshift=0.5ex}}
\pgfplotsset{every y tick label/.append style={font=\tiny}}
\begin{axis}[%
width=0.9\fwidth,
height=0.4\fheight,
at={(0\fwidth,0\fheight)},
scale only axis,
xmin=2.5,
xmax=21,
xlabel style={font=\color{white!15!black}\tiny},
xlabel={Time [s]},
ymin=0,
ymax=,
ylabel style={font=\color{white!15!black}\tiny},
ylabel={Throughput [Gbit/s]},
axis background/.style={fill=white},
title style={font=\bfseries},
xmajorgrids,
ymajorgrids,
ylabel shift = -5 pt,
yticklabel shift = -2 pt,
legend style={at={(0.5,0.5)}, anchor=south, legend cell align=left, align=left, draw=white!15!black}
]
\addplot [color=blue, dotted]
  table[row sep=crcr]{%
2	3.456106496\\
2.5	0.698351616\\
3	1.795162112\\
3.5	1.39460608\\
4	1.262485504\\
4.5	1.39460608\\
5	1.7825792\\
5.5	1.528823808\\
6	0\\
6.5	0.694157312\\
7	0\\
7.5	0\\
8	0.696254464\\
8.5	0\\
9	0\\
9.5	1.390411776\\
10	0.696254464\\
10.5	0\\
11	0.694157312\\
11.5	0\\
12	0\\
12.5	0.696254464\\
13	0\\
13.5	1.176502272\\
14	0\\
14.5	0.696254464\\
15	0\\
15.5	0.694157312\\
16	0\\
16.5	0\\
17	0\\
17.5	1.757413376\\
18	0.694157312\\
18.5	0.698351616\\
19	1.461714944\\
19.5	0.696254464\\
20	1.68820736\\
20.5	1.396703232\\
21	1.891631104\\
21.5	0.696254464\\
22	0.790429861\\
};
\addlegendentry{iPerf BW}

\addplot [color=red]
  table[row sep=crcr]{%
2	0.04792704\\
2.25	0.075268224\\
2.5	0.044555328\\
2.75	0\\
3	0.03580768\\
3.25	0.062052992\\
3.5	0.065030592\\
3.75	0\\
4	0.045132096\\
4.25	0.069260224\\
4.5	0.044603392\\
4.75	0\\
5	0.06022656\\
5.25	0.065078656\\
5.5	0\\
5.75	0.064453824\\
6	0.075268224\\
6.25	0.075268224\\
6.5	0.075316288\\
6.75	0.07017344\\
7	0\\
7.25	0\\
7.5	0\\
7.75	0\\
8	0\\
8.25	0\\
8.5	0\\
8.75	0\\
9	0.03917216\\
9.25	0\\
9.5	0.067340032\\
9.75	0.075268224\\
10	0.075268224\\
10.25	0.075316288\\
10.5	0.075268224\\
10.75	0.063252224\\
11	0\\
11.25	0\\
11.5	0\\
11.75	0\\
12	0\\
13.5	0.032779648\\
13.75	0.073105344\\
14	0.075316288\\
14.25	0.075268224\\
14.5	0.075268224\\
14.75	0.075316288\\
15	0.060512576\\
17	0.039124096\\
17.75	0.037203904\\
18	0.075316288\\
18.25	0.075268224\\
19.25	0.05479296\\
19.5	0.068058624\\
20	0.041286976\\
20.25	0.06897184\\
20.5	0.049650112\\
21	0.04998656\\
21.25	0.07017344\\
21.75	0.03917216\\
};
\addlegendentry{LTE throughput}

\addplot [color=black, dashed]
  table[row sep=crcr]{%
2	2.617193664\\
2.25	2.873938816\\
2.5	0.824730176\\
2.75	0.477131328\\
3	0.717694016\\
3.25	1.95932896\\
3.5	2.226132224\\
3.75	0.705723712\\
4	0.879763456\\
4.25	2.691920448\\
4.5	1.11051872\\
4.75	0.779069376\\
5	1.7008912\\
5.25	2.14629792\\
5.5	0.728361856\\
5.75	0.505104576\\
6	0.75676768\\
6.25	0.750375168\\
6.5	0.640981504\\
6.75	0.845397696\\
7	0.231332032\\
7.25	0.273580288\\
7.5	0.173559104\\
7.75	0.181393536\\
8	0.17519328\\
8.25	0.224747264\\
8.5	0.160485696\\
8.75	0.176298752\\
9	0.267478528\\
9.25	0.307946048\\
9.5	0.78104\\
9.75	0.62507232\\
10	0.680442048\\
10.25	0.651411392\\
10.5	0.747058752\\
10.75	0.552303424\\
11	0.255364032\\
11.25	0.24224256\\
11.5	0.162792768\\
11.75	0.141212032\\
12	0.196629824\\
12.25	0.193745984\\
12.5	0.229842048\\
12.75	0.24080064\\
13	0.142702016\\
13.25	0.149142592\\
13.5	0.177452288\\
13.75	0.51043968\\
14	0.52077344\\
14.25	0.558407552\\
14.5	0.720623552\\
14.75	0.665061568\\
15	0.675635648\\
15.25	0.2331104\\
15.5	0.15404512\\
15.75	0.168512384\\
16	0.159284096\\
16.25	0.201003648\\
16.5	0.154189312\\
16.75	0.112421696\\
17	0.229027328\\
17.25	0.252384064\\
17.5	0.280789888\\
17.75	0.34437856\\
18	0.953637824\\
18.25	2.796411584\\
18.5	0.250076992\\
18.75	0.327940672\\
19	0.695822528\\
19.25	1.468691648\\
19.5	2.421752704\\
19.75	0.686257792\\
20	0.84280224\\
20.25	2.497068992\\
20.5	1.376601024\\
20.75	0.763929216\\
21	1.291383552\\
21.25	2.539028864\\
21.5	0.790124096\\
21.75	0.753979968\\
22	2.019216704\\
};
\addlegendentry{mmWave throughput}

\end{axis}
\end{tikzpicture}%
  	\caption{\gls{balia}}
  	\label{fig:balia-th}
\end{subfigure}
\begin{subfigure}[b]{\columnwidth}
\setlength{\belowcaptionskip}{0cm}

	\centering
	\setlength\fwidth{0.85\columnwidth}
  	\setlength\fheight{0.45\columnwidth}
%
%
\begin{tikzpicture}
\tikzstyle{every node}=[font=\scriptsize]
\pgfplotsset{every x tick label/.append style={font=\tiny, yshift=0.5ex}}
\pgfplotsset{every y tick label/.append style={font=\tiny}}
\begin{axis}[%
width=0.9\fwidth,
height=0.4\fheight,
at={(0\fwidth,0\fheight)},
scale only axis,
xmin=2.5,
xmax=21,
xlabel style={font=\color{white!15!black}\tiny},
xlabel={Time [s]},
ymin=0,
ymax=,
ylabel style={font=\color{white!15!black}\tiny},
ylabel={Throughput [Gbit/s]},
axis background/.style={fill=white},
title style={font=\bfseries},
xmajorgrids,
ymajorgrids,
ylabel shift = -5 pt,
yticklabel shift = -2 pt,
legend style={at={(0.5,0.5)}, anchor=south,legend cell align=left, align=left, draw=white!15!black}
]
\addplot [color=blue, dotted]
  table[row sep=crcr]{%
2	4.1418752\\
2.5	2.785017856\\
3	2.787115008\\
3.5	2.7787264\\
4	2.782920704\\
4.5	2.793406464\\
5	2.088763392\\
5.5	2.29638144\\
6	0.694157312\\
6.5	0.694157312\\
7	0.694157312\\
7.5	0.694157312\\
8	0.694157312\\
8.5	0.696254464\\
9	0.694157312\\
9.5	0.694157312\\
10	0.694157312\\
10.5	0.696254464\\
11	0.694157312\\
11.5	0.696254464\\
12	0.696254464\\
12.5	0.694157312\\
13	0.694157312\\
13.5	0.694157312\\
14	0.696254464\\
14.5	0.694157312\\
15	0.696254464\\
15.5	0.696254464\\
16	0.696254464\\
16.5	0.696254464\\
17	0.694157312\\
17.5	0.696254464\\
18	0\\
18.5	0.694157312\\
19	0\\
19.5	3.542089728\\
20	0.696254464\\
20.5	2.814377984\\
21	2.281701376\\
21.5	2.08666624\\
22	1.262104522\\
};
\addlegendentry{iPerf BW}

\addplot [color=red]
  table[row sep=crcr]{%
2	0.04792704\\
2.25	0.075268224\\
2.5	0.075268224\\
2.75	0.075316288\\
3	0.075270592\\
3.25	0.075268224\\
3.5	0.075316288\\
3.75	0.075268224\\
4	0.075268224\\
4.25	0.075316288\\
4.5	0.075268224\\
4.75	0.075268224\\
5	0.075318656\\
5.25	0.046670144\\
5.5	0.05599456\\
5.75	0.075268224\\
6	0.075316288\\
6.25	0.075268224\\
6.5	0.075268224\\
6.75	0.075316288\\
7	0.075268224\\
7.25	0.075268224\\
7.5	0.075316288\\
7.75	0.075268224\\
8	0.075268224\\
8.25	0.075316288\\
8.5	0.075268224\\
8.75	0.075268224\\
9	0.075318656\\
9.25	0.075268224\\
9.5	0.075268224\\
9.75	0.075316288\\
10	0.075268224\\
10.25	0.075268224\\
10.5	0.075316288\\
10.75	0.075268224\\
11	0.075268224\\
11.25	0.075268224\\
11.5	0.075316288\\
11.75	0.075268224\\
12	0.075268224\\
12.25	0.075316288\\
12.5	0.075268224\\
12.75	0.075268224\\
13	0.075316288\\
13.25	0.075268224\\
13.5	0.075268224\\
13.75	0.075316288\\
14	0.075268224\\
14.25	0.075268224\\
14.5	0.075316288\\
14.75	0.075268224\\
15	0.075268224\\
15.25	0.075316288\\
15.5	0.075268224\\
15.75	0.075268224\\
16	0.075316288\\
16.25	0.075268224\\
16.5	0.075268224\\
16.75	0.075316288\\
17	0.075268224\\
17.25	0.075268224\\
17.5	0.075318656\\
17.75	0.075268224\\
18	0.065030592\\
19.5	0.06392512\\
19.75	0.075316288\\
20	0.050563328\\
20.25	0.040662144\\
20.5	0.072288256\\
20.75	0.075268224\\
21.25	0.048160128\\
21.5	0.075316288\\
21.75	0.065318976\\
};
\addlegendentry{LTE throughput}

\addplot [color=black, dashed]
  table[row sep=crcr]{%
2	2.28358144\\
2.25	2.957666304\\
2.5	2.957666304\\
2.75	2.957666304\\
3	2.957668672\\
3.25	2.896432768\\
3.5	2.957666304\\
3.75	2.957666304\\
4	2.956224384\\
4.25	2.956080192\\
4.5	2.955311168\\
4.75	2.957666304\\
5	2.95752448\\
5.25	1.851521408\\
5.5	2.084487616\\
5.75	0.385521344\\
6	0.766764992\\
6.25	0.761862464\\
6.5	0.647614336\\
6.75	0.937296064\\
7	0.501451712\\
7.25	0.613440832\\
7.5	0.711251072\\
7.75	0.717931968\\
8	0.662946752\\
8.25	0.62987872\\
8.5	0.607192512\\
8.75	0.867603264\\
9	0.572105792\\
9.25	0.673859648\\
9.5	0.844099968\\
9.75	0.630888064\\
10	0.68467168\\
10.25	0.651843968\\
10.5	0.749942592\\
10.75	0.729467328\\
11	0.622572992\\
11.25	0.730140224\\
11.5	0.611085696\\
11.75	0.660735808\\
12	0.537884224\\
12.25	0.647998848\\
12.5	0.52365728\\
12.75	0.697120256\\
13	0.83294912\\
13.25	0.646316608\\
13.5	0.695918656\\
13.75	0.499769472\\
14	0.522071168\\
14.25	0.558551744\\
14.5	0.721969344\\
14.75	0.640404736\\
15	0.82405728\\
15.25	0.684046848\\
15.5	0.973247936\\
15.75	0.452138048\\
16	0.759795712\\
16.25	0.772244288\\
16.5	0.590033664\\
16.75	0.75748864\\
17	0.455021888\\
17.25	0.511739776\\
17.5	0.686786496\\
17.75	0.779165504\\
18	0.842850304\\
18.25	0.519956352\\
18.5	0.453628032\\
18.75	0.447908416\\
19	0.38427168\\
19.25	0.48352384\\
19.5	2.75502848\\
19.75	2.957666304\\
20	1.669887552\\
20.25	1.081007424\\
20.5	2.747194048\\
20.75	2.94704416\\
21	1.035010176\\
21.25	1.50824832\\
21.5	2.956560832\\
21.75	2.4380464\\
22	1.015303936\\
22.25	1.029819264\\
};
\addlegendentry{mmWave throughput}

\end{axis}
\end{tikzpicture}%
  	\caption{Uncoupled Cubic}
  	\label{fig:cubic-th}
\end{subfigure}
\caption{Throughput with \gls{mptcp}.}
\label{fig:mptcpsim}
\end{figure}
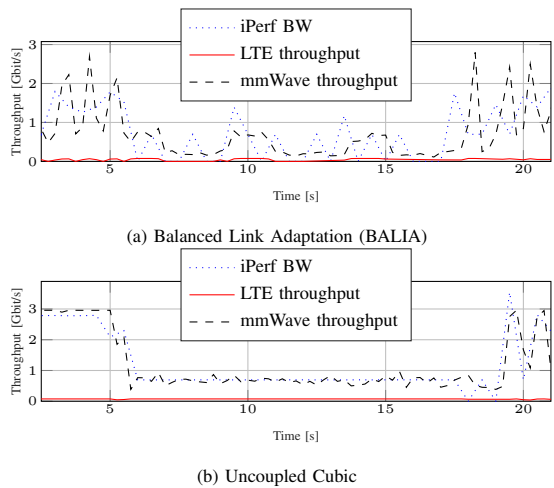

An example is in the file \texttt{dce-example-mptcp-mmwave}, which creates the scenario shown in Figure~\ref{fig:mptcpscenario}. The application used is iPerf, and the mobile device creates two uplink subflows to a remote server, the first on mmWave, and the second on \gls{lte}. The \gls{ue} moves along the y-axis and switches from a \gls{los} to a \gls{nlos} condition, and then returns to \gls{los}. Figure~\ref{fig:mptcpsim} shows the output of a simulation, with the \gls{tcp} throughput for two different congestion control algorithms for \gls{mptcp}, together with the per-subflow Radio Access Network throughput. In particular, Figure~\ref{fig:balia-th} shows the performance of \gls{balia}~\cite{balia}, which is a coupled congestion control algorithm, i.e., it tries to adapt the congestion window of each \gls{mptcp} subflow according to the congestion experienced on all links. In Figure~\ref{fig:cubic-th}, instead, the congestion controls on the \gls{lte} and on the mmWave subflows are uncoupled, i.e., each subflow is independent, and \gls{tcp} Cubic is used. The first observation is that the \gls{lte} subflow has, as expected, a much smaller throughput compared to the mmWave subflow, and thus the total throughput measured with iPerf is similar to that of the mmWave connection. The second is that the uncoupled solution manages to reach a more stable throughput in \gls{nlos} conditions, compared to the coupled solution, as was observed in~\cite{polese2017mptcp, polese2017tcp}, showing that the current coupled congestion control algorithms are not well suited for a deployment over these kinds of links.

\section{Potential uses and future extensions}
\label{sec:uses}
Given the full-stack nature of the simulation framework introduced in this paper, the 5G mmWave research community can leverage this tool to bring and test innovation at every layer. Each module can be easily extended while maintaining full backward compatibility. The fundamental components are in the form of functions, classes and modules, which can be implemented to design novel algorithms, procedures, and, more in general, architectures. For example, the \emph{scheduling} and \emph{allocation} strategies proposed in \cite{7929424,7841752,hotmobile17} can be readily integrated and tested in our framework with some simple tweaks. Similarly, due to the importance of coping with \emph{mobility} and frequent \emph{handovers}, innovative approaches like the one proposed in \cite{DBLP:journals/corr/SemiariSB17}, which exploits \emph{caching}, can take advantage of the modular structure of the ns--3 framework to test flexible and reprogrammable logics. Additionally, as previously mentioned, several papers already fully exploit the capabilities of this simulator to capture the performance of \emph{\gls{tcp}} in mmWave networks, and propose some novel approaches to mitigate the limiting effects of congestion control procedures with intermittent multi-Gbps mmWave links \cite{tcp-asilomar,tcp-mobicom,kim2017enhancing,mengleiBufferbloat,zhang2016transport}.

As part of our \textbf{future work}, we aim at expanding the code to include additional components such as:
\begin{itemize}
\item \gls{3gpp}-inspired \emph{signaling/beamtracking} procedures to better accommodate novel techniques like those proposed in \cite{DBLP:journals/corr/BeltranDW16,hotmobile17};
\item multi connectivity based on Carrier Aggregation~\cite{khan2014carrier};
\item novel applications such as \emph{virtual \& augmented reality}, to ultimately test  key 5G metrics as done in \cite{Hou2017}, where the authors leverage our mmWave module to run a performance analysis of traditional video delivery over mmWaves, and in~\cite{drago2018reliable}, where ns-3 mmWave is used to assess the performance of network coding and multi connectivity for reliable video streaming over mmWave;
\item \emph{multi-hop} architectures for both the access and the backhaul links, as presented in \cite{Drozdy2017};
\item \emph{vehicular} channel and traffic models to test and capture the end-to-end performance of mmWave communications for high-mobility scenarios     \cite{DBLP:journals/corr/TassiEPN17,DBLP:journals/corr/MavromatisTPN17a,DBLP:journals/corr/MavromatisTPN17}; 
\item \emph{public safety} scenarios, including aerial communications and robotics, where the propagation environments and the performance requirements differ from those of traditional cellular networks, as detailed in \cite{access-critical}.
\end{itemize}

Moreover, we plan to address the challenge of \emph{scalability}. 5G networks will likely comprise a large number of nodes, with high mobility, and thus channel states must be updated frequently. In addition, the use of low-latency applications requires that packet timelines must be scheduled at very fast interaction times. To do so, we will explore the following two design options:
\begin{itemize}
\item \emph{Low-rank channel modeling:} We will develop
computationally simple low-rank models that approximate the end-to-end phased
array system well.  
\item \emph{Migration to cluster computing:} Further scaling will be achieved
by investigating the deployment of the simulator onto open-source large cluster
platforms such as Amazon Web Services (AWS) \cite{aws}. 
\end{itemize}

Finally, we plan to officially merge our mmWave module into the main ns--3 release, in order to regularly maintain its compatibility with every new update.  

\section{Conclusions}
\label{sec:conclusions}
In this tutorial paper, we have presented the current status of the ns--3 framework for simulation of mmWave cellular systems. The code, which is publicly available at GitHub \cite{github-ns3-mmwave}, is highly modular and customizable to allow researchers to test novel \gls{5g} protocols.
We have shown some performance trends based on the mmWave channel models available. A detailed explanation of our configurable physical and \gls{mac} layers is provided, along with a corroborating set of simulation results for varying configurations. Implementations of advanced \gls{5g} architectural features, such as dual connectivity, are also available, and we have reported different representative results. We have also shown that the module can be interfaced with the higher-layer protocols and core network models from the ns--3 \gls{lte} module to enable full-stack simulations of end-to-end connectivity, along with the simulation of real applications through the implementation of direct code execution. The module is demonstrated through several example simulations showing the performance of our custom mmWave stack as well as custom congestion control algorithms, specifically designed for efficient utilization of the mmWave channel.

\renewcommand{\arraystretch}{1}
\setlength{\glsdescwidth}{0.75\columnwidth}
\printglossary[style=index]

\bibliographystyle{IEEEtran}
\bibliography{bibl}

\end{document}